\documentclass[epjc3,preprint]{svjour3}  
\RequirePackage{graphicx}
\RequirePackage[T1]{fontenc}

\RequirePackage{mathptmx}      
\RequirePackage{flushend}
\RequirePackage[numbers,sort&compress]{natbib}
\RequirePackage[colorlinks,citecolor=blue,urlcolor=blue,linkcolor=blue]{hyperref}
\def\makeheadbox{{%
\hbox to0pt{\vbox{\baselineskip=10dd\hrule\hbox
to\hsize{\vbox{}\hfil\vrule}\hrule}%
\hss}}}
\usepackage{mathtools}
\usepackage{mathbbol}
\usepackage{comment}
\usepackage{feynmp-auto}
\usepackage{epsfig}
\usepackage{xcolor}
\usepackage{natbib}
\usepackage{graphicx}
\usepackage{xifthen}
\usepackage{bm}
\usepackage{latexsym}
\usepackage[symbol]{footmisc}
\usepackage{siunitx}

\newcommand{\norm}[1]{\left\lVert#1\right\rVert}
\renewcommand{\vec}[1]{\bm#1}

\newcommand{\ket}[1]{\left| #1 \right\rangle}
\newcommand{\bra}[1]{\left\langle #1 \right|}
\newcommand{\braket}[2]{\langle #1 \vert #2 \rangle}

\newcommand{\rr} {\boldsymbol{r}}
\def\Xint#1{\mathchoice
   {\XXint\displaystyle\textstyle{#1}}%
   {\XXint\textstyle\scriptstyle{#1}}%
   {\XXint\scriptstyle\scriptscriptstyle{#1}}%
   {\XXint\scriptscriptstyle\scriptscriptstyle{#1}}%
   \!\int}
\def\XXint#1#2#3{{\setbox0=\hbox{$#1{#2#3}{\int}$}
     \vcenter{\hbox{$#2#3$}}\kern-.5\wd0}}

\def\dashint{\Xint-}

\newcommand{\eg}{{\textit{e.g., }}}
\newcommand{\ie}{{\textit{i.e., }}}

\begin{document}

\title{Pushing the Limits of the Periodic Table -- A Review on Atomic Relativistic Electronic Structure Theory and Calculations for the Superheavy Elements$^*$}
\dedication{In memoriam to two of the pioneers in this field, Jean-Paul Desclaux (Grenoble) and Sigurd Hofmann (Darmstadt)}

\author{O. R. Smits\thanksref{e3,addr3}    
        \and
P. Indelicato\thanksref{e1,addr1}
        \and
W. Nazarewicz\thanksref{e2,addr2}
        \and \\
M. Piibeleht\thanksref{addr3}
        \and
P. Schwerdtfeger\thanksref{e4,addr3}}
\thankstext{e3}{e-mail: smits.odile.rosette@gmail.com}
\thankstext{e1}{e-mail: paul.indelicato@lkb.upmc.fr}
\thankstext{e2}{e-mail: witek@frib.msu.edu}
\thankstext{e4}{e-mail: peter.schwerdtfeger@gmail.com}

\institute{
Centre for Theoretical Chemistry and Physics, The New Zealand Institute for Advanced Study, Massey University Auckland, 0632 Auckland, New Zealand\label{addr3}
\and
Laboratoire Kastler Brossel, Sorbonne Universit\'e, CNRS, ENS-PSL Research University, Coll\`ege de France, Case\ 74;\ 4, place Jussieu, F-75005 Paris, France \label{addr1}
\and
Facility for Rare Isotope Beams and Department of Physics and Astronomy, Michigan State University, East Lansing, Michigan 48824, USA \label{addr2}
}

\date{Received: date / Accepted: date}

\maketitle
\tableofcontents
\vspace{1cm}

\begin{abstract}
  We review the progress in atomic structure theory with a focus on superheavy elements and the aim to predict their ground state configuration and element's placement in the periodic table. To understand the electronic structure and correlations in the regime of large atomic numbers, it is important to correctly solve the Dirac equation in strong Coulomb fields, and also to take into account quantum electrodynamic effects. We specifically focus on the  fundamental difficulties encountered when dealing with the many-particle Dirac equation. We further discuss the possibility for future many-electron atomic structure calculations going beyond the critical nuclear charge \(Z_{\rm crit}\approx 170\), where levels such as the \(1s\) shell dive into the negative energy continuum (\(E_{n\kappa}<-m_ec^2\)). The nature of the resulting Gamow states within a rigged Hilbert space formalism is highlighted.
\end{abstract}

\section{Introduction}\label{sec:Intro}
\begin{figure}[b!]
  \begin{center}
\includegraphics[width=1.0\textwidth]{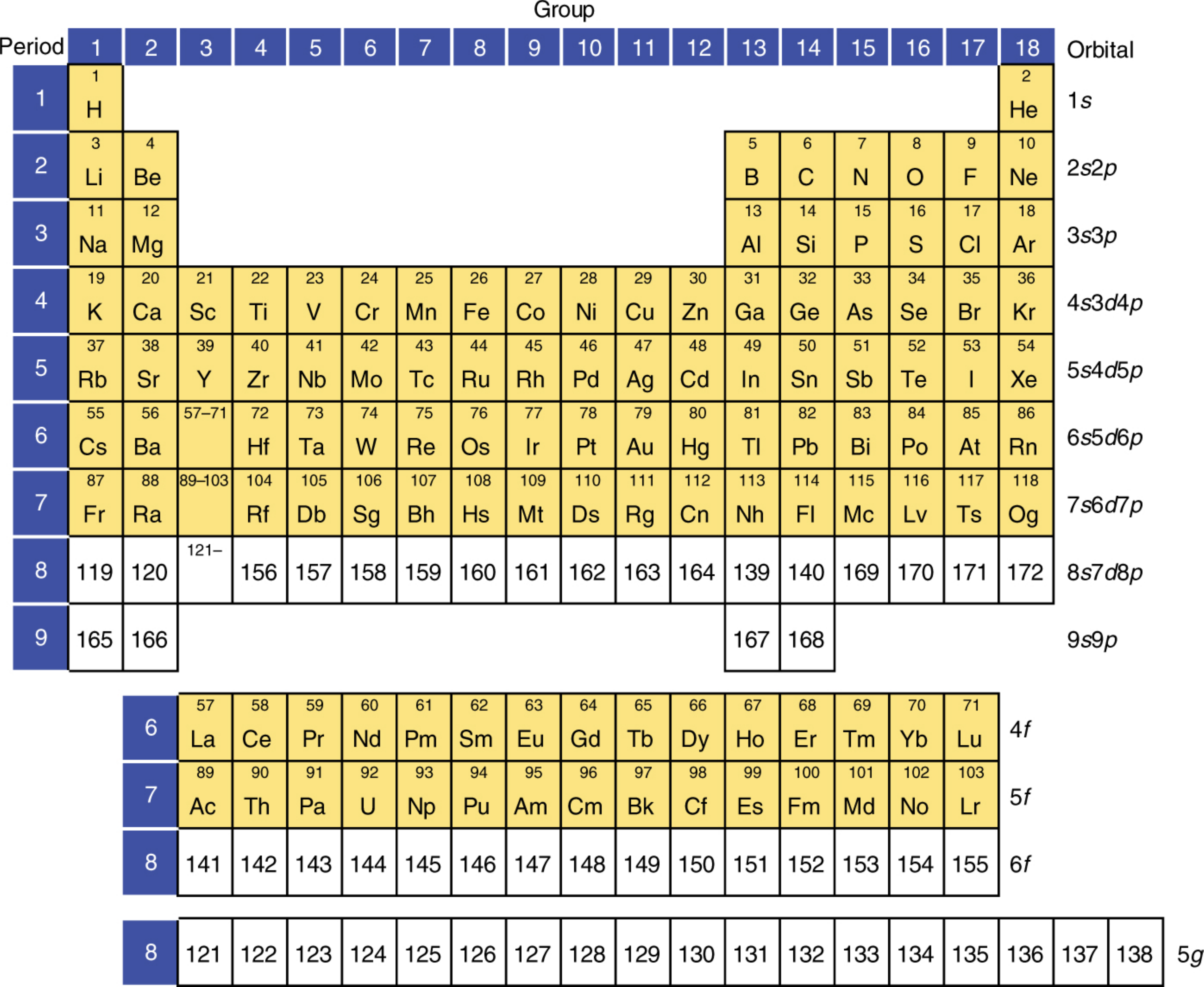}
  \caption{Pyykk\"o's periodic table extended to $Z=172$ (with permission from PCCP \cite{pyykko2011PT}).}
  \label{fig:PT}
  \end{center}
  \end{figure}

The periodic table (PT) of the elements, introduced by Dmitri Mendeleev and Lothar Meyer, is based on the Pauli and Aufbau (building-up) principle \cite{Schwerdtfeger2020}. Arguably, the PT is the most important and useful tool concerning the electronic structure of atoms and molecules \cite{Scerri2012periodic,Pyykko2012PT,Cao2021,SchwarzPT2022}. Chemical and physical similarities between the elements within a group or period obtained from their measurable properties is often hailed as a building block of the PT, but these patterns also follow from the underlying electronic shell structure of the atoms. Despite many controversies concerning the PT, for example, the starting and ending points of the $f$-block elements, the placement of the lightest elements hydrogen and helium, observed anomalies in chemical behavior or even the shape and visual representation \cite{restrepochallenges2019,Shaik2019,Steinhauser2019,Cao2021}, it is still going strong after 150 years. Furthermore, with the  nuclear synthesis of the $7p$ block elements up to oganesson with nuclear charge $Z=118$ \cite{Oganessian_2007,oganessian2011synthesis}, the full  $7^{\textrm{th}}$ period of the PT is now complete. Hence, what remains to be solved is how the PT can successfully be extended both theoretically and experimentally into the superheavy element region beyond $Z=118$ \cite{Fricke1971,fricke1976chemical,scerri2013cracks,ibj2011,Pyykkoe2019}. A progress in this direction has been made by placing the unknown elements up to nuclear charge $Z=172$ into the Periodic Table \cite{Nefedov2006,pyykko2011PT}, see for example Fig.~\ref{fig:PT}.

The existence and properties of new superheavy elements beyond oganesson depends on
 both nuclear and electronic structure properties \cite{giuliani2018}. There are, however, a number of open questions and major challenges to both electronic and nuclear structure theory concerning the accurate prediction of physical and chemical properties of the superheavy elements.\footnote{Here we define the starting point of the superheavy element region at the transactinides, $Z\ge 103$} For example, to correctly place an element into the PT and predict its basic properties, one should gain knowledge of its atomic shell structure, such as ground and excited electronic states and underlying dominant configurations  \cite{Fricke1971,fricke1976chemical}. In the case of dense spectra, which are prominent in open-shell systems as well as in the superheavy element region where high principal quantum number and angular momentum states are occupied, detailed knowledge of low-lying excited electronic states are required within a window of a few \si{\electronvolt}. This is often a very challenging task as both relativistic and electron correlation effects play a major role requiring sophisticated multi-reference methods at the relativistic Dirac-Coulomb-Breit level of theory. Currently, the heaviest element for which it is possible to compare theory and  experiment is lawrencium ($Z=103$) \cite{Sato2015,Sato2018}.

Moreover, the Dirac-Coulomb Hamiltonian has its limits in strong Coulomb fields as beyond the critical nuclear charge of $Z_{\mathrm{crit}}\approx 170$ for finite-size nuclei, the $1s$ electron level dives into the negative energy continuum below $E=-m_ec^2$  \cite{Pomeranchuk1945,Reinhard1971,Zeldovich_1972,Muller1972,Popov1974,Reinhardt-1977,Reinhardt1981,greinerrafelski1985,thaller1992,Gitman_2013,Schwerdtfeger2015,shabaev2019qed}. At the single-particle level of theory,  the correct description and interpretation of the resulting resonances can be given in terms of Gamow states \cite{Gamow1928,Gamow1929,Siegert1939,bohm1989,Civitares2004}, but how such diving states can correctly and accurately be described within a multi-electron framework, and how the PT can be  extended beyond the critical nuclear charge, are open questions.
  
At high nuclear charge, the PT is ultimately limited by the nuclear stability, not by its electronic shell structure \cite{Nazarewicz2018,giuliani2018}.
For nuclear structure theory and corresponding predictions of nuclear stability of isotopes see for example Refs.~\cite{Nazarewicz_2016Challenges, Nazarewicz2018,giuliani2018} and references therein.
Here we focus solely on the discussion of relativistic electronic structure theory in the superheavy element region \cite{isbd2007,ibj2011,Lackenby2018,lackenby2019Ds}. 

 The outline of this Review is as follows. We first discuss the Dirac equation and its  peculiarities compared to the non-relativistic Schr{\"o}dinger equation, specifically for electrons in strong Coulomb fields. We discuss the critical nuclear charge in detail to clarify the region of validity of the Dirac-Coulomb Hamiltonian and discuss how states embedded in the negative energy continuum should be interpreted. The process of spontaneous pair creation in a supercritical field is analyzed including most recent references.
 The importance of quantum electrodynamics (QED) effects and how these can be treated in strong Coulomb fields is outlined. 
 The major problem of correctly describing electron correlation for the accurate prediction of electronic spectra in the superheavy element region is addressed. We review the current status of electronic structure calculations for the transactinides and discuss the placement of the elements beyond oganesson into the PT based on quantum theoretical predictions. The literature on this topic is vast \cite{BetheSalpeter1951,greiner2000relativistic,grant2007relativistic,greinerrafelski1985}, including a rigorous mathematical treatment of the Dirac equation and its generalizations \cite{richtmyer1978principles,bagrov1990exact,thaller1992,gitman2012self,sargsjan2012sturm,bagrov2014dirac}.

\section{The Dirac Equation in Strong Coulomb Fields}
\label{sec:dirac-strong-field}
\subsection{The QED Lagrangian}

Electronic structure theory is based on the QED sector of the Standard Model of particle physics.
Within the Standard Model, electrons are spin-1/2 Dirac fermions, and their dynamics is described by the QED Lagrangian density
\begin{align}
\begin{aligned}
\label{eq:qed-L}
  \mathcal{L}_{\rm QED} &=& i\hbar c\bar\psi(x)\gamma^\mu\partial_\mu\psi(x)
  -m_ec^2\bar\psi(x)\psi(x) \\
  &&-\frac{1}{4} F_{\mu\nu} F^{\mu\nu}
  -e\bar\psi(x)\gamma^\mu A_\mu(x)\psi(x),
\end{aligned}
\end{align}
where $\psi(x)$ is the field operator and \(\gamma^\mu\) are the Dirac matrices.
The first two terms in \eqref{eq:qed-L} are the kinetic and mass terms describing the  free electrons with mass $m_e$, whereas the third term describes the photon field \(A^\mu = (\phi, \vec{A})\), corresponding to the electromagnetic scalar and vector potentials (with \(F_{\mu\nu} = \partial_\mu A_\nu - \partial_\nu A_\mu\)).
The last term corresponds to the interaction between electrons and photons, with the elementary charge \(e\) acting as the coupling constant. The  interaction picture
represented by Eq.~\eqref{eq:qed-L} has been  extensively used in quantum field theory and it has been demonstrated to work to astonishingly high accuracy.

It would be highly desirable to treat the QED Lagrangian for a many-electron system in an external Coulomb field to avoid divergencies that appear in perturbative treatments  \cite{Magnifico2021}. Such a direct treatment could in principle be performed through lattice gauge theory which is mathematically well defined  \cite{dyson1952,Heinzl2021}.  However, the long-range nature of the Coulomb potential, related to the zero rest-mass of the photon, currently prevents any accurate computational treatment using lattice gauge theory in finite boxes \cite{Kogut1987}. Treating the required large boxes is currently computationally too demanding.
However, progress in this field has recently been made on the nuclear length scale.  For instance, a combined lattice QCD+QED approach  has been used to successfully calculate hadron and meson mass differences, such as the proton-to-neutron mass splitting, and its dependence on both the strong and electromagnetic coupling constants  \cite{borsanyi2015,Sinclair2021}.

\subsection{The Many-Electron Dirac-Coulomb-Breit Hamiltonian}
Atomic physics calculations are performed in the Hamiltonian formalism derived from the Langrangian  \eqref{eq:qed-L} by a Legendre transformation~\cite{fdw1972,des1973,Sucher1980,fischer2016}. The resulting first-quantized $N$-particle Hamiltonian can be written in atomic units (\ie $\hbar=1,e=1,m_e=1$) as
 \cite{grant1983,grant2007relativistic,johnson2007book}:
\begin{align}
\begin{aligned}
\label{eq:QEDHamiltonian}
H_{\rm D}&=\sum_{k=1}^N h_k+\sum_{k<l}^N V^{\mathrm{ee}}\left(r_{kl}\right)+H_{\rm QED}+H_{\rm other},\\
h_k&= -i c\vec{\alpha}_k\cdot\vec{\nabla}_k + \beta_k m_ec^2+V(r_k),\\
\end{aligned}
\end{align}
where $\vec{\alpha} = \gamma^0\vec{\gamma}$, $\beta = \gamma^0$, $r_{kl}=|r_k-r_l|$ is the inter-electronic distance, and
$h_k$ is  the single-particle Dirac Hamiltonian with an external potential $V(r)$, which can  be the physical nuclear potential (accounting for the finite extent of atomic nuclei), or an effective potential also including electron screening, providing a better starting point for perturbative calculations~\cite{cheng2008}. The full electron-electron interaction $V^{\mathrm{ee}}\left(r_{kl}\right)$, as derived from QED, will be discussed in Sec. \ref{subsubsec:two-elec-qed}.
The electron-electron interaction is often approximated by 
\begin{equation}
   V^{\mathrm{ee}}\left(r_{kl}\right)=\frac{1}{r_{kl}}-\frac{1}{2r_{kl}}\left[ \vec{\alpha}_k\cdot\vec{\alpha}_l + \frac{(\vec{\alpha}_k\cdot\vec{r}_{kl})(\vec{\alpha}_l\cdot\vec{r}_{kl})}{r_{kl}^2}\right],
\label{eq:Breit}  
\end{equation}
were the first term is the classical Coulomb interaction, which is the dominant contribution. The frequency-independent Breit interaction (second term) contains magnetic interactions and retardation effects up to order $1/c^2$ and is an important correction to the fine structure in atoms. Together, Eqs. \eqref{eq:QEDHamiltonian} and \eqref{eq:Breit} form the Dirac-Coulomb-Breit Hamiltonian, the starting point of most applications in relativistic electronic structure theory.
The importance of the effect of the Breit contribution to the $1s$ shell energy of superheavy elements has been pointed out quite early \cite{ind1986}. QED effects, represented by $H_{\rm QED}$, which are the focus of Sec.~\ref{sec:qed}, are often included using effective Hamiltonians~\cite{igd1987,iad1990,Flambaum2005,ShabaevTupitsyn2013,pyykkoe2003}.
The Hamiltonian may also include additional terms,
represented by $H_{\rm other}$, such as the ones arising for example from the hyperfine structure \cite{johnson2007book,pilkuhn2008book}, the nucleus-electron Breit term \cite{HardekopfSucher1985}, or from weak interactions \cite{greiner1996weak}.

It is worth mentioning that the  Hamiltonian \eqref{eq:QEDHamiltonian} is not Lorentz invariant, as the Breit operator accounts for magnetic interactions and retardation effects only to order $1/c^2$ \cite{Mourad_1995}.
However, the corresponding deviations are supposed to be small compared to other sources of errors, such as from the approximate treatment of electron correlation~\cite{Gorceix1988,Pasteka2017}. For inner shells, the all-order retardation contribution may not be negligible. Including the Breit operator in a self-consistent process to obtain its contribution to all-orders can also have a strong effect \cite{ind1995}.

\begin{figure}[t]
  \begin{center}
\includegraphics[width=0.47\textwidth]{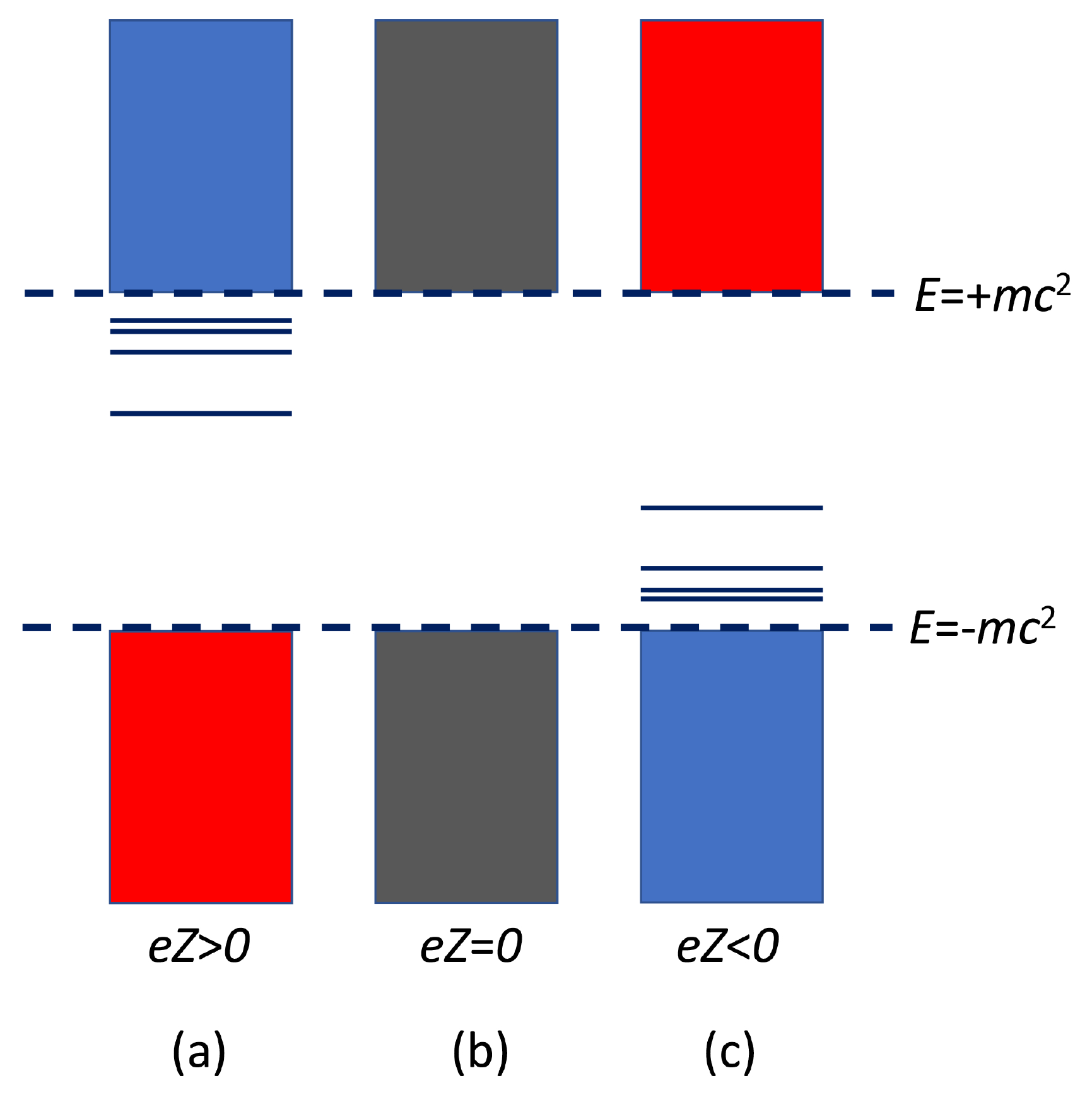}
  \caption{Schematic spectrum of the one-particle Dirac operator with potential (in SI units) $V(r)=-\frac{e^2}{4\pi\epsilon_0}\frac{Z}{r}$ showing the discrete $\{\phi^d\}$ and positive $\{\phi^c_+\}$ and negative $\{\phi^c_-\}$ energy continuum states. (a) negatively charged particle of mass $m$ in a Coulomb potential with $eZ>0$, (b) free particle ($eZ$=0), and (c) the charge conjugated case of an antiparticle of charge $+e$ and mass $m$ in a Coulomb potential with $eZ<0$.}
  \label{fig:DS}
  \end{center}
  \end{figure}

In order to describe electrons in the field of high nuclear charges, one first requires a detailed understanding of the spectrum of the Dirac or Dirac-Coulomb-Breit Hamiltonian in strong Coulomb fields~\cite{Reinhard1971,Muller1972,Reinhardt-1977,greinerrafelski1985}.
One of the major differences between the (many-particle) Dirac operator and its non-relativistic counterpart, is that the Dirac operator is not bounded from below and features a continuum of negative-energy  states, as shown in Fig.~\ref{fig:DS}. This gives rise to difficulties with variational approaches that have plagued the atomic physics and quantum chemistry communities for a long time~\cite{Brown1951,Wallmeier1982,Kutzelnigg1984,Brown_1987,ind2013,hllm1986,gra1987,lhlm1987,Dolbeault2000}.
This is now seen, however, as more of a technical problem than a fundamental one\footnote{We distinguish between problems of fundamental nature as those where knowledge to solve a particular problem is not yet available (such as problems involving physics beyond the standard model, the foundation of quantum field theory and Haag's theorem \cite{haag1955}, etc.) and those where knowledge is in principle available but the solution of the problem can be very hard to obtain (such as electron correlation and QED to all orders) or can be solved based on existing theory (such as resonant states embedded in the scattering continuum).} 
discussed in more detail in Sec.~\ref{sec:MCDHF}.

\subsection{The one-particle Dirac equation}
\label{subsec:one-part-dirac}
In order to solve the many-electron problem, one must first understand the  single-particle case. Thus, in the following,  we  consider the stationary Dirac equation for a single particle.

\subsubsection{Point nucleus and self-adjointness}

In strong Coulomb fields, a difficulty arises for the Dirac equation modelled with a point nuclear charge (PNC). To illustrate this, it suffices to consider the radial form of the one-particle Dirac-Coulomb equation:
\begin{equation}
\label{eq:Dirac}
\left(
\begin{array}{cc}
m_ec^2+V(r)-E_{n\kappa} & c\left( -\frac{d}{dr}+\frac{\kappa}{r}\right)\\
c\left( \frac{d}{dr}+\frac{\kappa}{r}\right) & -m_ec^2+V(r)-E_{n\kappa}
\end{array}
\right)
\left(
\begin{array}{cc}
P_{n\kappa}(r) \\
Q_{n\kappa}(r)
\end{array}
\right)
=0,
\end{equation}
with the corresponding four-component orbital spinor
\begin{equation}
\label{eq:orbital}
\psi_{n\kappa\mu}(r)=\frac{1}{r}
\left[
\begin{matrix}
P_{n\kappa}(r) \chi_{\kappa\mu}(\theta,\phi)\\
iQ_{n\kappa}(r) \chi_{-\kappa\mu}(\theta,\phi)
\end{matrix}
\right],
\end{equation}
where $\kappa=\pm(j+\frac{1}{2})$ for $j=\ell\mp\frac{1}{2}$.
The bound state eigenvalues for the {\it point} nuclear charge, $V(r)=-Z/r$ are \cite{Darwin1928,Darwin1928a,Gordon1928}
\begin{equation}
E_{n\kappa}(Z) = m_ec^2\left(1+\frac{(Z\alpha)^2}{\left[n-|\kappa| +\sqrt{\kappa^2 -(Z\alpha)^2}\right]^2}\right)^{-1/2},
\label{Hlike}
\end{equation}
where $\alpha$ is the fine-structure constant ($\alpha^{-1}=137.035999206(11)$ \cite{Morel2020}). The solution (\ref{Hlike}) is known as the Sommerfeld fine-structure formula \cite{sommerfeld1916}. A historical overview is given in Weinberg's book on the quantum theory of fields \cite{weinberg1995}.

\begin{figure}[tb!]
  \begin{center}
\includegraphics[width=0.8\textwidth]{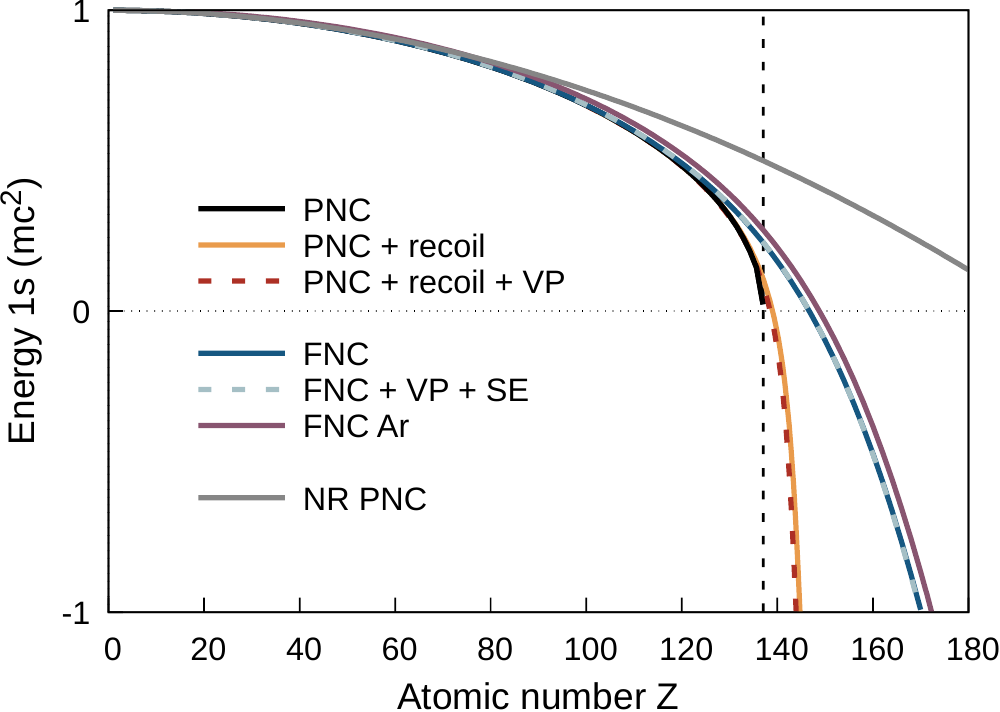}
  \caption{Nuclear charge dependence of the $1s$ energy levels for hydrogen-like atoms at various levels of theory using the Dirac equation. If not otherwise stated the results are from Ref.~\cite{Schwerdtfeger2015}. 
  The models considered are : PNC - point nuclear charge; FNC - finite nuclear charge distribution; recoil - nuclear recoil effects according to Eq.~(\ref{recoil}) \cite{Aleksandrov2016}; recoil+VP - includes the Uehling vacuum polarization term \cite{Aleksandrov2016}; VP+SE - includes major QED corrections from vacuum polarization and self-energy; NR - nonrelativistic results. Results are also shown for the Ar-like system with FNC.}
  \label{fig:diving}
  \end{center}
  \end{figure}

It is  apparent that a problem occurs when $Z > Z_{\mathrm{c}_\mathrm{p}} = |\kappa |/\alpha$, as $E_{n\kappa}(Z)$ becomes imaginary \cite{Schiff1940}. The range of such large $Z$-values  is usually referred to as the \textit{critical nuclear charge region}.  At the onset of the imaginary solutions, Eq.~(\ref{Hlike}) simplifies to
\begin{equation}
E_{n,\kappa}(Z_{\mathrm{c}_\mathrm{p}})=m_ec^2(n-|\kappa|)\left( n^2-2|\kappa|n+2\kappa^2 \right)^{-1/2} \ge 0,
\label{Hlike1}
\end{equation}
and one obtains $E_{1,-1}=0$ for $1s$, $E_{2,-1}=E_{2,1}=m_ec^2/\sqrt{2}$ for $2s$ and  $2p_{1/2}$ at $Z_{\mathrm{c}_\mathrm{p}}=1/\alpha\simeq 137.036$, and $E_{2,-2}=0$ for $2p_{3/2}$ at $Z_{\mathrm{c}_\mathrm{p}}=2/\alpha\simeq 274.072$. The difference between the behaviour of the nonrelativistic and relativistic $1s$ energies with increasing nuclear charge is shown in Fig.~\ref{fig:diving}.

The presence of the critical charge distinguishes the Dirac equation from the standard Schr{\"o}dinger equation with a Coulomb potential of a point nuclear charge, where all values $Z\ge 1$ are allowed, although one would run into similar problems with the Schr{\"o}dinger equation for potentials of the form $V(r)=-Z/r^n$ with $n\ge 2$ \cite{alliluev1972}. 

To treat atoms with nuclear charges beyond a certain critical charge, $Z>Z_\mathrm{nsa}$, where the Dirac operator becomes non-self-adjoint (nsa), one has to carefully choose an appropriate self-adjoint extension to the basic Dirac-Coulomb operator together with the correct operator domain \cite{Schmincke1972, sch1972a,Hogreve_2012, Gitman_2013, Gallone2017, Case1950, richtmyer1978principles, thaller1992}.  For example, this can be done by adding additional operators such as the nuclear recoil and Uehling terms, discussed in section \ref{sec:recoil}, or by removing the problematic singularity in the Coulomb term at zero by working with a realistic finite-size nuclear charge distribution to regularize the Coulomb interaction.
The mathematical problem arises due to the singularity of the Coulomb operator $-Z/r$ at the origin. As a result, the Dirac operator is not (essentially) self-adjoint anymore in the critical nuclear charge region. In fact, $H_{\rm D}$ becomes non-self-adjoint \cite{Hogreve_2012} for a $j$-state at $Z\ge Z_{\mathrm{nsa}}=\sqrt{j(j+1)}/\alpha$. For the $1s$ level this corresponds to $Z\ge\sqrt{3}/(2  \alpha) \simeq 118.677$ \cite{Schmincke1972,sch1972a,Esteban2007}, which lies just above the nuclear charge of oganesson ($Z=118$).  This was pointed out as early as in 1928 by Gordon \cite{Gordon1928}. For a more rigorous mathematical analysis on the self-adjointness of the point-charge Dirac-Coulomb operator we refer the reader to Sec.~\ref{sec:appA} and the literature cited therein.

On a historical note, the onset of imaginary solutions for the Dirac equation with the bare Coulomb operator led Feynman to the conclusion that elements above $Z=137$ should not exist. Hence,  the element with nuclear charge \num{137} is sometimes (jokingly) called Feynmanium \cite{schweber2020qed}.

\subsubsection{Nuclear Recoil and Uehling terms}\label{sec:recoil}
For a point-like nucleus, the nuclear recoil operator can be approximated by \cite{Aleksandrov2016}
\begin{equation}
H_{\rm NRB}= -\frac{1}{2M}\Delta + i\frac{(Z\alpha)}{2Mr}\left[\vec{\alpha}\cdot \vec{\nabla} + \frac{1}{r^2} ( \vec{\alpha} \cdot \vec{r})(\vec{r}\cdot\vec{\nabla}) \right],
\label{recoil}
\end{equation}
where $M$ is the mass of the nucleus (for a more concise QED treatment see \cite{Adkins2007}).
This recoil operator can be added to the one-particle Dirac-Coulomb operator. For a more detailed discussion of nuclear recoil effects see Refs.~\cite{Breit1948,shabaev2001relativistic}.

In Ref.~\cite{Aleksandrov2016}, both the recoil correction and the Uehling potential $V_U$ for a point nucleus were included in the Dirac equation to see how that would change the   $Z_{\mathrm{c}_\mathrm{p}}$. A value of $Z_{\mathrm{c}_\mathrm{p}}(1s)=144$ is then obtained. These additional operators do not necessarily secure the self-adjointness of $H_{\rm D}+H_{\rm NRB}$ in the critical charge  region, hence,  a careful analysis of the $1s$ eigenfunction at the origin is still required. More details on vacuum polarization (VP) and on the Uehling potential can be found in Sec.~\ref{sec:qed}.

\begin{figure}[tb!]
  \begin{center}
\includegraphics[width=1.0\textwidth]{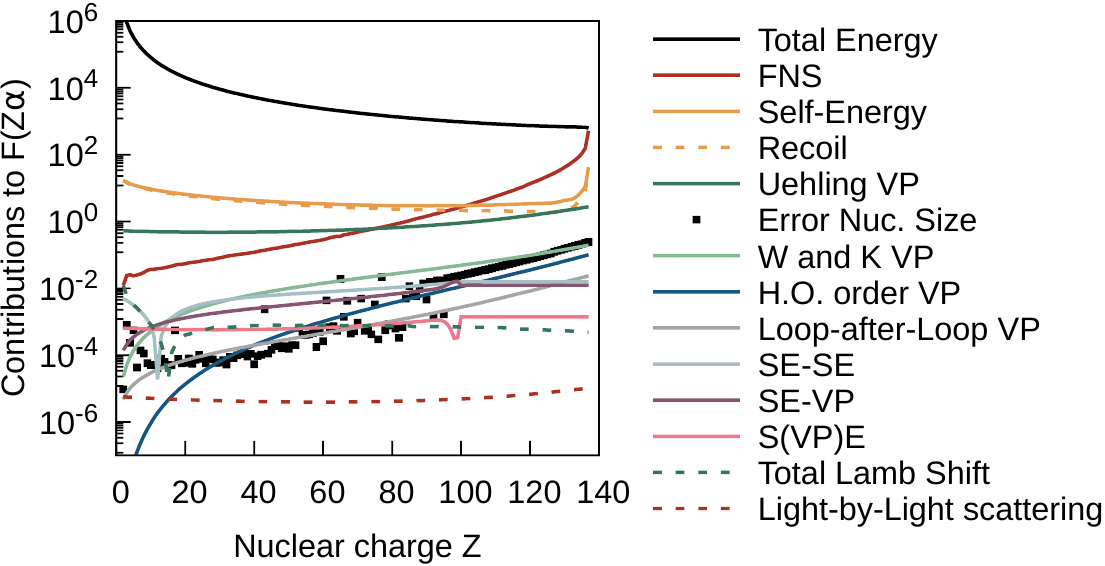}
  \caption{Different contributions to the $1s$ energy level for hydrogen-like atoms, evaluated using the MCDFGME code \cite{mdgme}. Higher-order VP includes the Wichmann and Kroll (WK) correction (order $\alpha (Z\alpha)^3$) as well as approximation to the $\alpha (Z\alpha)^5$ and $\alpha (Z\alpha)^7$ potential contributions. Two-loop self-energy corrections SE-SE, SE-VP and S(VP)E are from Refs.~\cite{yis2005,yis2005a,yis2007,yis2008,yer2009,yer2010,yer2018}. Loop-after-loop VP is approximated by solving the Dirac equation including the Uelhing potential. Finite nuclear size correction and uncertainties on nuclear size are from Ref.~\cite{Angeli2013}. See also \cite{yas2015,ind2019} and references therein.
  }
  \label{fig:lambshift}
  \end{center}
  \end{figure}

Nevertheless, numerical calculations show that the critical charge for the $1s$ state before reaching the onset of the negative energy continuum is $Z_{\mathrm{c}}(1s)\approx 144.75$ due to the nuclear recoil, and $Z_{\mathrm{c}_\mathrm{p}}(1s) \approx 143.95$ due to the nuclear recoil-plus-Uehling term as shown in Fig.~\ref{fig:diving} \cite{Aleksandrov2016}. Moreover, the diving of the 2$p_{1/2}$, $2s$ and $3s$ levels comes at nuclear charges of $Z_{\mathrm{c}_\mathrm{p}}\approx 146$, \num{165}, and \num{193} respectively. The lifting of the $2s/2p_{1/2}$ level degeneracy due to the nuclear recoil becomes thus quite sizable at high-$Z$ values. 
For $Z\alpha<1$, the results including nuclear recoil and Uehling terms are close to the point nuclear charge (PNC) case, as is the steep descent of the energy levels towards the critical nuclear charge. 
On the other hand, Fig.~\ref{fig:lambshift} demonstrates that around $Z=120$ the finite nuclear size correction becomes more important than that originating from the nuclear recoil and Uehling terms. This is addressed in the following section.

\subsubsection{Finite nuclear charge distributions}

By considering a finite nuclear charge  distribution, $\rho_N(\vec{r})$, the problematic singularity at zero is removed. As a result $H_{\rm D}$ becomes self-adjoint for $Z>Z_{\mathrm{c}_\mathrm{p}}$ with real eigenvalues and real radial functions for the discrete spectrum, and thus represents the most natural self-adjoint extension to the PNC Dirac Hamiltonian. This was already realized by Schiff, Snyder and Weinberg as early as in 1939 \cite{Schiff1940}: \textit{In all these cases where the energy cannot be brought to diagonal form, one must take into account either existing deviations from the assumed potential, such as the breakdown of the Coulomb law at small distances, or the reaction of the pair field itself on the external field}.

The potential for an electron interacting with a nuclear charge distribution is given by
\begin{equation}
V(\vec{r})= -\int d{\vec{R}}\frac{\rho_N(\vec{R})}{|\vec{r}-\vec{R}|}.
\label{eq:Coulomb}
\end{equation}

The nuclear charge densities should in principle  be obtained using nuclear density functional theory (DFT) based on realistic energy density functionals, see Sec.~\ref{sec:HFB}.
To obtain the nuclear charge density from computed proton and neutron density distributions, several corrections have to be considered \cite{Friar1975,Friedrich1986,Reinhard2021a}. 
The nucleon structure is taken into account  by folding with the intrinsic form factor of the free nucleons expressed in terms of the Sachs form factors \cite{sac1962}.
The spurious center-of-mass motion can be corrected by an unfolding with the width of the centre-of-mass vibrations. Finally,  one should include  the contribution from the spin-orbit currents \cite{Bertozzi1972}. 
Note that, for the deformed nuclei, the spin-orbit contributions change  gradually as the single-particle spin-orbit strength becomes highly fragmented by deformation and nucleonic pairing (nucleonic superconductivity)
\cite{Reinhard2021a}. Precise nuclear charge densities are essential for interpreting atomic experiments searching for new physics \cite{sbdk2018,Hur2022} or for studying effects related to fundamental symmetry violations \cite{PREX}.

\begin{figure}[tb]
\centering
\includegraphics[width=0.8\linewidth]{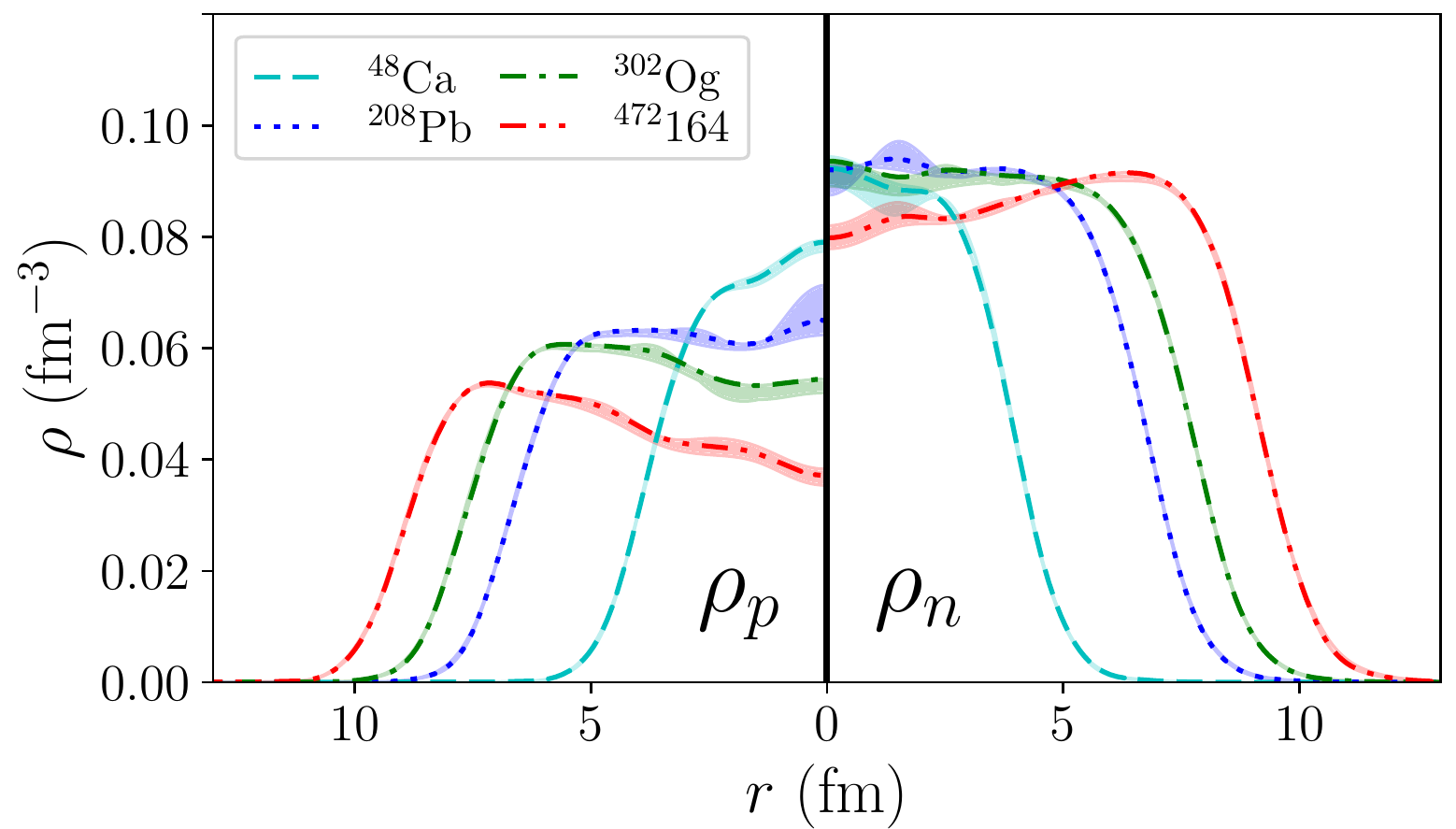}
\caption{\label{fig:densities}
Radial proton (left) and neutron (right) densities of doubly-magic nuclei
$^{48}$Ca, $^{208}$Pb, $^{302}$Og,
and $^{472}$164 obtained in nuclear DFT with three
different energy density functionals.
 The shaded areas indicate the spread of DFT predictions. (Modified from \cite{Schuetrumpf2017a}.)}
\end{figure}

Realistic nuclear modeling of charge densities is particularly important for the superheavy nuclei, the existence of which depends on the interplay between the short-ranged attractive nuclear force and long-ranged electrostatic repulsion, which rapidly grows with  $Z$.
Since the Coulomb
repulsion minimizes the total binding energy of the nucleus by increasing the average distance between protons, the total energy is
significantly lowered by pushing protons toward the nuclear surface.
This mismatch between interaction ranges  in superheavy nuclei results in Coulomb frustration effects  \cite{Nazarewicz2018,giuliani2018}, which   are expected to produce  exotic topologies of nucleonic densities, such as voids (bubbles) or tori. Figure~\ref{fig:densities}  shows the proton and neutron density distributions of several nuclei predicted by nuclear DFT \cite{Schuetrumpf2017a}. The superheavy nuclei such as $^{302}$Og,
and $^{472}$164  show a clear central depression in the proton density distributions resulting in a semi-bubble structure. The properties of Coulomb-frustrated superheavy
nuclei, including their characteristic density distributions and shell structure, have been investigated in numerous studies, see  Refs.~\cite{Afanasjev2005,Schuetrumpf2017a,Agbemava2021} and references cited therein.

In the absence of predictions based on realistic nuclear models, schematic approximations for $\rho_N$  are often applied. These are sufficient for most applications in heavy element research. There is a range of nuclear charge models in use and, for several of these models, analytical expressions for the integral \eqref{eq:Coulomb} in terms of standard functions can be found in Ref.~\cite{Andrae2000}. Most implementations in numerical atomic structure programs apply the (spherical) Fermi two-parameter model \cite{Hofstadter1956, Hofstadter1958} 
\begin{equation}
\rho_N(R)= \frac{\rho_0}{1+e^{(R-R_0)/a}} \, ,
\label{fermi}
\end{equation}
where $R_0$ is the half-density radius, $a$ is the diffuseness parameter, and $\rho_0$ is a normalization constant such that $\int \rho_N(\vec{R}) d\vec{R}=Z$. 
For many nuclei, this model reasonably agrees   with nuclear  DFT calculations. It is to be noted, however, that a simple model like (\ref{fermi}) is bound to fail for superheavy nuclei that exhibit  appreciable Coulomb frustration effects, see Fig.~\ref{fig:densities}. Nevertheless, for the valence shell this nuclear charge model should perform reasonably well even for the superheavy elements. For example, the Fermi charge distribution have been used for electronic structure calculations of Ref.~ \cite{ibj2011} in the superheavy element region up to $Z=173$.

For the homogeneous nuclear charge distribution analytical expressions for the radial Dirac components of the wave function exist. In that category of nuclear models, the simplest one is the uniformly charged spherical shell or top slice (TS) model, with the nuclear charge being smeared out over a spherical nuclear surface at radius $R_0$ \cite{Hill-Ford-1953,Hill-Ford-1954,Andrae2000} 
\begin{equation}
\rho(r)=\frac{Z}{4\pi r^2}\delta(r-R_0) \, .
\end{equation}
It results in a potential of the form $V(r)=-Z/R_0$ for $0\le r\le R_0$ together with the usual Coulomb term  $V(r)=-Z/r$ at $r>R_0$. This approximation cuts off the problematic singularity of the Coulomb potential at nuclear radius $R_0$ and therefore secures the self-adjointness in the region $|E| < m_ec^2$ of the discrete spectrum \cite{Pomeranchuk1945}. To express the radial Dirac components analytically, one divides the solution of the Dirac equation into the two regions $[0,R_0]$ and $[R_0,\infty)$ with an additional boundary condition at $R_0$ to match the two wave functions (see also Sec. \ref{sec:Analytcont}) \cite{greinerrafelski1985}.

The potential $V(r)$ for the TS model is, however, discontinuous in its first derivative and is therefore often extended to the homogeneously charged sphere (HCS) model of the form \cite{Breit1958}
\begin{equation}
\rho(r)=\rho_0\Theta(1-r/R_0),
\end{equation}
where $\rho_0=3Z/4\pi R_0^3$ and $\Theta(x)$ is the Heaviside step function \cite{Hill-Ford-1954,Andrae2000}. The resulting HCS potential is of the form $V(r)=V_0+V_2r^2$, where $V_0=-3Z/2R_0$ and $V_2=V_0/2R_0^2$. This potential is discontinuous in its second derivative.\footnote{Because of the discontinuity in the potential at $R_0$, one has to set one of the grid points in numerical program packages at the nuclear boundary to avoid numerical instabilities \cite{Visscher1997}.} It is clear that by choosing $V_0=-Z/R_0$ and $V_2=0$ the TS model is recovered. This results in a special case of a Fuchs-type differential equation for which analytical solutions to the Dirac equation can be formulated similarly to the procedure used for the TS model \cite{Pieper-1969}. The HCS model was recently used to study isotope shifts using a modified nuclear parameter $\delta \langle r^2 \rangle \rightarrow \delta \langle r^{2\gamma}\rangle$, such that the electronic structure factor $\tilde{F}_i$ becomes isotope independent \cite{FlambaumGeddes2018, Lackenby2019a, Flambaum2019}. Note that this approximate expression for the energy shift is valid only when $\alpha Z R_0 \ll 1$ \cite{shabaev1993finite}, where $R_0$ is expressed in atomic units. For nuclear charge $Z>137$ this expression is manifestly wrong as  $\gamma$ becomes imaginary.

The HCS model can be further generalized by using a Taylor expansion for the nuclear density around the origin \cite{Andrae2000}
\begin{equation}
\label{eq:rhonucTaylor}
\rho(x)=\Theta(1-x)\sum_{i=0}^n a_i x^i \, ,
\end{equation}
with $x=r/R_0$ resulting in a power series for $V(r)$. Breit introduced the simple potential $V(r)=V_0+V_2r^n$, where $V_0=-(n+1)Z/nR_0$ and $V_2=Z/nR_0^{(n+1)}$ \cite{Breit1958}.
For these nuclear models one can derive the radial Dirac wave function from a polynomial expansion. \cite{Pieper-1969,Andrae2000}. In the region $r<R_0$ for $\kappa>0$ the radial wave function is expressed as \cite{maartensson2003}
\begin{align}
\begin{aligned}
P_{n\kappa}(r)&=N_{n\kappa}r^{\kappa} \left\{ r - \left[ \frac{(E_{n\kappa}-V_0)(E_{n\kappa}+2c^2-V_0)}{2c^2(3+2\kappa)} + \frac{V_2(1+2\kappa)}{(E_{n\kappa}+2c^2-V_0)(3+2\kappa)}\right]r^3 + \cdots \right\} \\
Q_{n\kappa}(r)&=N_{n\kappa}r^{\kappa} \left\{ \frac{c(1+2\kappa)}{(E_{n\kappa}+2c^2-V_0)} - \frac{E_{n\kappa}-V_0}{2c}r^2 + \cdots\right\} \, ,
\label{eq:FNCsolutions_kapp}
\end{aligned}
\end{align}
and for $\kappa<0$
\begin{align}
\begin{aligned}
P_{n\kappa}(r)&=N_{n\kappa}r^{|\kappa |} \left\{ 1 - \frac{(E_{n\kappa}-V_0)(E_{n\kappa}+2c^2+V_0)}{2c^2(1+2|\kappa|)}r^2 + \cdots \right\}  \\
Q_{n\kappa}(r)&=N_{n\kappa}r^{|\kappa |} \left\{ -\frac{E_{n\kappa}-V_0}{c(1+2|\kappa|)}r + \left[ \frac{(E_{n\kappa}-V_0)^2(E_{n\kappa}+2c^2+V_0)}{2c^3(1+2|\kappa|)(3+2|\kappa|)} + \frac{V_2}{c(3+2|\kappa|)}\right]r^3 + \cdots\right\} \, .
\label{eq:FNCsolutions_kapm}
\end{aligned}
\end{align}
Since the exponents of the $r^{|\kappa |+i}$ terms in \eqref{eq:FNCsolutions_kapp} and \eqref{eq:FNCsolutions_kapm} are integers, there is no problem at the origin and the derivative norm exists. Furthermore, the wave function is locally absolutely continuous, unlike for the PNC case. 
To show this more rigorously, one applies the Weyl-Weidmann limit point - limit circle theorem \cite{Weidmann1982} and shows that the Dirac operator is self-adjoint in the range $E_{n\kappa}\in[-m_ec^2,m_ec^2]$, with the Sobolev space $\mathcal{W}_{1,2}(\mathbb{R}_+)^2$ as the natural domain of the Dirac operator. Hence, for the Dirac equation with a finite-size nuclear charge distribution, the only critical charge is at the onset of the negative energy continuum at $E=-m_ec^2$.
Full analytic expressions for the Dirac wave functions for TS and uniformly charged nucleus have been derived for $s_{1/2}$ and $p_{1/2}$ orbitals and used in the evaluation of the self-energy with finite size contribution \cite{mas1993,mps1998}.

\begin{figure}[t]
  \begin{center}
\includegraphics[width=0.67\textwidth]{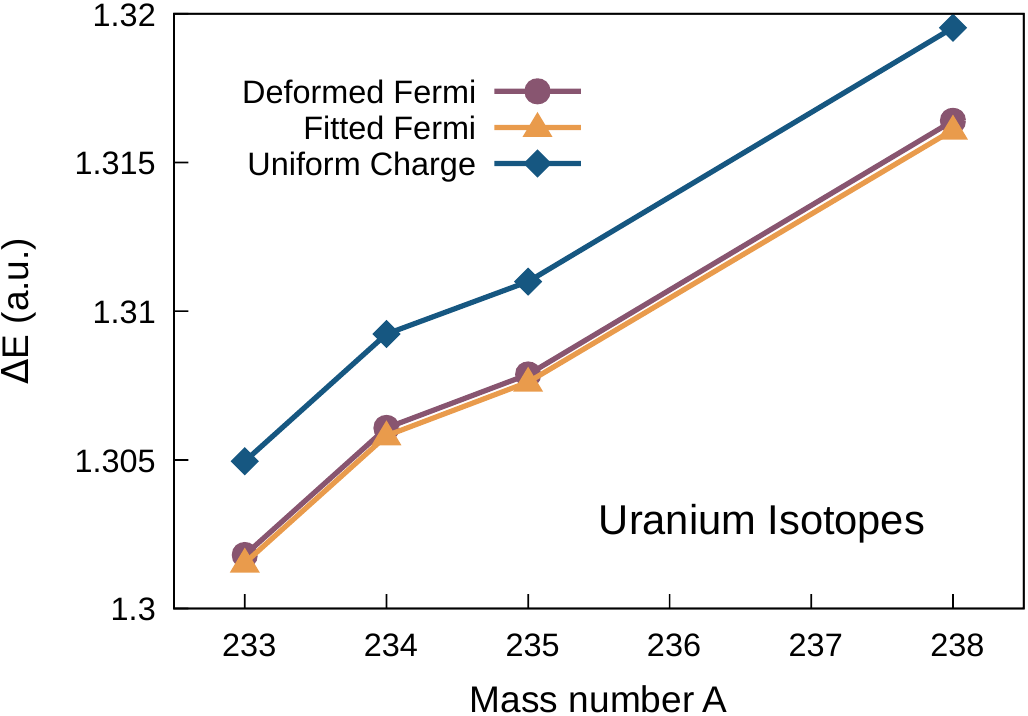}
  \caption{The nuclear-size contributions to the ground-state energies of the Li-like uranium isotopes using a deformed Fermi model for $\rho_N$, a fitted Fermi model, and a uniform charge distribution. (From \cite{Ynnerman1994}.)}
  \label{fig:uranium}
  \end{center}
  \end{figure}

Shifting from a point nucleus to a model that accounts for the finite nuclear charge distribution leads to a noticeable contribution to the total electronic energy. The difference in energy originating from the use of different nuclear charge models is far smaller \cite{JohnsonSoff1985}. For example, 
Fig. \ref{fig:uranium} shows  the calculated ground state energy shift in Li-like uranium due to the finite nuclear charge distribution for the Fermi and uniform charge distributions \cite{Ynnerman1994}.

When introducing a finite nuclear charge into the Dirac equation, the degeneracy between the states of the same $(nj)$ but with different $\kappa$ quantum numbers is lifted. 
This is most prominently seen between the $2s_{1/2}$ and $2p_{1/2}$ levels. This lifting of degeneracy  already appears at the nonrelativistic level between levels of same $n$ but different $\ell$ quantum numbers, but to a much smaller extent compared to the relativistic case \cite{Andrae2000}.

\begin{figure}[t]
  \begin{center}
\includegraphics[width=0.67\textwidth]{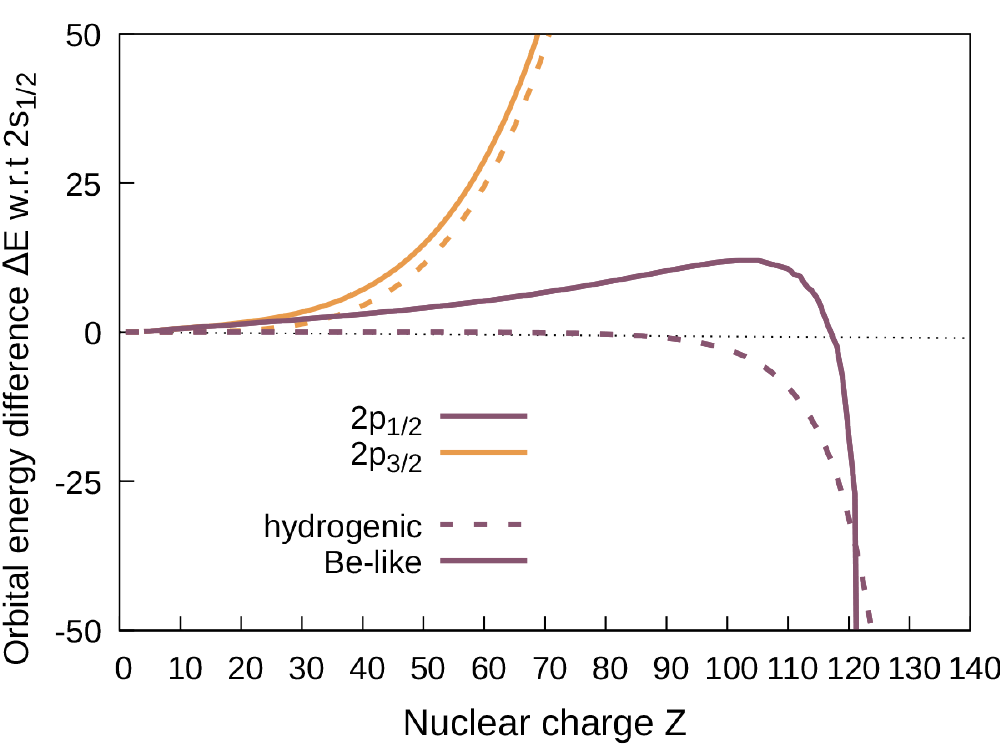}
  \caption{Orbital energy difference $\Delta E$ (in a.u.) of the $2p_{1/2}$ (purple) and $2p_{3/2}$ (orange) states relative to the $2s_{1/2}$ state (in a.u.). The dashed lines are hydrogenic energy differences. The solid lines are multi-reference energies for Be-like $J=0$ states systems involving the major configurations $1s^2\;2s^2, 1s^2\;2p^2_{1/2}, 1s^2\;2p^2_{3/2}$.}
  \label{fig:2s2p}
  \end{center}
  \end{figure}

Figure \ref{fig:2s2p} shows the energy difference $\Delta E$ between the 2p$_{1/2}$ and $2p_{3/2}$  orbitals and  the $2s_{1/2}$ orbital for the hydrogen-like and the Be-like state \cite{isbd2007}. 
The lifting of degeneracy by the finite size of the nuclear charge for the hydrogen-like system can be qualitatively explained by perturbation theory. However, in the small region inside the nucleus, the perturbing potential is so large that a first-order calculation for high nuclear charges is insufficient \cite{Schawlow1955}. In contrast to the hydrogen-like energy difference, in multi-electron systems the $2s$ shell lies below the $2p$ shell for nuclear charges up to about $Z=120$. This comes from the different effective screening of the nucleus for these two shells, which gave rise in the early history of quantum theory to the Slater rules \cite{Slater1930}.  For nuclear charges beyond $Z=120$, the $1s^2\;2p^2_{1/2} \; J=0$ configuration lies below the $1s^2\;2s^2$ configuration, as demonstrated for the Be-like $J=0$ state in Fig.~\ref{fig:2s2p} and in Ref.  \cite{isbd2007}. This is because, in strong Coulomb fields, the Coulomb operator starts to dominate over the electron-electron repulsion and the atom behaves more hydrogen-like.
As a result of this effect, the $2p_{1/2}$ level dives into the negative energy continuum at a far earlier stage at $Z_{\mathrm{c}}\left(2p_{1/2}\right) \approx 218$ compared to the $2s$ level at $Z_{\mathrm{c}}\left(2s\right)\approx 247$ \cite{Schwerdtfeger2015}, see discussion in Sec.\,\ref{sec:Analytcont} for more details.

\subsubsection{$1s$ energy level reaching the negative energy continuum}

Figure \ref{fig:diving} shows the $1s$ energy level as a function of nuclear charge for hydrogen-like systems in the FNC variant, computed using the relativistic atomic program package GRASP \cite{DyaGraJoh89}. The calculations predict a critical charge of $Z_{\mathrm{c}}(1s)=170.161$ (170.017 including QED effects) before diving into the negative energy continuum \cite{Schwerdtfeger2015}.

\begin{figure}[tb]
  \begin{center}
\includegraphics[width=0.85\textwidth]{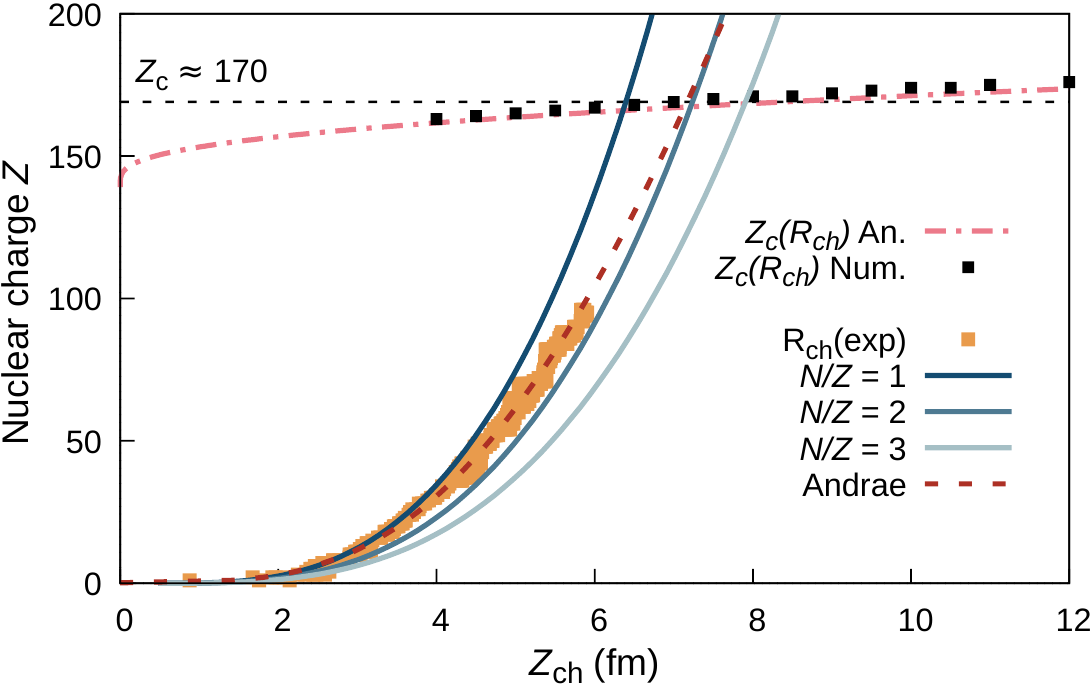}
  \caption{The nuclear charge radius $R_{\rm ch}$ as a function of $Z$, using  phenomenological expressions with different neutron/proton ratios (solid lines) and the expression by Andrae \cite{Andrae2000} (dashed line). Experimentally known charge radii \cite{Angeli2013} are marked by orange squares. The critical charge as a function of nuclear radius obtained with the analytical expression of Ref.~\cite{kuleshov2015vs} is shown by a dash-dotted line. The $Z_{\mathrm{c}}\left(R_{\rm ch}\right)$ Numerical values (black square) have been obtained using the MDFGME code \cite{desclaux1975, ibj2011} with a Fermi nuclear charge model.}
  \label{fig:ZversusM}
  \end{center}
  \end{figure}

Using different models of nuclear charge distribution, the predictions for the critical nuclear charge can vary widely between \numrange[parse-numbers=false]{Z_{\mathrm{c}}=164}{174} for the $1s$ level \cite{Graf1991,Greiner1998}, but more realistically between \numrange{168}{172} using the uniform nuclear charge distribution and neutron numbers varying between $N=Z$ and $N=3Z$.
This is demonstrated in Fig.~\ref{fig:ZversusM}, which shows the relation between the rms nuclear charge radius $R_{\rm ch}$ and the proton number $Z$.
The  critical charge as a function of $R_{\rm ch}$ has been computed using the analytical expressions of Ref.~\cite{kuleshov2015vs}. 
Filled squares mark the experimental charge radii \cite{Angeli2013}. The lines denote the relation between $R_{\rm ch}$ and $Z$ for three different neutron to proton ratios \cite{Bethe1940,Present1941} and the semi-empirical relation \cite{Andrae2000}.
From the intercept between the dashed and dash-dotted lines, an estimate for the critical charge is  $Z_{\mathrm{c}}(1s) = 170.26 $, with a  nuclear charge radius of \SI[parse-numbers=false]{R_{\mathrm{ch}} = 7.19}{fm}. 

In the context of the above discussion, it is interesting to notice   that because of the mass scaling $\sqrt m$ of the Dirac equation (\ref{eq:Dirac})  the critical charge for muonic atoms  ($m_{\mu}/m_e=206.7682830(46)$ 
\cite{mohr2016codata}) for a point nucleus is more than an order of magnitude larger  $Z_{\mathrm{c}_{\mathrm{p}}}^{\mu} (1s) \approx1966$ compared to the electronic case.
Taking into consideration the finite nuclear radius, the critical value shifts to $Z_{\mathrm{c}}^{\mu} (1s)\approx 2200$ \cite{soff1974precise}. As in the free-particle case, the small component becomes large and takes over for $E\rightarrow -m_ec^2$.

\subsection{Electron states in the super-critical region}
In 1969, Pieper and Greiner \cite{Pieper-1969} analyzed in detail the analytical solutions for FNC models  as the limit $E_{n\kappa}=-m_ec^2$ is approached for different $(n\kappa)$ states. The coefficients in the $r$-expansion in \eqref{eq:FNCsolutions_kapp} and \eqref{eq:FNCsolutions_kapm} do not exhibit any pathological behavior, but the radial functions and eigenvalues become complex in the critical region $E_{n\kappa}<-m_ec^2$ and thus lie outside the natural domain of the self-adjoint Dirac operator. 
As a result, the Dirac-Hamiltonian  eigenstates  embedded in the continuum cannot readily be reached by standard  atomic structure theory. 
In the following, we discuss some of the approaches to deal with this problem.
  
\subsubsection{Energy-projected Dirac equation}\label{EPDE}
The relation between the absence of self-adjointness and the appearance of the negative energy continuum in the spectrum of the Dirac operator was studied by restricting the Hilbert space to the subspace defined by the positive energy continuum states.
This can be effectively achieved by means of the projection technique, analogous to the Feshbach projection technique \cite{Feshbach1958,Feshbach1962} used in the context of open quantum systems. Effectively, in this method,  the negative-energy continuum space is removed \cite{HardekopfSucher1985}.  The resulting single-particle Dirac Hamiltonian, the so-called \emph{no-pair external field Dirac Hamiltonian}, becomes:
\begin{equation}\label{eq:diracfem-projection-h}
  \hat{h}^{+} = \Lambda^+ (h_{\mathrm{D}} + V_{\mathrm{ext}}) \Lambda^+
\end{equation}
where $\Lambda^+$ is the projection operator onto the free-particle positive energy subspace of the free-particle Dirac Hamiltonian $H_{D}^{\rm FP}$.
As long as $H_D^{\rm FP}$ has no zero eigenvalues, the operator $\Lambda^+$ can be written as
\begin{equation}
\Lambda^+ = \frac{1}{2}\left( 1 + \frac{H^{\rm FP}_{\rm D}}{|H^{\rm FP}_{\rm D}|}\right) =  \frac{ \vec{\alpha}\cdot\vec{p} + \beta mc}{\sqrt{\vec{p}^2+m^2c^2}}
\label{Casimir}
\end{equation}
where the quotient $H_{\rm D}^{\rm FP}/|H^{\rm FP}_{\rm D}|$ is called the sign operator.\footnote{
The Hamiltonian $H_D^{\rm FP}$, while similar to, is not the same as the no-pair Hamiltonian often used in relativistic quantum chemistry to avoid the continuum dissolution.
In that case, the projection operator is usually constructed from the positive energy eigenstates of the full external-field Dirac Hamiltonian, and does not span quite the same space as that of free-particle states. 
Furthermore, the corresponding projection operators depends on the nuclear charge distribution \cite{Sucher1980,hllm1986,ind1995,dyall2007book}.}
The eigenvalues of the free-particle Dirac Hamiltonian $h_{\mathrm{D}}$ are $|E| \geq m_ec^2$ \cite{thaller1992}.
The projected Dirac Hamiltonian~\eqref{eq:diracfem-projection-h} can be traced back to Bethe and Salpeter~\cite{BetheSalpeter1951,bethe2012quantum}, and is therefore sometimes referred to as the Bethe-Salpeter operator~\cite{Evans1996}. 
As discussed in \cite{hardekopf1984}, the projection operator effectively removes the pair creation and annihilation terms from the Dirac Hamiltonian, \ie  removes the coupling to the pair creation/annihilation channel.

Intuitively one would expect that  various mathematical problems with the Dirac equation might disappear if the negative-energy continuum states are projected out. However, if the external field $V_{\mathrm{ext}}(r)$ is the simple $1/r$ potential corresponding to a point nucleus, $\hat{h}^+$ also has a critical charge at which it becomes non-self-adjoint, just like the standard Dirac operator.
In fact, the critical charge of $\hat{h}^+$,
\begin{equation}
  Z_{c} = \left(\frac{2}{\pi} + \frac{\pi}{2}\right) \alpha^{-1} \approx 124.16,
\end{equation}
is lower than $\alpha^{-1} \approx 137$  \cite{HardekopfSucher1985,Evans1996}.\footnote{
As discussed in Sec.~\ref{sec:appA}, the Dirac equation with a $1/r$ potential has another critical nuclear charge at $Z_{\mathrm{c}} = (\sqrt{3}/2) \alpha^{-1} \approx 118.68$, when the condition $|| H_\mathrm{D} \phi ||_2 < \infty $ is imposed  to guarantee self-adjointness. The projected equation exhibits a similar critical charge at $Z_{\mathrm{c}} = (3/4) \alpha^{-1} \approx 102.78$ ~\cite[Eq.~(2.9)]{HardekopfSucher1985}.}
The no-pair approach based on the free Dirac Hamiltonian has therefore been criticized in Ref.~\cite{hllm1986}, where it is shown that it does not prevent continuum dissolution and that projection operators from the bound Dirac Hamiltonian must be used instead. The necessity to use projection operators for correlation orbitals is shown in Ref. \cite{ind1995}.

Unlike the Dirac equation, the no-pair operator has a lower bound in the sub-critical region.
This result was further refined in Refs.~\cite{tix1997lower,tix1998}, which demonstrated that the operator's eigenvalues are strictly positive~\cite{tix1997lower,tix1998}, in contrast to the point nucleus Dirac equation, for which the eigenvalues go to zero for increasing nuclear charge up to $Z\alpha=1$.

The projection equation with a finite nuclear potential was initially thought to remove all the problems with the negative energy continuum. Table~\ref{tab:u91} benchmarks
 the no-pair approximation  against Dirac-Coulomb calculations for the $1s_{1/2}$ ionization potential and transition energies of $^{238}$U$^{91+}$. Such highly charged atoms are important for precision tests of QED \cite{Gumberidze2005}, and QED results agree with experiments to a few \si{\electronvolt} \cite{ind2019}. Unlike in the Dirac-Coulomb variant, the results of the free-particle projected approach shown in Table~\ref{tab:u91} compare poorly with experiment. This indicates that the projected Dirac Hamiltonian appears to be a far worse starting point than the standard Dirac equation for further QED refinements. The reasonable choice of projection operators  for the whole range of nuclear charges $Z$ remains a challenging problem. At this stage, keeping the physically relevant negative-energy continuum and dealing with directly it seems to be a better solution. However, this requires to correctly describe resonance states with $E\le -m_ec^2$ as discussed in Secs.~ \ref{sec:HFB}-\ref{sec:Gamow}.
\begin{table}[tb]
\label{tab:u91}
  \centering
\begin{tabular}{|r|r|r|r|r|r|r|}
\hline
& \multicolumn{2}{c|}{Ionization potential} & \multicolumn{2}{c|}{$1s_{1/2} \rightarrow 2p_{1/2}$} & \multicolumn{2}{c|}{$1s_{1/2} \rightarrow 2p_{3/2}$} \\ \cline{2-7}
& \multicolumn{1}{c|}{$E$} & \multicolumn{1}{c|}{$\Delta E_{\mathrm{exp}}$}&\multicolumn{1}{c|}{$E$} & \multicolumn{1}{c|}{$\Delta E_{\mathrm{exp}}$}& \multicolumn{1}{c|}{$E$} & \multicolumn{1}{c|}{$\Delta E_{\mathrm{exp}}$}\\ \hline
DC / PNC & $132,279.93$ & $454.83$ & $98,064.45$ & $458.84$ & $102,630.10$ & $451.98$ \\
DC / FNC & $132,083.55$ & $258.45$ & $97,872.42$ & $266.81$ & $102,433.71$ & $255.59$ \\
PDC/ FNC & $140,474.30$ & $8,649.20$ & $105,686.10$ & $8,080.49$ & $110,767.33$ & $8,589.21$ \\ \hline
Exp. & \multicolumn{2}{c|}{$131,825.10 \pm 4.20$} & \multicolumn{2}{c|}{$97,605.61 \pm 16.00$} & \multicolumn{2}{c|}{$102,178.12 \pm 4.33$} \\ \hline
\end{tabular}
  \caption{Comparison of Dirac-Coulomb calculations with experimental values for the $1s_{1/2}$ ionization potential and transition energies of $^{238}$U$^{91+}$.
    All energies in eV. The rows correspond to the standard hydrogenic Dirac-Coulomb (DC) equation with point nucleus (PNC), finite nucleus (FNC), and the free-particle projected Dirac-Coulomb (PDC) equation with a FNC approximation. The experimental values are taken from Refs. \cite{Gumberidze2005,ind2019} by picking the values with the lowest uncertainty. The homogeneous uniformly charged sphere model was used. The difference with experiment and calculation for the FNC value is due to QED corrections which are not included here.}
\end{table}

\subsubsection{Hartree-Fock-Bogoliubov equation analogy}\label{sec:HFB}

It is instructive to make an analogy between  the one-particle Dirac-Coulomb Eq.~(\ref{eq:Dirac})
and one-quasiparticle Hartree-Fock-Bogoliubov (HFB; or Bogoliubov-de
Gennes)  equation used in  the density
functional theory (DFT) of superconductors and atomic nuclei.

The  HFB equation in the
coordinate representation \cite{bulgac1999hartree,Dobaczewski1984} can be written as:
\begin{eqnarray}
  \left[
\begin{array}{cccc}%
 \displaystyle h-\lambda& {\hspace{0.7cm} } \Delta \vspace{2pt}\\
  \displaystyle -\Delta^*& -h^* +\lambda \\
\end{array}
\right]\left[
\begin{array}{clrr}%
u_i \\ \vspace{2pt} v_i\\
\end{array}
\right]=E_i\left[
\begin{array}{clrr}%
 u_i  \vspace{2pt}\\ v_i
\end{array}
\right],\label{HFB}
\end{eqnarray}
where $h$ is the single-particle Hamiltonian; $\Delta$ is the pairing mean-field; $\lambda$ is the chemical potential (or Fermi level);
 $E_i$ is the quasi-particle energy; 
 and $u_i(\rr,\sigma)$ and $v_i(\rr,\sigma)$ are the upper and lower components of quasi-particle
wave functions, respectively, that depend on the spatial coordinates $\rr$ and spin $\sigma$.
 The main DFT ingredient is the energy density functional (EDF) that depends on  the particle and pair densities and currents. 
The mean-fields $h$ and $\Delta$ are determined self-consistently from the one-body densities and the assumed EDF.

\begin{figure}[htb]
  \begin{center}
\includegraphics[width=0.50\textwidth]{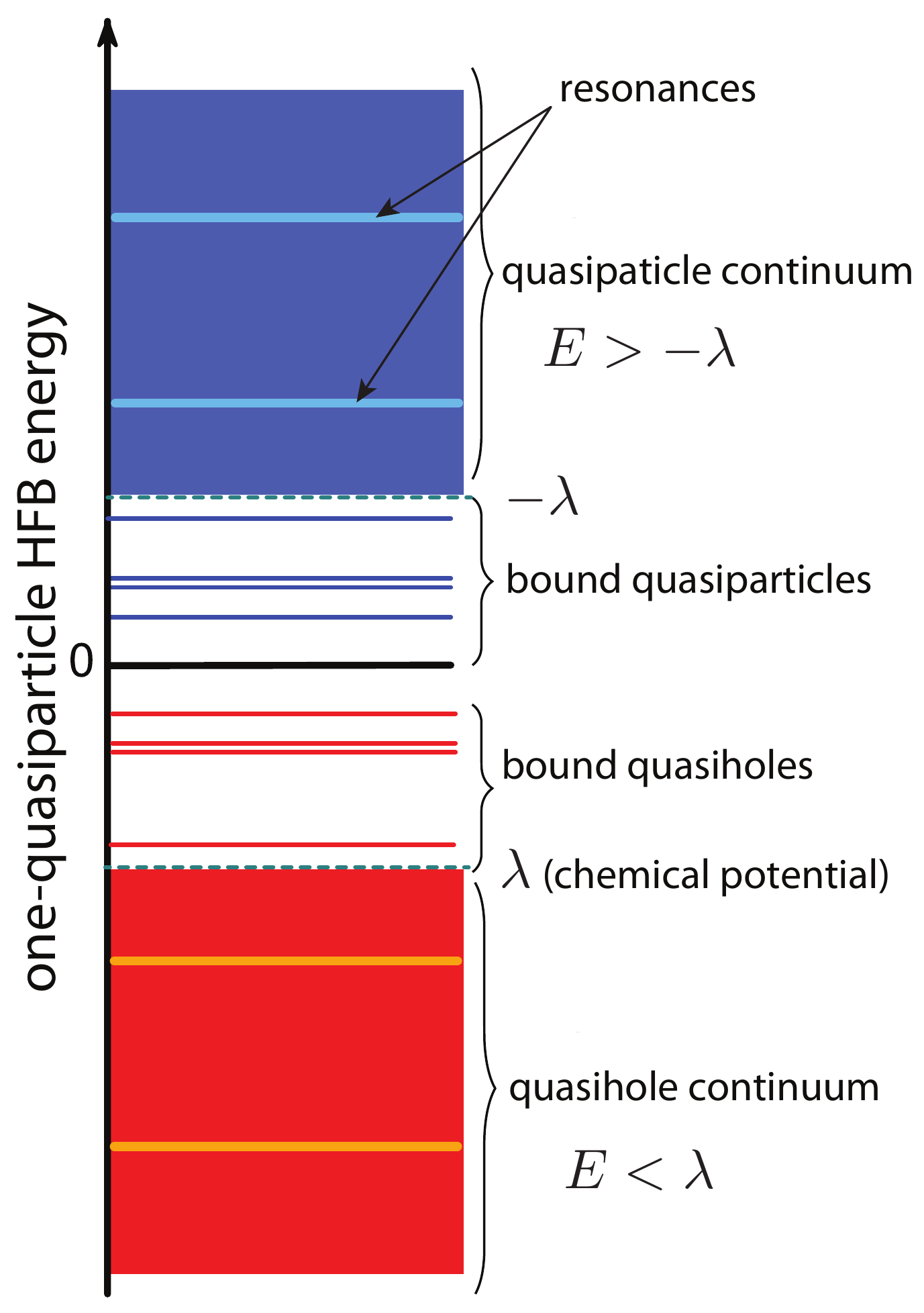}
  \caption{One-quasiparticle HFB spectrum.
The bound states exist  in the
energy region $|E|<-\lambda$, where $\lambda$ is the chemical potential (negative for a particle-bound system).}
  \label{fig:HFBspectrum}
  \end{center}
  \end{figure}

The quasiparticle vectors are
two-component wave functions $u_i(\rr,\sigma)$ and $v_i(\rr,\sigma)$, which acquire
specific asymptotic properties \cite{bulgac1999hartree,Dobaczewski1984,Belyaev1987,Dobaczewski1996}
determining the asymptotic behavior of local densities. As shown
in Fig.~\ref{fig:HFBspectrum},
the quasiparticle energy spectrum $E_i$ of HFB consists of discrete bound
states, resonances, and non-resonant continuum states. The bound HFB solutions exist only in the
energy region $|E|<-\lambda$. The quasiparticle continuum
with $|E| > -\lambda$ consists of non-resonant (scattering) continuum and
quasiparticle resonances.

The HFB equation (\ref{HFB}) possesses the {\it quasiparticle-quasihole
symmetry}. Namely, for each quasiparticle state
$(u_i,v_i)$  and energy $E_i$ there exists
a conjugate quasihole state
$(v^*_i,u^*_i)$ 
of opposite energy $-E_i$.
That is, the spectrum is
composed  of pairs of states with opposite energies, see Fig.~\ref{fig:HFBspectrum}. 
The conjugate states can be related through a discrete symmetry, such as time reversal \cite{Frauendorf2001}.
In the HFB vacuum, corresponding to even number of fermions, all negative-energy eigenstates
are occupied by quasiparticles.  This set of
quasihole states is referred to as the Bogoliubov sea \cite{Dobaczewski2013,Bertsch2009}.
It follows from the projection property of the generalized HFB density matrix that if a positive-energy  one-quasiparticle state is occupied,  its conjugated negative-energy partner is empty
\cite{Valatin1961}, and vice-versa. 

The Bogoliubov sea is infinitely deep, in a
full analogy with the sea of negative-energy states of the Dirac
equation.
In practice, since infinite sums over the Bogoliubov sea cannot be carried out
when computing local HFB densities,
the number of HFB-active states must be truncated.
Two different ways of achieving this goal are most often
implemented, namely, solution of the HFB equations in a finite
Hartree-Fock  space \cite{Gall1994} and truncation of  the
quasiparticle space. The second method corresponds to truncating directly the quasiparticle space and using a renormalization or regularization technique to account for the truncated states  \cite{Dobaczewski1984,Dobaczewski1996,Bulgac2002,Borycki2006,Pei2011,Li2012,Pei2015}. 

The proper treatment of nuclear quasi-particle HFB continuum  is important
for accurate description of ground-state  properties and  excitations \cite{Dobaczewski1996,Dobaczewski2013,Pei2011,Terasaki2005,Mizuyama2009}.  Within the real-energy HFB framework, the HFB equations must be solved
by imposing the scattering boundary conditions on the quasiparticle vectors.
If the outgoing  boundary conditions are imposed, the  unbound HFB eigenstates
have complex energies;  within such Gamow HFB (GHFB) approach \cite{Michel2008}  the imaginary energies are related to the particle decay width. 

The quasi-particle HFB continuum
can also be treated in an approximate way by means of a discretization method.  The commonly used approach is to impose  the box boundary
conditions \cite{Dobaczewski1996,Grasso2001,Pei2011,Zhang2013}, in which HFB
eigenvectors  $(u_i,v_i)$ are spanned by a basis of 
$\mathcal{L}^2$-integrable orthonormal functions
defined on a lattice in  coordinate space and
vanish at box boundaries. In this approach, referred to as the  $\mathcal{L}^2$ discretization, quasi-particle continuum of HFB  is
represented by a  finite number of box  states. The structure of the discretized continuum depends on the  size and geometry of the box \cite{Chen2022}. 
In the context of the Dirac equation, scalar confinements at the level of strong Coulomb fields need to be explored, for example within a finite element approach \cite{styp2004,Grant_2009}.\footnote{Confinement potentials need to be introduced in scalar form, i.e. added to the mass term. Adding a confinement to the potential term causes the spectrum to become completely continuous \cite{Plesset1932,thaller1992,greinerrafelski1985}.}

There are two kinds of quasiparticle HFB resonances. 
The {\it  particle resonances} represent metastable states that have large particle (upper) component, i.e., the normalization of $u_i$ is much larger than that of
$v_i$. The {\it deep-hole resonances} are associated with excitations of low-lying hole states of the s.p. Hamiltonian $h$.
For those states, the lower component $v_i$ dominates. The deep-hole resonances  acquire decay width through the coupling to the pairing channel \cite{Belyaev1987,Dobaczewski1996}.

Quasiparticle resonances can be directly calculated  using coordinate-space Green's function technique
\cite{Oba2009,Zhang2011} and  GHFB \cite{Michel2008}. For
approaches based on the   $\mathcal{L}^2$-discretization, 
approximate methods have been developed to deal with HFB resonances.
Since the HFB quasiparticle resonances are highly-localized states
whose energies are weakly affected by the box size,
the  stabilization method based on box solutions with different box sizes
\cite{Zhang2008,Pei2011} can be used to obtain
 the resonance energies and widths.
Besides the
stabilization method, a straightforward smoothing and fitting technique
that utilizes the smoothed occupation numbers obtained from the dense spectrum  of box states has been successfully used
\cite{Pei2011}.

Summarizing this section, there are many similarities between the single-particle Dirac problem and one-quasiparticle HFB problem:
\begin{itemize}
\item 
The corresponding equations have a similar two-component form.
\item
In both cases, the energy spectra are symmetric with respect to zero energy. In the Dirac case, this is related  to charge conjugation. In the HFB case, this is due to the quasiparticle-quasihole symmetry. For a recent discussion of 
particle–hole symmetries of multi-fermion systems (such as band insulators or superconductors)  and the charge-conjugation symmetry of relativistic Dirac fermions, see
Ref.~\cite{Zirnbauer2021}.
\item
In both cases, the resonances can be divided into particle resonances with the upper component dominating over the lower component  and the hole resonances, for which the lower component dominates. At $Z\approx Z_{c}$, the diving states resemble hole resonances of HFB.
\item
In both cases, one deals with spectra that are partly discrete and partly continuous. The continuum space contains metastable states (resonances) that are embedded in the non-resonant background. 
\item
The Dirac and HFB spectra are  bound neither from above nor from below.
This leads to a variational collapse (Dirac) and difficulties with the use of the imaginary time method (for both Dirac and HFB), see, e.g.,  Ref. \cite{Tanimura2013} for a remedy.
\item
In both cases, one has to deal with  continuum-space truncations.
\end{itemize}
Those analogies can be helpful when tackling similar problems or interpreting similar phenomena with the Dirac equation. See also Refs.~\cite{popov1973,Zirnbauer2021} for relevant examples.

\subsubsection{Perturbative approach}\label{sec:pert}
For narrow resonances with energies close to $E=-m_ec^2$, the energy eigenstates can be obtained perturbatively. 
To this end, one can employ the two-potential approach \cite{Goldberger1964} to the decay of a metastable state \cite{Gurvitz1987,Gurvitz2004}. 
Within this method, the potential describing the decaying system can be decomposed into $V=V_0+V'$, where $V_0$ represents the bound-state potential of a closed quantum system and $V'$ is the closing potential. When applied to the diving states, one can assume $V_0$ in the form of the Coulomb potential of the finite nuclear charge distribution
with $Z_0< Z_{\rm c}$ and $V'=(Z'/Z_0)V_0$, where 
$Z>Z_{\rm c}$ and $Z'=Z-Z_0$  \cite{Muller1972,greinerrafelski1985}.
 This decomposes the overcritical Dirac Hamiltonian into
$H_{\rm D}= H_{\rm D_0}+V'$. 
Seeking for an expression of the discrete ${1s}$ state as a solution to the overcritical Hamiltonian, the approximate eigenvector is chosen to be
\begin{equation}
\psi_E(\vec{x})=a(E)\psi^0_{1s}(\vec{x}) + \int^{-m_ec^2}_{-\infty} dE'~ b(E',E) \psi^0_{E'}(\vec{x}),
\label{ansatz}
\end{equation}
where $\psi^0_{1s}(\vec{x})$, the $1s$ bound state, and $\psi^0_{E'}(\vec{x})$, a continuum state with energy $E'$, are the solutions of the total Dirac equation just before diving. $a(E)$ and $b(E',E)$ are coefficients to be determined \cite{fano1961effects}.
This leads to the perturbative expression for the  $1s$ state energy embedded in the continuum
\begin{equation}
\label{energy}
E_{1s}^{\rm cr}=E^0_{1s}+\Delta E_{1s}+F_{1s}(E),
\end{equation}
where
\begin{equation}
\label{1sdifference}
\Delta E_{1s}= \langle \psi^0_{1s}(\vec{x}) | V'(\vec{x}) | \psi^0_{1s}(\vec{x}) \rangle \propto  Z’
\end{equation}
and
\begin{equation}
\label{1sdifference1}
F_{1s}(E)= \dashint dE'~ \frac{|\langle \psi^0_{E'}(\vec{x}) | V'(\vec{x}) | \psi^0_{1s}(\vec{x}) \rangle|^2}{E-E'}  \propto -Z’^2\, .
\end{equation}
The dash in the integral \eqref{1sdifference1} indicates  the Cauchy principal value.
The function $F_{1s}(E)$ in \eqref{energy} introduces an energy distribution to $E^0_{1s}+\Delta E_{1s}$ with a width of
\begin{equation}
\label{Gamma}
\Gamma_E=2\pi |V_E|^2  \propto  \gamma Z’^2.
\end{equation}
The width can be interpreted in terms of the positron escape width \cite{Muller1972,greinerrafelski1985}.

\subsubsection{Analytical continuation}
\label{sec:Analytcont}

One-particle resonances embedded in the negative energy continuum can be found by
extending the Dirac eigenvalue problem into
 the complex domain. 
One approach is based on solving the Dirac-equation  eigenproblem with the {\it incoming boundary condition}. The resulting discrete resonant (Gamow) states have complex energies  $E=E_0+i\Gamma/2$ with a  positive imaginary part $\Gamma$, termed \textit{supercritical} in the remainder of this review. This interpretation  differs from the usual
complex-energy description of \textit{ decaying Gamow states} for which $E=E_0-i\Gamma/2$. Indeed, the supercritical negative-energy electron resonances
can be interpreted in terms of resonances in scattering of  positive-energy positron  propagating \textit{backwards in time} according to the Feynman-St{\"u}ckelberg interpretation \cite{stuckelberg1942,Feynman1949}.
 
 Complex-energy solutions for the differential equation are obtained using the appropriate boundary conditions, analogous to states in the discrete region. This has, for example, been studied for the spectrum of the Dirac equation with a spherical well potential \cite{rafelski1978fermions,Szpak_2008,Krylov2020} and for a Coulomb cut-off potential \cite{popov1973,greinerrafelski1985,kuleshov2015vs,Godunov2017,Krylov2020}, for which the solutions can be analytically expressed.
Alternatively, solutions to the Dirac equation can be analytically continued into the complex plane by complex scaling  or by introducing a complex absorbing potential \cite{Seba1988,riss1993calculation,ackad2007supercritical,ackad2007numerical,popov2020access}.
In the following, we discuss  the complex-energy solutions following the analytical continuation approach of Ref.~\cite{Godunov2017}, in which the nuclear potential is assumed to be constant inside the sphere of radius $R_0$.

At distances up to a cut-off radius $R_0$, the  solution to the radial  Dirac equation is given by the Bessel functions:
\begin{equation}
\label{Gshort}
\biggl(\begin{matrix}
P(r) \\ Q(r)
\end{matrix}
\biggr)
= C \sqrt{\beta r}
\biggl(\begin{matrix}
\mp J_{\mp(1/2+\kappa)} (\beta r) \\ 
J_{\pm(1/2-\kappa)}(\beta r) \frac{\beta}{E + m_ec^2 + \frac{Z \alpha}{R_0}}
\end{matrix}\biggr),
\end{equation}
where $\beta = \sqrt{(E+ Z\alpha/R_0)^2 - m^2c^4}$ and upper (lower) signs correspond to
$\kappa<0$ ($\kappa>0$).
For  $r > R_0$, the solutions are given by the Dirac equation with a Coulomb potential. A combination of exponential and confluent hypergeometric functions satisfy the boundary conditions \cite{Slater1960}:
\begin{equation}
\label{Glong}
\biggl(\begin{array}{cc}
P(E,r) \\ Q(E,r)
\end{array}
\biggr)
=
\biggl(
\begin{array}{cc}
\sqrt{m_ec^2+E}\\
- \sqrt{ m_ec^2- E}
\end{array}
\biggr)
e^{ikr} \rho^{i\tau} 
\biggl(\begin{array}{cc}
f_1(E, r)  \\
f_2(E, r)  
\end{array}
\biggr)
\end{equation}
Here, $\tau = \sqrt{(Z\alpha)^2 - \kappa^2}, \rho = -2ikr, -ik = \sqrt{(m_ec^2-E)(m_ec^2+E)}$, and the functions $f_{i}$ contain Kummer's confluent hypergeometric functions. The full analytical form can be found in Ref.~\cite{Godunov2017}. \footnote{A linear combination of the two-parameter Tricomi function and the exponential terms $e^{ikr}$ and $e^{-ikr}$, is given in Ref.~\cite{kuleshov2015vs}.}
The poles of the $S$-matrix correspond to the resonant states; these are found by matching the $P/Q$ ratio of (\ref{Gshort}) and (\ref{Glong}) at $R_0$. This results in real eigenenergies for solutions in the domain $E_0\in [m_ec^2,-m_ec^2]$.  Solutions with $E_0 \le -m_ec^2$ are embedded in the negative energy continuum, and are of the form $E=E_0+\frac{i}{2}\Gamma$ with real energies $E_0<-m_ec^2$ and widths $\Gamma>0$.
The states in the continuum diverge as $r\rightarrow \infty$ and are identified as Gamow wave functions, see Sec.~\ref{sec:Gamow} below for a detailed description.
Note that at the critical energy $E=-m_ec^2$ the upper Dirac component $P$ in Eq.~\eqref{Glong} vanishes at large distances. This means that close to $Z_{\mathrm{c}}$ the diving resonances resemble the deep-hole HFB states  discussed in Sec.~\ref{sec:HFB}.
\begin{figure}[tb]
  \begin{center}
\includegraphics[width=0.8\textwidth]{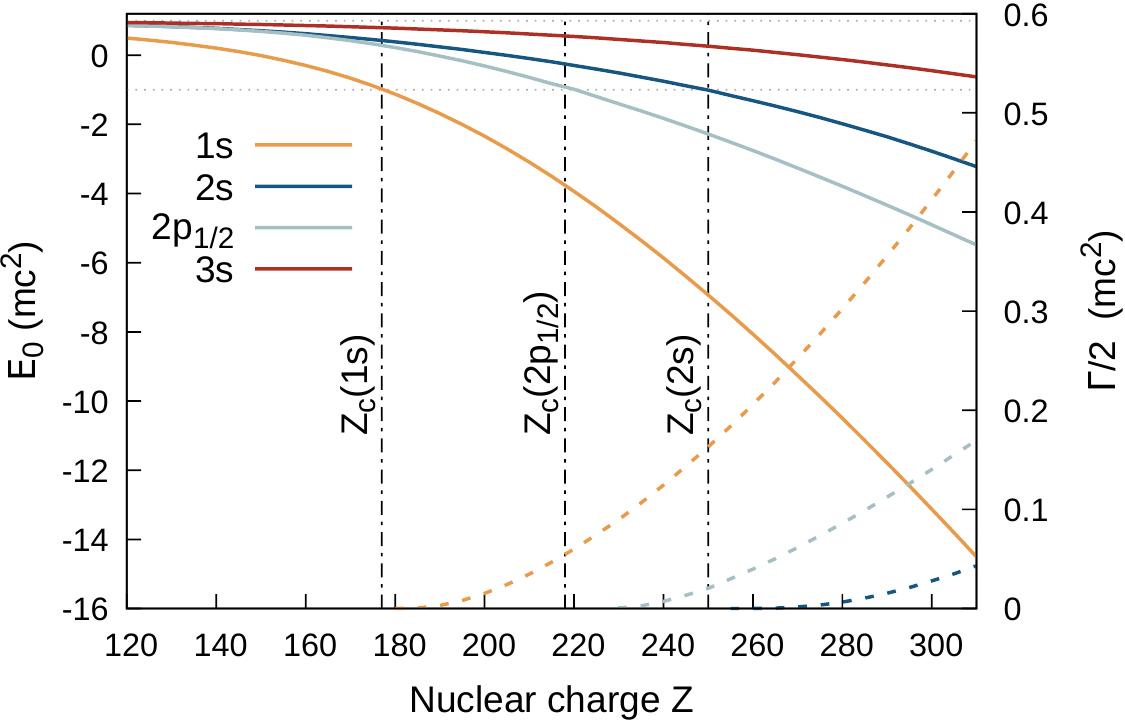}
  \caption{Single particle energy levels as a function of the nuclear charge $Z$. Solid lines corresponds to the real part of the energy $E_0$, the dashed lines to the complex contribution $\frac{i}{2} \Gamma$. Energies are obtained by analytical continuation for a nuclear cut-off of $R_{\mathrm{cut}}= 0.031$ \, in units of $\hbar/(mc)$. The critical charges are highlighted with a vertical dash-dotted line ($Z_{\mathrm{c}}(1s_{1/2}) \approx 177$, $Z_{\mathrm{c}}(2p_{1/2}) \approx 218$, $Z_{\mathrm{c}}(2s_{1/2}) \approx 247$).  }
  \label{fig:Greiner}
  \end{center}
  \end{figure}

Energies of  several single-particle  states, obtained by the exact approach as detailed above are shown in Fig.~\ref{fig:Greiner}. The energies are similar to the perturbative  result of Sec.~\ref{sec:pert} at close vicinity to $Z_{\mathrm{c}}$ but deviate at larger $Z$ values as expected.  
\begin{figure}[tb]
  \centering
  \includegraphics[width=0.95\textwidth]{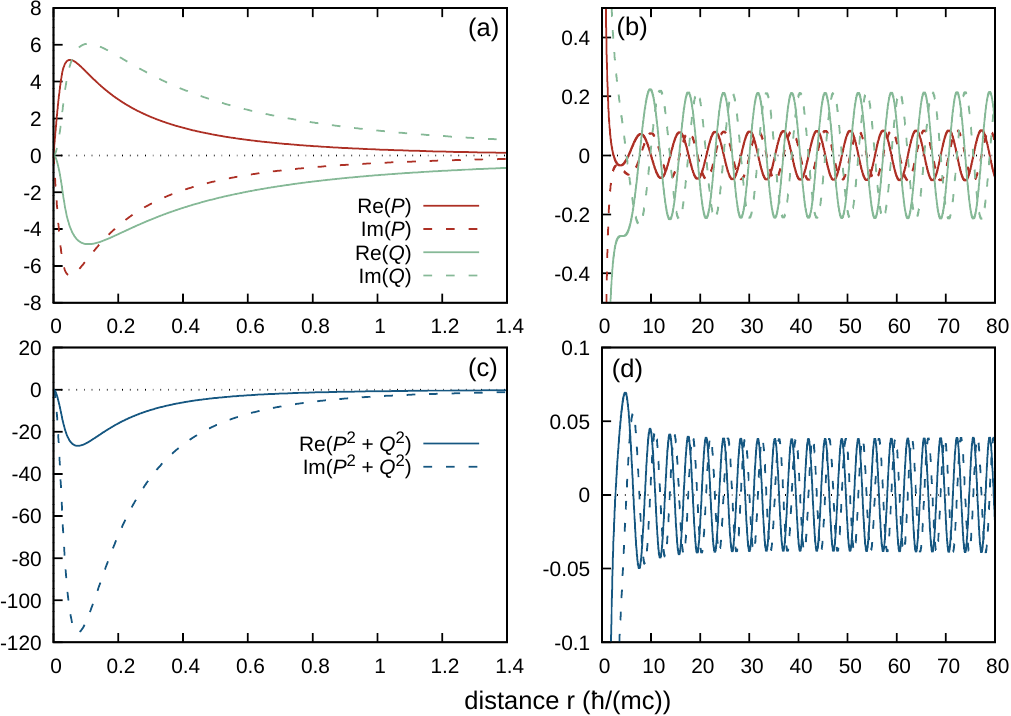}
\caption{Individual unnormalized $P$ and $Q$ components (a), (b) and the unnormalized density of the $1s$ wave function (c), (d)  for an atom with $Z=185$ and $R_{\mathrm{cut}}=0.031$ in units of $\hbar/(mc)$.}
\label{fig:PQZ185}
\end{figure}
\begin{figure}[tb]
  \centering
  \includegraphics[width=0.95\textwidth]{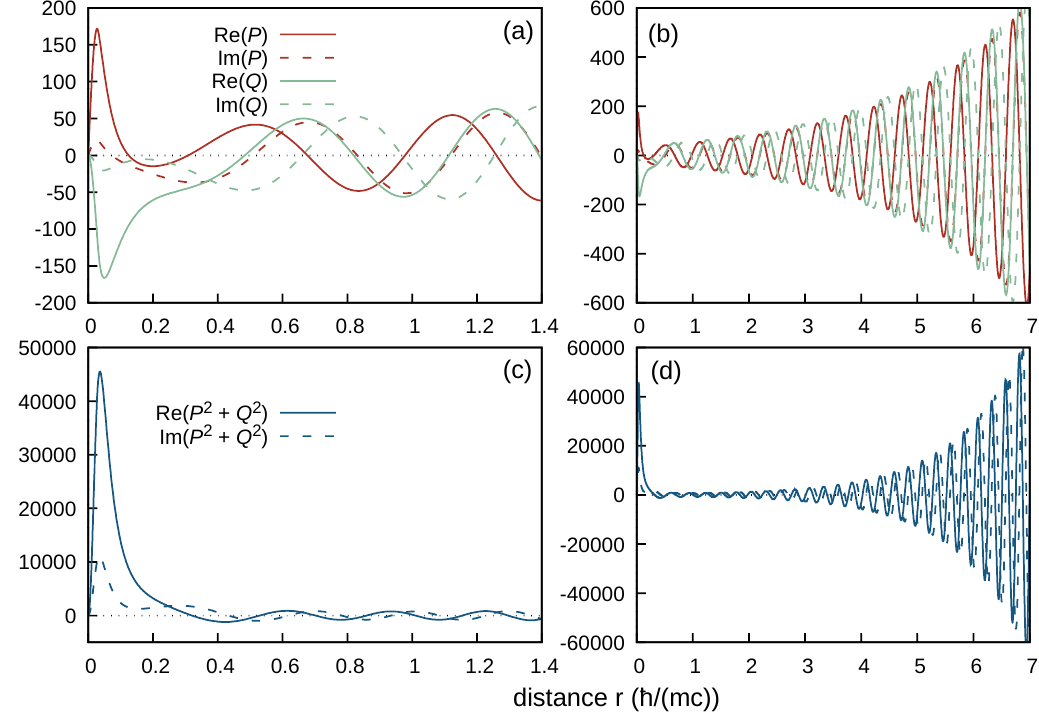}
\caption{Similar as in Fig.~\ref{fig:PQZ185} but for an atom with $Z=300$.}
\label{fig:PQZ300}
\end{figure}

Figures~\ref{fig:PQZ185} and \ref{fig:PQZ300} show the (outgoing Gamow) wave function  of a $1s$  resonant state  embedded in the negative energy continuum for (hypothetical) nuclei with charges 
$Z=185$ and $Z=300$, respectively. The wave function at short range is localized close to the nucleus.  At large distances from the nucleus (panels (b) and (d)) the wave function is dominated by the term $e^{ikr}$ and shows an exponential increasing oscillatory behaviour.

\subsubsection{Gamow states}
\label{sec:Gamow}
The narrow resonances embedded in the continuum are essentially Gamow  resonant states.
Gamow states are generalized eigenfunctions of linear operators with complex eigenvalues, which do not belong to the natural domain of a self-adjoint operators in the standard Hilbert space formalism. The mathematical foundation lies in a rigged Hilbert space (RHS) formalism
\cite{bohm1989}, which is outlined in Sec. \ref{sec:appB}. 
In scattering theory, Gamow states describe  capturing or decaying states corresponding to the poles of the scattering matrix
in the complex-momentum space. 

Gamow states have been extensively used in nuclear and atomic physics for describing resonances and other quasi-stationary states \cite{Humblet1961,Berggren1968,Berggren1982,Berggren1993,Lind1993,Bollini1996,Tolstikhin1997,Civitares2004, Michel2008, kato2001, hinze2013electron}. They were originally introduced in 1928 as resonance states by Gamow to describe $\alpha$ decay of nuclei \cite{Gamow1928,Gamow1929} and by Siegert \cite{Siegert1939}\footnote{Gamow states are sometimes also called Siegert states.} to describe scattering cross sections. For a detailed discussion of Gamow  states in nuclear physics see Refs.~\cite{Michel_2008,GSMbook}. 

\begin{figure}[tb]
\centering
\includegraphics[width=0.8\textwidth]{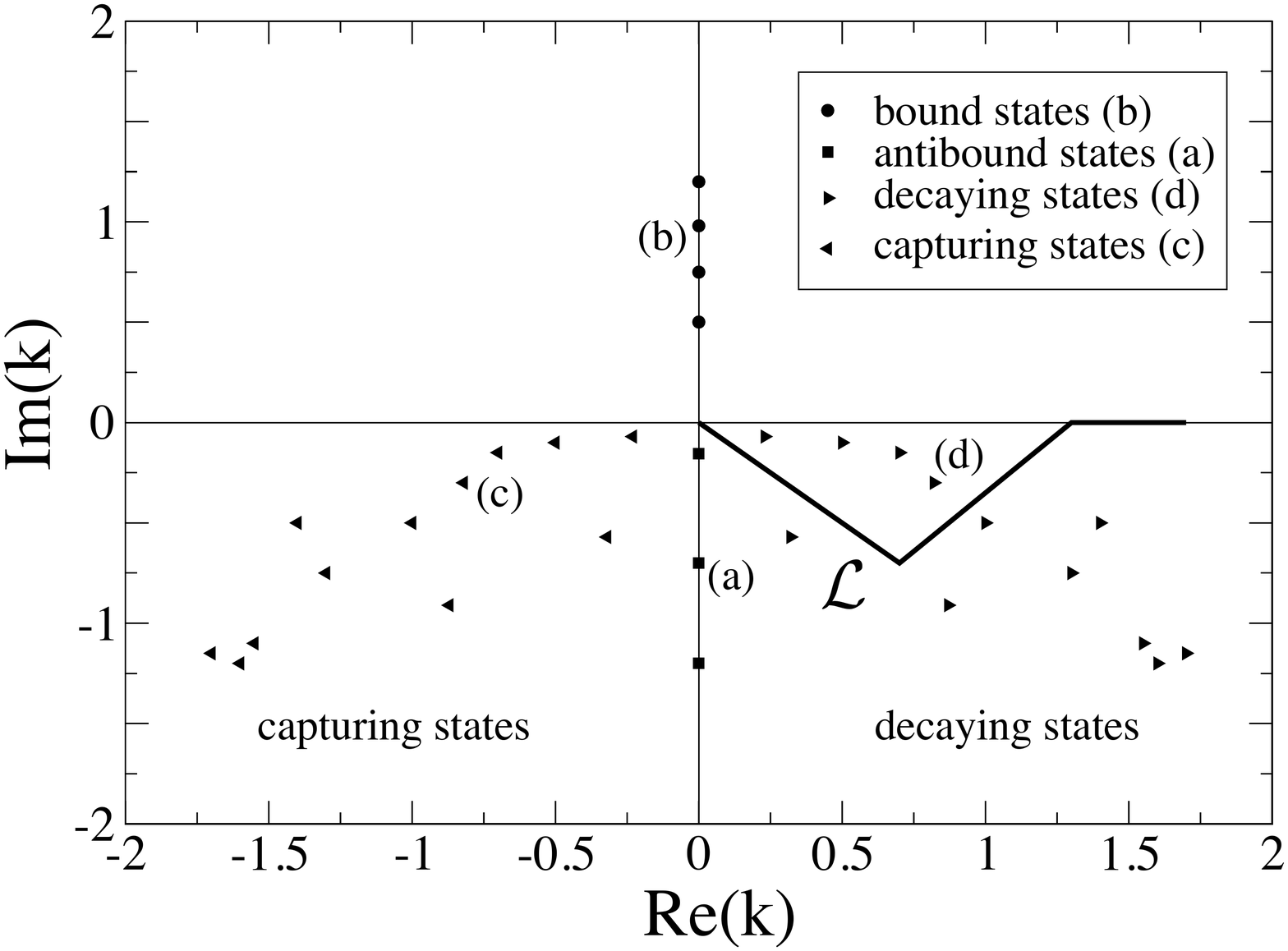}
\caption{\label{L-contour}
Location of resonant states  in the  complex momentum plane.
The Berggren completeness relation, Eq.~ \eqref{eq:Berggren},
used in the decay context involves
the bound states (b) lying on the imaginary $k$-axis, 
scattering states on the  $\cal L$ contour (solid thick line),
and resonant decaying states  (d) in the fourth quadrant of the complex $k$-plane
 lying between the real axis and  $\cal L$. For problems involving capture, the capturing resonant states (c) need to be considered and the scattering contour needs to be moved to the third quadrant. The antibound
states (a)  can be included in the generalized
completeness relation, see 
For  general expansions of the resolvent, see Refs.~\cite{Berggren1993,Lind1993}. The antibound (virtual)
states (a)  can be included in the generalized
completeness relation; in this case the scattering contour has to be slightly
 deformed \cite{Michel2006,Mao2018}.}
\end{figure}

Asymptotically, the resonant states $u_n(E_n,r)$ obey the outgoing (or incoming)
boundary condition
\begin{equation}
u_n(E_n,r)_{ \overrightarrow{\small{r\rightarrow \infty}}} \, O_l(k_n r)\sim e^{ik_n r}
\end{equation}
where $k_n= \gamma_n - i \kappa_n$ (for details see Ref.\cite{Berggren1968}). 
As shown in Fig.~\ref{L-contour}, the  bound states   with
$k_n =i\kappa_n~~(\kappa_n>0)$
lie on the positive imaginary $k$-axis while the antibound (or virtual)  states with $\kappa_n<0$ lie on the negative imaginary $k$-axis.
The decaying resonant states with
$(\kappa_n,\gamma_n>0)$ lie in the fourth quadrant of the complex $k$-plane while the capturing resonant states with
$(\kappa_n>0,\gamma_n<0)$ lie in the third quadrant. The resonant-state trajectories  in complex $k$-plane near the continuum thresholds $E=\pm m_ec^2$ have been analysed in Ref.~\cite{Krylov2020}.

The single-particle resonant states, augmented by complex-energy 
scattering continuum states $u(k,r)$ lying on the contour $\cal L$ obey the Berggren
completeness relation  \cite{Berggren1968}:
\begin{eqnarray}\label{eq:Berggren}
\sum_{n} | u_n \rangle \langle u_n | + 
\int_{\cal L} |u(k) \rangle \langle u(k)| dk = 1.
\end{eqnarray}

Since  complex-energy continuum states belong to the RHS,  the
metric  has to be
 generalized by introducing a biorthogonal basis
for the radial wave functions. In particular, contrary
to  the Hilbert space situation,
no complex conjugation appears in the radial wave functions of bra vectors \cite{Berggren1968,GSMbook}. That is why the radial densities of $1s$ states shown in Figs.~\ref{fig:PQZ185} and \ref{fig:PQZ300} are defined through squared upper and lower Dirac  components \cite{Michel2008}.
Moreover,  the
radial integrals must be regularized as
the Gamow states  with $\mathrm{Im}(k)<0$ exponentially diverge 
as $r\rightarrow \infty$, see Figs.~\ref{fig:PQZ185} and \ref{fig:PQZ300}.
This can be done  by various techniques
\cite{Zeldovich1960,Mur2003,Romo1968}, including the external complex scaling method \cite{Gyarmati1971}.
The very reason for the asymptotic growth of the Gamow state wave
function at large $r$ is the fact that such a state represents 
the stationary approach to the intrinsically time-dependent problem of decay. Indeed, 
the exponential temporal  decrease of the wave function amplitude must be complemented by
its exponential spatial increase, and this
assures that the particle number is conserved ~\cite{Baz1969}.

It is important to note, that due to the charge conjugation property of the Dirac equation, the appearance of 
the electron Gamow state in the negative-energy continuum results in the presence of a positron resonant state in the positive-energy continuum \cite{Godunov2017,Krylov2020}. This suggest an interpretation of diving electron states in terms of positron scattering resonances, see Sec.~\ref{sec:critical}.

As discussed in Sec.~\ref{sec:HFB}, resonances can also be described within the real-energy framework of standard quantum mechanics. The commonly used approach is based on the dense continuum discretization, elimination of the smooth non-resonant background, and  fitting the resonance peaks \cite{Pei2011}. Another approach is
the  stabilization method, in which resonances are extracted from phase shifts obtained  from box solutions obtained by assuming different box sizes \cite{Bacic1982,Mandelshtam1994,Zhang2008,Pei2011}. For very narrow resonances, perturbative methods, such as the two-potential method of Sec.~\ref{sec:pert} can also be used.

Despite some work on resonances embedded in the continuum \cite{greinerrafelski1985,kuleshov2015vs,Godunov2017,Krylov2020}, a direct utilization  of Dirac Gamow states in atomic many-body calculations is practically nonexistent. The basic mathematical formulation rests on the rigged Hilbert space structure which comes with its own challenges. To be of use in atomic structure calculations of the superheavy elements, Dirac Gamow states need be studied within a multi-electron framework. Computing  Dirac Berggren ensemble defined in Eq. \eqref{eq:Berggren}, which can be used in a numerical atomic structure program packages,  will offer many exciting avenues.

\subsection{Positron production in the super-critical regime}\label{sec:critical}

The QED vacuum is unstable in the presence of a strong electromagnetic field above the Schwinger field limit, $E_S=m_e^2c^3/e\hbar =$ \SI{1.32e-18}{\volt\per\meter} (or the equivalent intensity of $I_S=$\SI{2.3e29}{\watt\meter^{-2}}) \cite{Koga2020}, and decays by emitting electron-positron pairs \cite{Klein1929,Sauter1931,Schwinger1951}. In the case of a potential barrier, it results in the much discussed and debated Klein's paradox. 

As pointed out in Ref.~\cite{Hansen1981},  pair production cannot be described within a one-body Dirac theory: it requires quantum field theoretical treatment within a time-dependent rigged Fock-space formalism that dynamically couples particles (electrons) and holes (positrons) in the Dirac continuum. A close non-relativistic analogy to this problem is a two-nucleon nuclear decay of a Gamow resonance \cite{Wang2021}. A concise mathematical treatment in terms of incoming and outgoing electron/positron states is given by Rumpf, where the outgoing basis may be connected with the ingoing one by a unitary Bogoliubov transformation \cite{Rumpf1979I,Rumpf1979II,Rumpf1979III}. For further details see Ref.~\cite{Soffel1982}. 

As discussed in Sec. \ref{sec:Gamow}, the resonance states of the supercritical Dirac equation are the Gamow states. The  physical interpretation of an electron state embedded in the continuum was extensively studied by the Frankfurt group \cite{greinerrafelski1985}. According to these works, if an empty level is embedded in the negative energy continuum, the initially neutral vacuum  can  spontaneously decay into a positron and  a bound electron with a supercritical energy. In such a case, an empty level in the Dirac sea is interpreted as a positronic state, with the positron escaping the supercritical field. After two positrons are emitted, the supercritical $K$-shell has been successively filled with two electrons, and the Pauli principle prevents further decay \cite{Pieper-1969,Gershtein1969,popov1971positron, Zeldovich_1972,Tomoda1982,greinerrafelski1985,Reus1988, Ackad2008}.\footnote{We could, in principle, excite an electron from the filled Gamow $1s_{1/2}$ state in the continuum into one of the discrete states above $-m_ec^2$. This creates another hole in the state embedded in the negative energy continuum and the possibility for yet another pair creation.} The resonance's width has been interpreted as the positron escape width with the characteristic time $\tau_E=\hbar/\Gamma$ for the pair creation process. 

This picture of pair creation was debated \cite{kuleshov2015vs, Kuleshov2017, Krylov2020} on the basis of the unitarity of the $S$-matrix. 
Indeed, the unitarity of the partial scattering matrix is equivalent to the absence of inelastic channels, in particular, the absence of spontaneous electron-positron creation.
in which it  has been proven that for a \textit{static} external field the probability of pair creation is exactly zero \cite[p. 298]{thaller1992}. However, the probability for pair creation does not go exactly to zero as the time derivative of the external field approaches zero. Instead, in the adiabatic limit one observes a sudden jump in the probability of adiabatic pair creation for critical fields which may be defined as spontaneous pair creation \cite{thaller1992}. That is, one requires only a weak time dependence to trigger pair creation. Consequently, rather than to talk about ``spontaneous pair creation'', it has been recommended to use  ``adiabatic pair creation''  \cite{Pickl_2008,Pickl2008a}. Recently, the vacuum polarization energy decline and spontaneous positron emission in QED under Coulomb supercriticality were explored within the Dirac-Coulomb problem with an external static or adiabatically slowly varying spherically symmetric Coulomb potential created by a uniformly charged sphere \cite{Grashin2022}. It was found that in the supercritical region the vacuum polarization energy is a decreasing function of the Coulomb charge, resulting in a decay, with a  vacuum polarization energy $\mathcal{E_{\mathrm{VP}}}^{\mathrm{ren}}\sim -Z^4/R(Z)$, which provides the required energy for positron emission (\cite{Grashin2022} Eq.~(104)). Here $R(Z)$ is the nuclear radius. The vacuum polarization and its effect on the value of supercritical $Z$ are also studied in \cite{kas2022}.
This debate could, however, have been avoided by referring to Thaller's  work on scattering operators \cite{thaller1992}.

In principle, one could initiate pair creation using intensive laser fields above the Schwinger limit $I_S$ (see Ref.~\cite{gbmm2022} for a recent review). One proposal is by using multiple\footnote{An electromagnetic plane wave that fulfills $E^2=B^2$ and $\vec{E}\cdot \vec{B}=0$ cannot produce electron-positron pairs.} focused beams from x-ray free electron lasers \cite{Alkofer2001}. Repeated cycles of particle creation and annihilation can take place in tune with the laser frequency and the production of a few hundred particle pairs per laser period can occur. 
As an analogous approach, Ref.~\cite{Klar2019} proposed a model of the quantum Dirac field realized by ultra–cold fermionic atoms in an optical lattice. Here, numerical simulations demonstrate the effect of spontaneous pair creation in the optical analogue system. Yet another possibility is to use a strong laser beam coupled to an atomic or molecular system with a strong Coulomb field as found for example in graphene \cite{Geim2007,Castro2009,Fillion2015,Kuleshov2017}. A Schwinger-like production of hot electron-hole plasma in semi-metallic graphene has been claimed to be observed for the first time only very recently \cite{Geim2022}.

\subsection{Experimental perspective: heavy-ion collisions}

\begin{figure}[htb]
  \begin{center}
\includegraphics[width=0.67\textwidth]{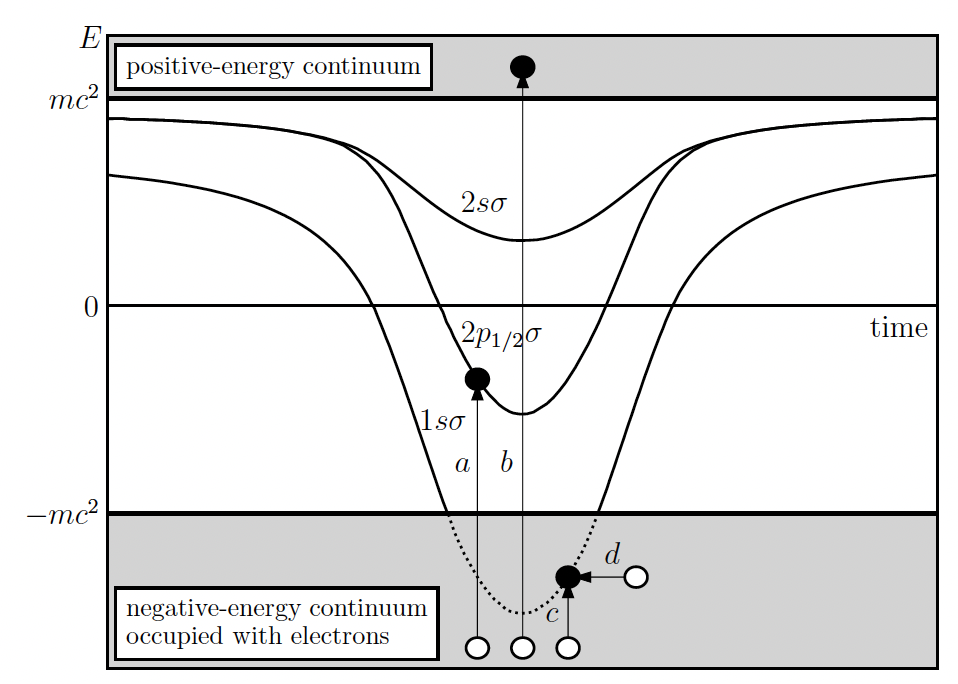}
  \caption{ The low-lying energy levels formed by the collision of two uranium nuclei as functions of time. The arrows $a$, $b$, and $c$ denote different dynamical pair-creation mechanisms and the arrow d indicates the spontaneous pair creation. The $1s$ state dives into the negative-energy continuum for about \SI{1E-21}{\second}. Figure taken from \cite{popov2020access}; see also \cite{Szpak_2008}. }
  \label{fig:SpontaneousDynamical}
  \end{center}
  \end{figure}

It was proposed, that pair creation should occur in the collision between two bare nuclei with total charge number exceeding the critical value, such as the case for two U$^{92+}$ ions with a combined nuclear charge of $Z=184$ \cite{Reus1988,greinerrafelski1985}. The collision system will have a supercritical regime time for \SI{\sim 2.3e-21}{\second} for the U$^{92+}$+U$^{92+}$ collision at center-of-mass energy of $E_{\mathrm{cm}}=$\SI{740}{\mega\electronvolt} \cite{Ackad2008} as shown in Fig. \ref{fig:SpontaneousDynamical}. The expected lifetime of the supercritical resonance state is  \SI{\sim 392E-21}{\second}, which is two orders of magnitude shorter than the time required for vacuum decay. The probability of pair production is therefore estimated to be around $1\% $ for the 1s level \cite{popov2020access}. Early attempt to observe this effect \cite{tbbc1992} using ions without a $1s_{1/2}$ hole failed. The use of cooled U$^{92+}$ ions in the ESR storage ring of GSI/FAIR \cite{dijk2019} could allow to observe this effect for the first time. A test experiment using collisions of a Xe$^{54+}$ beam on a Xe gas jet target is underway \cite{gkzt2020}.

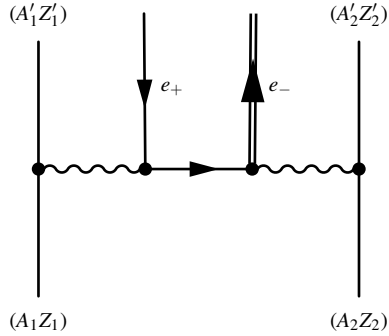
\begin{figure}[tb]
\vspace{-1pt}
\centering
\begin{tabular}{@{\extracolsep{8mm}}lll}
\begin{fmffile}{figure}
	\fmfframe(5,7)(0,7){
\begin{fmfgraph*}(120,120)
    \fmfbottom{i1,i2}
    \fmftop{o1,o2,o3,o4}
    \fmf{plain}{i1,v1,o1}
    \fmf{plain}{i2,v2,o4}
    \fmffreeze
	\fmfdot{v1}
	\fmfdot{v2}
	\fmfdot{g1}
	\fmfdot{g2}
    \fmf{photon}{v1,g1}
    \fmf{fermion}{g1,g2}
    \fmf{fermion,label=$e_+$,tension=0}{o2,g1}
    \fmf{dbl_plain_arrow,label=$e_-$,tension=0,label.side=right}{g2,o3}
    \fmf{photon}{g2,v2}
    \fmfv{label=($A_1Z_1$),label.angle=-90}{i1}
	\fmfv{label=($A_2Z_2$),label.angle=-90}{i2}
	\fmfv{label=($A_1'Z_1'$),label.angle=90}{o1}
	\fmfv{label=($A_2'Z_2'$),label.angle=90}{o4}
\end{fmfgraph*}
	}
\end{fmffile}
\end{tabular}
\caption{Schematic Feynman diagram for the dynamical pair creation for the (inelastic) collision of two heavy nuclei with mass numbers and nuclear charges $(A_1,Z_1)$ and $(A_2,Z_2)$ respectively, where the outgoing nucleus binds an electron ($Z^{'}_2 + e^-$). Two colliding nuclei create a strong electromagnetic field, strong enough to generate an electron positron pair \cite{Artemyev2012}.  Colliding nuclei are represented by normal lines while  wavy lines refer to  virtual photons and the  lines with arrows correspond to leptons (electrons and positron). The double line represents a bound electron.}
\label{fig:creation}
\end{figure}

The spontaneous emission is not the only process that can occur during the collision.\footnote{One should  not forget  possible  weak decay processes such as the electron capture that is a common decay mode  of proton rich nuclei, albeit the time frame for weak decays is much longer than for nuclear or electronic transitions  \cite{Heenen2015}. Take for example the work on relativistic quantum dynamic calculations of the probability of K-vacancy production  in the Xe-Xe$^{54+}$ collision at 30\,MeV \cite{Kozhedub2015}.}
It is generally masked by a dynamical positron emission, which is induced by the time-dependent potential of the colliding nuclei above the Coulomb barrier \cite{szpak2012optical,Lee2016, maltsev2017pair, Maltsev2015,popov2020access}. In this mechanism, the two colliding nuclei create a strong electromagnetic field, strong enough to generate electron-positron pairs. The pair creation in heavy atom collisions is visualized by the Feynman diagrams in Fig. \ref{fig:creation} \cite{Szpak_2008}.

While the spontaneous pair creation works only in the supercritical regime, the dynamical pair creation takes place in both subcritical \cite{Lee2016} and supercritical modes if the collision energy is high enough \cite{Khriplovich2016, Khriplovich2017}.
Experimental verification of spontaneous pair creation and the distinction from the dynamical process is however challenging as the energy-differential spectra of emitted positrons by spontaneous vacuum decay are indistinguishable from the spectra of positrons emitted by the dynamical process. There are however a range of different approaches that should make vacuum decay observable.
One example is by collisions with nuclear sticking, in which nuclei are bound to each other for some period of time by nuclear forces allowing for few nuclear rotations. In this very short time frame, typically of the order of \SIrange{1e-21}{1e-20}{\second} \cite{duRietz2013,Simenel2020}, there is an increase in pair creation probability that can only be explained with the spontaneous pair creation mechanism \cite{Reinhardt1981, Maltsev2015,Reus1988}. 
Additionally, it has been shown that the pair-production probability varies as a function of nuclear collision velocities in the supercritical and subcritical region, allowing for the detection of vacuum decay experimentally \cite{Maltsev2019, popov2020access}.
The impact of the vacuum polarization on the value of  $Z_\mathrm{c}$ in the case of heavy ions collisions is considered in Ref.~\cite{kas2022}. Moreover, it has been argued \cite{popov2020access} that the positron spectra for symmetric collisions of heavy ions with $83\le Z\le 96$ as a function of the collision energy should show a  signature of the transition to the supercritical regime.

\section{Multi-Configuration Dirac-Hartree-Fock}\label{sec:MCDHF}

With very few exceptions \cite{Nakatsuji2005,Nakatsuji2012}, one treats the multi-electron Dirac equation within mean-field theory, that is either at the D-HF (Dirac-Hartree-Fock) level or by using D-DFT (Dirac density functional theory) \cite{Savin1983,Engel1995}, with the latter method being more popular in molecular calculations. It is fair to say that the accuracy of current density functional approximations cannot compete with wave-function-based methods (for a recent critical analysis on DFT see Ref. \cite{teale2022}), especially when QED effects need to be included. At an early stage of atomic structure calculations, however, DFT in the form of D-HF-Slater theory did play an important role as electron correlation is approximately included in such a scheme \cite{Mann1973}. 
Here, we focus on modern multi-reference D-HF electronic structure theory for static correlation describing correctly the states of a given $J^\pi$ symmetry, with $J$ being the total angular momentum and $\pi$ the parity. Dynamic electron correlation and its effects on atomic structure is described in Sec.~\ref{sec:elcor} below.

Like in the nonrelativistic HF case, to obtain the correct ground state symmetry and low-lying electronic transitions in open-shell cases, one requires the correct description of static correlations. In finite basis-set calculations this requires a set of Slater determinants in a multi-reference treatment within a nonrelativistic or relativistic coupling scheme. In relativistic atomic numerical program packages such as GRASP \cite{DyaGraJoh89,fischer2016,Jonsson2013,FroeseFischer2019,Grant2022} or MDFGME \cite{desclaux1975, ibj2011}, this is done through linear combinations of multi-shell configurational state functions (CSF’s) within a $jj$-coupling scheme \cite{grant2007relativistic}:
\begin{equation}
\label{configstatefunc}
\Psi_i(J^\pi,M_J)= \sum_r c_{ri} \Phi_r\left( \gamma_\nu J^\pi,M_J\right),
\end{equation}
where the $\Phi_r$ wavefunctions share the same overall total angular momentum $J$, corresponding $M_J$, and parity $\pi$. The quantity $\gamma_\nu$ stands for all other values such as angular momentum recoupling and seniority numbers \cite{grant2007relativistic}. Each CSF $\Phi_r$ is a linear combination of Slater determinants 
\begin{equation}
\label{configstatefunc1}
\Phi_r\left( \gamma_\nu J^\pi,M_J\right)=\sum_i d_i 
\begin{vmatrix}
\phi_1^i(r_1) & \dots & \phi_N^i(r_1) \\
\vdots & \ddots & \vdots\\
\phi_1^i(r_N) & \dots & \phi_N^i(r_N) 
\end{vmatrix},
\end{equation}
where $\phi$ are the Dirac four-component orbital spinors defined in Eq. \eqref{eq:orbital}, and the coefficients $d_i$ are determined such that the CSF
is an eigenstate to both $J^2$ and $J_z$. The eigenvalues and eigenvectors (configuration mixing coefficients $c_{ri}$) are then obtained by diagonalizing the Hamiltonian matrix  $H_{ij}=\bra{\Psi_i(J^\pi,M_J)}  H_D  \ket{\Psi_j(J^\pi,M_J)}$. 

Multi-reference methods (including complete active space SCF) used in the quantum chemistry community have been reviewed extensively \cite{Hirao1999,Fleig2012,li2020}. A comprehensive account on MCSCF theory in relativistic atomic structure calculations (usually termed MCDHF) has been provided in a textbook \cite{grant2007relativistic} and several publications \cite{grant1970,ddei2003}. The construction of these multi-reference functions can be a formidable task if many high angular momentum open-shell $j$-states  are involved \cite{grant2007relativistic}. The multi-reference treatment, therefore, provides a challenge for superheavy element calculations where the electronic spectrum becomes very dense and, as a result, the multi-reference space becomes huge.\footnote{This is similar to the strong correlation problem in solid state physics to describe, for example, metallic systems.} In addition, SCF convergence problems can arise for nearly-degenerate states. Nonetheless, for few-electron systems, high-accuracy in excitation energies can be achieved if both QED and dynamic correlation effects are included, see Secs.~\ref{subsec:qed-res} and \ref{sec:elcor}, respectively. 
High-accuracy atomic structure calculations are also required, for example, in the search of physics beyond the standard model (BSM) \cite{Karshenboim2005,ind2019,kpd2019,chcj2020,fbdf2022,sbdk2018,bbdf2018,bdgs2020,myas2021,adfa2021,mdf2022,hcck2022}. 

Numerical program packages, such as MCHF for the nonrelativistic \cite{froesefischer1977} case or GRASP and MDFGME for the relativistic case, apply the finite difference method (FDM) \cite{hartree1957book}.
Alternatively, the finite element method (FEM) employing, for example, B-splines (piecewise polynomials) \cite{Johnson1988,saj1996,styp2004,faz2009} can be used, as implemented for example in the program AMBiT \cite{Berengut2019}. 
The use of B-splines has certain advantages in relativistic atomic structure calculations  \cite{saj1996}. As the radial wave functions are restricted to an interval  $[0,R_\mathrm{c}]$, the atoms are spherically confined within a radius $R_\mathrm{c}$ set large enough (usually around \SI{40}{a.u.}) to achieve accurate numerical results. 
This discretizes the positive and negative real-energy continuum.   It thus allows for an easy implementation of projection operators \cite{ind1995}. This method could therefore be well suited to approximately describe diving occupied levels with $E<-m_ec^2$ at charges $Z>Z_{\mathrm{c}}$.\footnote{The accuracy of such a discretization procedure has been shown to be poor, when it comes to the description of narrow resonance states \cite{Pei2011}. In such a case, the preferred method to deal with these resonances is the Gamow-state framework.}  The virtual space created can be used for a successive electron correlation procedure, such as configuration interaction or coupled cluster or MCDF. In all these numerical procedures one usually chooses exponentially spaced grid points (called knots in FEM) with $r=r_0e^t, t>0$, to describe the radial wave function $\phi(r)$ accurately in the near nuclear region. We note that the correct description of the wave function in the inner core region is mandatory for the accurate treatment of relativistic effects \cite{Schwarz-1990,Schwarz-Wezenbeek-1989}. B-splines have also been used to create basis sets to perform many-body perturbation theory \cite{jbs1988,saj1996,jbs1988,jbs1988a} or to do MCDF calculations, as they can be used to implement projection operators with the nucleus and electrons average potential, and obtain correlation orbitals \cite{ind1995}. More recently an improved method, the dual kinetic balance \cite{styp2004}, has been proposed to obtain basis sets free of spurious states.

The systems of coupled integro-differential equations obtained in multi-configuration methods are intrinsically very non-linear. In particular exchange potentials for correlation orbitals are inversely proportional to the square of the configuration weight, and can then be huge. Initial configuration state functions for an SCF calculation are usually obtained from either the Thomas-Fermi model or from single-particle Dirac-Coulomb solutions using screened nuclear charges \cite{grant2007relativistic}. However, severe convergence problems can be experienced when, for example, diffuse orbitals are involved such as for high angular momentum functions or negatively charged atoms, or when doing correlation calculations with highly-excited configurations. In such cases, choosing the right initial guess becomes important. Convergence issues within the MCDF procedure have been discussed in Refs.~\cite{Chantler2000,grant2007relativistic,Chantler2010,isbd2007}. In some cases the problem occurs due to the relativistic nature of the atom or ion being studied. When going to very high-$Z$ the angular coupling goes from $LSJ$ coupling to almost pure $JJ$ coupling. In that case the weight of some of the configurations contributing to a given $LSJ$ level becomes very small and severe convergence problems are observed \cite{isbd2007}. 

When the four components of the spinor in relativistic methods are each allowed to vary independently, the matrix representations of the Dirac operator will fail to give the right formal nonrelativistic limit, resulting in an energy below the true numerical value, known as variational collapse or finite basis set disease \cite{Kutzelnigg1984}. It arises whenever one wants to expand wave functions in a given basis replacing operators by their matrix representation. To prevent such an unwanted effect, certain boundary conditions such as the kinetic balance (which is automatically considered in numerical calculations) have to be imposed which ensures the correct relation between the large and small component \cite{Kutzelnigg1984,grant2007relativistic,Dolbeault2000}. In finite basis set treatments of the D-HF equations, using for example Slater or Gaussian type basis sets, small errors may nevertheless occur due to variational problems (prolapse). Since the kinetic balance condition implicitly projects onto the positive energy states, it is possible that, due to the incompleteness of the basis set, the total energy lies below the one obtained from numerical DHF calculations \cite{dyall2007book}. This can  be avoided by freezing the inner core functions such that core orbitals are sufficiently well described, or by restricting the size of the $s$ and $p$ basis sets, or by making use of specifically derived prolapse-free Gaussian basis sets \cite{Tatewaki2003,Tatewaki2004,DeMacedo2007,Teodoro2014}.

Upon inclusion of the Breit operator in Eq.~\eqref{eq:QEDHamiltonian}, coupling of the positive and negative continuum states  occurs due to  electron-electron interaction, leading  to the non-existence of a discrete spectrum. This is known as continuum dissolution or the Brown–Ravenhall disease \cite{Brown1951}, and can be avoided by removing all Slater determinants containing negative-energy orbitals using a projection operator, effectively eliminating electron-positron pair contributions \cite{Sucher1980}. The projection operator is usually constructed from the positive energy eigenstates of the full external field Dirac Hamiltonian, leading to the no-pair Hamiltonian of Sec.~\ref{EPDE}. (For a recent discussion on this topic see \cite{Saue2016}.)  For the case where photon-matter field interactions are removed, a single Slater determinant (D-HF solution) automatically includes the HF projection operators on positive energy states \cite{Mittleman1981}, \ie the low-frequency Breit interaction has been shown to cause no variational failure when included in the iterative solutions of the
D-HF equations \cite{Quiney-1987a,Quiney-1987}. Thus, the Breit interaction has been successfully applied perturbatively \cite{Gorceix1988} as well as in variational treatments \cite{Ley-Koo1997,grant2007relativistic,Lindroth_1987,Quiney-1987,Gorceix_1987,ind1995,daSilva1996,Thierfelder2010}, where the solutions of the Dirac-Breit-HF equations serve as a starting point for further electron correlation and QED treatments.

An other issue with Dirac-Fock codes is the fact that for levels originating from the same $LS$ level, they may give wrong values.
It was shown in \cite{hkcd1982} that the $2p_{1/2}-2p_{3/2}$ fine structure energy in B-like ions and the $2p^5\,J=3/2-2p^5\,J=1/2$ one in F-like ions did not provide the right value for light elements. The non-relativistic limit obtained by setting the speed of light to a high value was not zero as it should have been. At the time the proposed solution was to remove the energy splitting obtained for $c\to \infty$ from the relativistic value. More recently it was shown that this effect could be handled by doing large scale correlation calculations to obtain those level  energies, including all single excitations, even the ones obeying the Brillouin theorem \cite{ild2005}. The same issue was also identified in the evaluation of forbidden transitions probabilities \cite{kpmi1998}.

\section{Quantum Electrodynamic Effects}
\label{sec:qed}

Besides the corrections stemming from relativistic electron correlation described in Sec. \ref{sec:MCDHF}, corrections issued from bound-state quantum electrodynamics must be added to get accurate predictions.  The need for such corrections was demonstrated by two famous experimental discoveries. The first discovery, made by Lamb and Retherford, was the non-degeneracy between the $2p_{1/2}$ and $2s_{1/2}$ states, in contradiction to the Dirac equation, which gives degenerate levels \cite{lar1947}. The second discovery, made by Kusch and Foley, was that the electron Land\'e $g$-factor  is not exactly equal to \num{2} in Na and Ga \cite{kaf1948}, later understood to be due to the anomalous magnetic moment of the electron. The experimental discoveries were followed by the theoretical work of Bethe \cite{bet1947}, Feynman \cite{Feynman1949a,Feynman1949}, Schwinger \cite{sch1948a, sch1949a,sch1949,Schwinger1951} and Tomonaga \cite{tat1948}, which lead to the foundation of QED, the principle of which remains unchallenged up to now \cite{ind2019}.

The derivation of the different QED contributions starts from the QED Lagrangian \eqref{eq:qed-L}. Several methods have been proposed to calculate all-order QED corrections which are necessary for applications to high-$Z$ elements. However, it is not trivial to define physical particle states in the presence of an external gauge field within the framework of gauge invariant quantum field theory \cite{Soffel1982}.
Pioneering works  on all-order vacuum polarization \cite{WichmannKroll1956,bls1959,bam1959} have led to the modern calculations. The first accurate all-order calculation of the $1s$ self-energy \cite{moh1974,moh1974a} showed that $Z\alpha$ expansions used up to that time were non-convergent at medium- and high-$Z$. A first attempt to evaluate the $1s_{1/2}$ state self-energy in superheavy elements was done in Ref.~\cite{caj1976}. It was followed by the calculation  of the self-energy contribution of the $1s_{1/2}$ level for finite nuclei up to $Z=170$ \cite{ssmg1982}. This evaluation has recently been extended to all states up to $n=5$ and $J=5/2$ \cite{mgst2022}.
The method  described in Ref. \cite{moh1974} is based on the $S$-Matrix formalism, which allow a full treatment of QED corrections in one-electron systems and to calculate corrections to the electron-electron interaction in few-electron systems beyond the no-pair approximation \cite{bmjs1993,mas2000}, provided there is a well isolated reference system.  A review of QED corrections in low-$Z$ one-electron systems can be found in Ref.~\cite{egs2001}.

The Bethe-Salpeter equation \cite{BetheSalpeter1951} is a real two-body equation that has been used to derive, for example, higher-order recoil corrections in hydrogen \cite{sal1952} beyond what can be obtained with the Breit equation (see, \eg  Ref.~\cite{egs2001} and references therein). The Bethe-Salpeter equation has, however, some fundamental problems \cite{Nakanishi1965,Nakanishi1965a} and it  becomes soon intractable for many-electron systems. For the efficient treatment of many-electron systems one requires a Hamiltonian approach (\eg the Dirac-Coulomb-Breit Hamiltonian as a starting point) with additional effective QED perturbation terms  that describe the multi-electron system to the required accuracy.  The Bethe-Salpeter equation can in principle be transformed into two independent equations that match the equations of Hamiltonian relativistic quantum mechanics \cite{Sazdjian1987}; there is also the quasipotential approach \cite{Ramalho2002}).

Three methods have been developed to deal with bound state QED (BSQED) calculations, in particular in heavy-elements. The original one is based on the $S$-matrix formalism.  A detailed description of the $S$-matrix formalism and review of QED calculations based on it can be found in Refs.\,\cite{mps1998,iam2017,jaa2021}. An overview of this method is given in subsection \ref{subsec:s-matrix}.
Approaches capable of  dealing with quasi-degenerate reference states have been proposed  by using  (i) a method  based on the two-times Green function \,\cite{sha2002,art2017a,art2017} (Subsec. \ref{subsec:two-times}) and (ii)  a covariant version of RMBPT based on the time-evolution operator, which allow to treat more easily degenerate and quasi-degenerate states \cite{sap1993,lin2000,lasm2001,lai2017} (Subsec. \ref{subsec:evol-op}). In practice, the complexity of the involved calculation is the main limitation to the use of any of these approaches, and approximate methods had to be devised.

BSQED is usually based on the Furry bound picture \cite{fur1951}. The unperturbed Dirac Hamiltonian $H_{D}$ contains the Coulomb field of the nucleus, such that the Coulomb potential is included to all orders. The electron-electron interaction is treated as a perturbation given by the potential
 \begin{equation}
V_{\epsilon,g}=gH_{I}e^{-\epsilon|t|},
  \label{eq:vpot}
 \end{equation}
 where $g$ is a formal expansion parameter and the interaction Hamiltonian is
 \begin{equation}
H_{I}=j^{\mu}A_{\mu}-\delta M(x),
\label{eq:hpot}
\end{equation}
which contains a mass renormalization term. As the electromagnetic interactions can act at an infinite distance, the term $e^{-\epsilon|t|}$ is added to turn off adiabatically the interaction at $t=\pm \infty$ to recover the unperturbed states before and after the interaction. 

The electron-positron field operators defined on an appropriate Fock-space are expanded in terms of electron and positron annihilation and creation operators, 
\begin{equation}
\psi(x)= \sum_{E_n > -m_ec^2} a_n \phi_n(x)+\sum_{E_m<-m_ec^2} b_m^{\dagger} \phi_m(x),
\label{eq:field-op-sbuc-gen}
\end{equation}
while the BSQED Hamiltonian is given as \cite{Soffel1982}, 
\begin{equation}
H_0 = \sum_{E_n > -m_ec^2}  E_n a_n^{\dagger}a_n \phi_n(x)-\sum_{E_m<-m_ec^2} E_m b_m^{\dagger} b_m \phi_m(x),
\label{eq:2nd_qant_ham-gen}
 \end{equation}
where $a_n$ an electron annihilation operator for an electron in state $n$, with energy $E_n>-m_ec^2$ and $b^{\dagger}$ 
a positron creation operator for a positron in state $m$ with energy $E_m<-m_ec^2$. For the Gamow states that dive into the negative energy continuum and have complex energies, the formalism has to be further extended. It should be noted that the formalism in Eqs.\eqref{eq:field-op-sbuc-gen} and \eqref{eq:2nd_qant_ham-gen} is the proper quantum-field theory replacement for the Hamiltonian with projection operators given in Eq. \eqref{eq:diracfem-projection-h}, which is based on the Dirac sea definition of the positrons. Yet, for practical applications in many-electron systems, the BSQED formalism is too difficult to use, and has not been used beyond second-order corrections. 

The expressions \eqref{eq:field-op-sbuc-gen} and \eqref{eq:2nd_qant_ham-gen} are usually formulated in terms of the positive ($E_n > 0$) and negative ($E_n < 0$) spectrum of the Dirac operator, loosely termed electronic and positronic states \cite{mps1998}. Such terminology originates from a free-particle QED formalism \cite{FurryOppenheimer1934,dys1949}, and was later adopted for Coulomb fields describing a point nuclear charge where the lower part of the discrete spectrum terminates at $E_n = 0$ at $Z\alpha=1$. As already pointed out, for the general case of a finite nucleus the energy can become negative and eventually the state can dive below $E=-m_ec^2$ for $Z\approx 170$. Hence the terminology of positive and negative energy states makes only sense if one shifts the spectrum up by $m_ec^2$ where the lower continuum starts then at $E<0$.

\subsection{$S$-matrix formalism}
\label{subsec:s-matrix}
The evaluation of the energy shift in QED for an isolated $q$-electron state 
with no real photons $\left|N_{q};0\right>=\left| 
n_{1},\ldots , n_{q};0\right>$, is made through the Gell-Mann and Low theorem \cite{gal1951,faw1971}, symmetrized by Sucher \cite{suc1957}
\begin{equation}
 \Delta E_{N_{q}}=
  \lim_{
   \stackrel{\epsilon \rightarrow 0}{ g \rightarrow 1}
       }
 \frac{i\epsilon g}{2} \frac {\partial}{\partial g}\log \left< N_{q};0 \right|  
 S_{\epsilon,g}\left| N_{q};0\right> \, ,
   \label{eq:gmlt}
\end{equation}  
where the adiabatic $S$-matrix is given by
\begin{equation}
S_{\epsilon,g}=\lim_{t \rightarrow \infty 
}U_{\epsilon,g}(-t,t),
\label{eq:s-matrix}
\end{equation}
and $U_{\epsilon,g}$ is the adiabatic evolution operator defined as
\begin{equation}
 U_{\epsilon,g}\left(t_{1},t_{2}\right)=T e^{-i \int^{t_{2}}_{t_{1}}dt 
 V_{\epsilon,g}(t)} \, ,
 \label{eq:evolop}
\end{equation}
where $T$ is the time ordering operator.

The next step is to expand the connected adiabatic $S$-matrix in power of the coupling constant $g$ as done in Refs.~\cite{dys1949,moh1989,moh1996,mps1998,iam2017} 
\begin{eqnarray}
\left. g \frac {\partial}{\partial g}\log \left<S_{\epsilon,g} 
\right>_C\right|_{g=1}  & = & \frac{
 \left<S_{\epsilon,1}^{(1)} \right>_C+2\left<S_{\epsilon,1}^{(2)} 
 \right>_C+3\left<S_{\epsilon,1}^{(3)} \right>_C+\cdots}
 {1+\left<S_{\epsilon,1}^{(1)} \right>_C+\left<S_{\epsilon,1}^{(2)} 
 \right>_C+\left<S_{\epsilon,1}^{(3)} \right>_C+\cdots}
 \nonumber  \\
  & = & \left<S_{\epsilon,1}^{(1)} \right>_C+2\left<S_{\epsilon,1}^{(2)} 
 \right>_C-\left<S_{\epsilon,1}^{(1)} \right>_C^{2}
 \nonumber  \\
  & + & 3\left<S_{\epsilon,1}^{(3)} \right>_C
  -3\left<S_{\epsilon,1}^{(1)} \right>_C\left<S_{\epsilon,1}^{(2)} \right>_C
  +\left<S_{\epsilon,1}^{(1)} \right>_C^{3}
 \nonumber  \\
  & + & 4\left<S_{\epsilon,1}^{(4)} \right>_C
  -4\left<S_{\epsilon,1}^{(1)} \right>_C\left<S_{\epsilon,1}^{(3)} \right>_C
  -2\left<S_{\epsilon,1}^{(2)} \right>_C^{2}
    \nonumber  \\
  & + & 4\left<S_{\epsilon,1}^{(1)} \right>_C^{2}\left<S_{\epsilon,1}^{(2)} \right>_C
  -\left<S_{\epsilon,1}^{(1)} \right>_C^{4} \, ,
 \label{eq:smatexp}
\end{eqnarray}
where the connected $S$-matrix is defined by
 \begin{equation}
 \left<S_{\epsilon,g} \right>_C  =  \left< N_{q};0 \right|  
 S_{\epsilon,g}\left| N_{q};0\right>_C =\sum_j\left<S_{\epsilon,1}^{(j)} \right>_C \, ,
\end{equation}
with
\begin{equation}
 \left<S_{\epsilon,1}^{(j)} \right>_C  =  \left< N_{q};0 \right|  
 S_{\epsilon,1}^{(j)}\left| N_{q};0\right>_C \, .
 \end{equation}

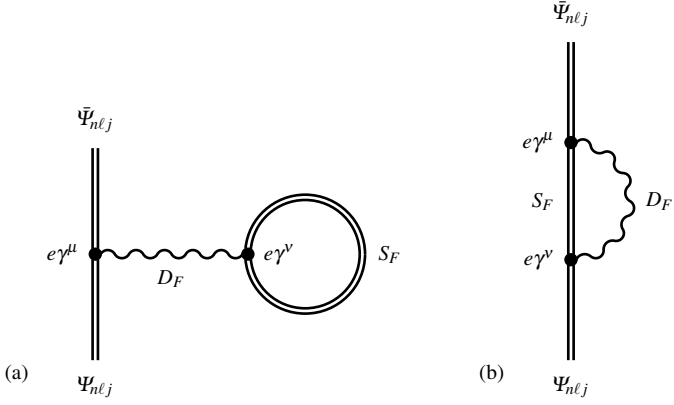
\begin{figure}[tb!]
\vspace{-10pt}
\centering
\begin{tabular}{@{\extracolsep{8mm}}lll}
\\
\\
(a)\quad\quad
\begin{fmffile}{vacuum_polarisation}
	\fmfframe(5,7)(5,7){
	\begin{fmfgraph*}(100,80)
		\fmfstraight
		\fmfbottom{in,i2,i3}
		\fmftop{out,o2,o3}
		\fmfright{right}
		\fmf{dbl_plain}{in,v1}
		\fmfdot{v1}
		\fmf{dbl_plain}{v1,out}
		\fmffreeze
		\fmf{photon,label=$D_F$,tension=0.6}{v1,v2}
		\fmfdot{v2}
		\fmf{dbl_plain,left,tension=0.4}{v2,right,v2}
		\fmfv{label=$\Psi_{n\ell j}$,label.angle=-90}{in}
		\fmfv{label=$e\gamma^\mu$,label.angle=180}{v1}
		\fmfv{label=$\bar{\Psi}_{n\ell j}$,label.angle=90}{out}
		\fmfv{label=$e\gamma^\nu$,label.angle=0}{v2}
		\fmfv{label=$S_F$}{right} 
	\end{fmfgraph*}
	}
\end{fmffile}
&
(b)\quad\quad
\begin{fmffile}{self_energy}
	\fmfframe(5,7)(0,7){
	\begin{fmfgraph*}(80,120)
		\fmfstraight
		\fmfbottom{in,i2,i3}
		\fmftop{out,o2,o3}
		\fmf{dbl_plain,tension=7}{in,v1}
		\fmfdot{v1}
		\fmf{dbl_plain,tension=5,label=$S_F$,label.side=left}{v1,v2}
		\fmf{photon,right,label=$D_F$}{v1,v2}
		\fmfdot{v2}
		\fmf{dbl_plain,tension=7}{v2,out}
		\fmfv{label=$\Psi_{n\ell j}$,label.angle=-90}{in}
		\fmfv{label=$e\gamma^\nu$,label.angle=180}{v1}
		\fmfv{label=$e\gamma^\mu$,label.angle=180}{v2}
		\fmfv{label=$\bar{\Psi}_{n\ell j}$,label.angle=90}{out}
	\end{fmfgraph*}
	}
\end{fmffile}
\end{tabular}
\caption{Bound state QED corrections of lowest order with the usual labelling of Feynman diagrams. (a) vacuum polarisation; (b) one-electron self-energy. Elementary charge $e$ is included for  clarity. $D_F$ and $S_F$ are the Dyson (photon) and Feynman (electron/positron) propagators respectively. The double line represents a propagator in the field of the nucleus. $\Psi_{n\ell j}$ represents a bound electron wave function.}
\label{fig:feynman_ord_alpha}
\end{figure}

\begin{figure}[tb]
  \begin{center}
\includegraphics[width=0.67\textwidth]{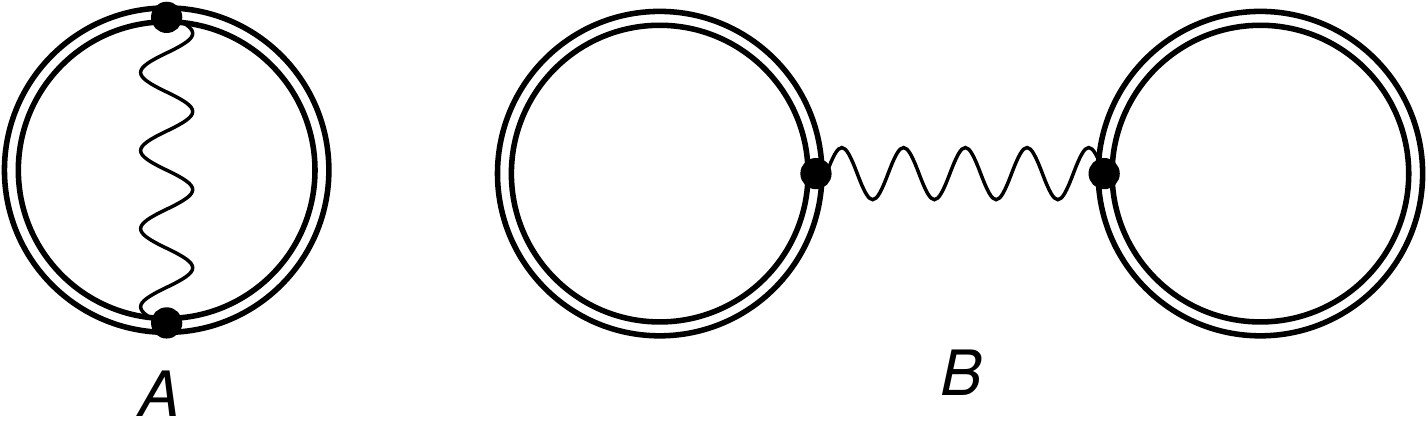}
  \caption{Example of disconnected diagrams of order $\alpha$, which only contribute to the vacuum energy. (a): one electron case, (b): two-electron case.}
  \label{fig:disconnected}
\end{center}
\end{figure}

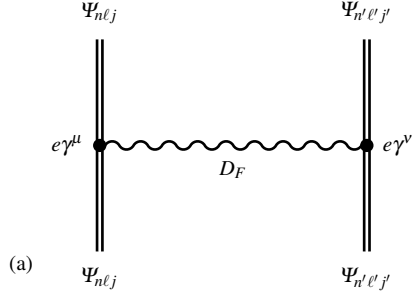
\begin{figure}[tb]
\vspace{-10pt}
\centering
\begin{tabular}{@{\extracolsep{8mm}}lll}
\\
\\
(a)\quad\quad
\begin{fmffile}{breit_interaction}
	\fmfframe(5,7)(5,7){
	\begin{fmfgraph*}(100,80)
		\fmfstraight
		\fmfbottom{i1,i2}
		\fmftop{o1,o2}
		\fmf{dbl_plain}{i1,v1,o1}
		\fmfdot{v1}
		\fmf{dbl_plain}{i2,v2,o2}
		\fmfdot{v2}
		\fmffreeze
		\fmf{photon,label=$D_F$}{v1,v2}
		\fmfv{label=$\Psi_{n\ell j}$,label.angle=-90}{i1}
		\fmfv{label=$\Psi_{n'\ell'j'}$,label.angle=-90}{i2}
		\fmfv{label=$e\gamma^\mu$,label.angle=180}{v1}
		\fmfv{label=$e\gamma^\nu$,label.angle=0}{v2}
		\fmfv{label=$\bar\Psi_{n\ell j}$,label.angle=90}{o1}
		\fmfv{label=$\bar\Psi_{n'\ell'j'}$,label.angle=90}{o2}
	\end{fmfgraph*}
	}
\end{fmffile}
\end{tabular}
\caption{Similar as in Fig. \protect \ref{fig:feynman_ord_alpha} but for the electron-electron interaction Feynman diagram.}
\label{fig:feynman_ee}
\end{figure}

\begin{figure}[tb]
\vspace{-10pt}
\centering
\begin{tabular}{@{\extracolsep{8mm}}lll}
\\
\\
(a)\quad\quad
\begin{fmffile}{Second_order_interaction}
	\fmfframe(5,7)(5,7){
	\begin{fmfgraph*}(100,80)
		\fmfstraight
		\fmfbottom{i1,i2}
		\fmftop{o1,o2}
		\fmf{dbl_plain,tension=5}{i1,v1,u1,o1}
		\fmfdot{v1}
		\fmf{dbl_plain,tension=5}{i2,v2,u2,o2}
		\fmfdot{v2}
		\fmfdot{u1}
		\fmfdot{u2}
		\fmffreeze
		\fmf{photon,label=$D_F$}{v1,v2}
		\fmf{photon,label=$D_F$}{u1,u2}
		\fmfv{label=$\Psi_{n\ell j}$,label.angle=-90}{i1}
		\fmfv{label=$\Psi_{n'\ell'j'}$,label.angle=-90}{i2}
		\fmfv{label=$e\gamma^\mu$,label.angle=180}{v1}
		\fmfv{label=$e\gamma^\nu$,label.angle=0}{v2}
		\fmfv{label=$e\gamma^\mu$,label.angle=180}{u1}
		\fmfv{label=$e\gamma^\nu$,label.angle=0}{u2}
		\fmfv{label=$\bar\Psi_{n\ell j}$,label.angle=90}{o1}
		\fmfv{label=$\bar\Psi_{n'\ell'j'}$,label.angle=90}{o2}
	\end{fmfgraph*}
	}
\end{fmffile}
&
(b)\quad\quad
\begin{fmffile}{Second_order_interaction_cross}
	\fmfframe(5,7)(5,7){
	\begin{fmfgraph*}(100,80)
		\fmfstraight
		\fmfbottom{i1,i2}
		\fmftop{o1,o2}
		\fmf{dbl_plain,tension=5}{i1,v1,u1,o1}
		\fmfdot{v1}
		\fmf{dbl_plain,tension=5}{i2,v2,u2,o2}
		\fmfdot{v2}
		\fmfdot{u1}
		\fmfdot{u2}
		\fmffreeze
		\fmf{photon,label=$D_F$}{v1,u2}
		\fmf{photon,label=$D_F$}{u1,v2}
		\fmfv{label=$\Psi_{n\ell j}$,label.angle=-90}{i1}
		\fmfv{label=$\Psi_{n'\ell'j'}$,label.angle=-90}{i2}
		\fmfv{label=$e\gamma^\mu$,label.angle=180}{v1}
		\fmfv{label=$e\gamma^\nu$,label.angle=0}{v2}
		\fmfv{label=$e\gamma^\mu$,label.angle=180}{u1}
		\fmfv{label=$e\gamma^\nu$,label.angle=0}{u2}
		\fmfv{label=$\bar\Psi_{n\ell j}$,label.angle=90}{o1}
		\fmfv{label=$\bar\Psi_{n'\ell'j'}$,label.angle=90}{o2}
	\end{fmfgraph*}
	}
\end{fmffile}
\end{tabular}
\caption{Similar as in Fig. \protect \ref{fig:feynman_ord_alpha} but for the second-order electron-electron interaction Feynman diagrams: (a) ladder diagram; (b) crossed diagram.}
\label{fig:feynman_lad_cross}
\end{figure}
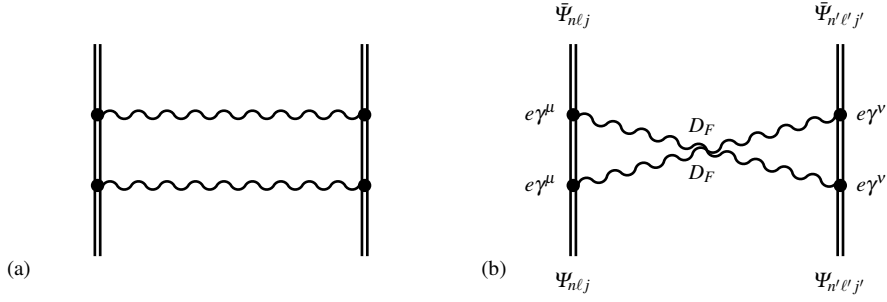

Connected diagrams are diagrams with external legs, which are bound-state wave functions like the ones in Figs. \ref{fig:feynman_ord_alpha}, \ref{fig:feynman_ee} and \ref{fig:feynman_lad_cross}. The disconnected diagrams, which have only closed loops, only contribute to the energy of the vacuum. Examples of disconnected diagrams for one- and two-electron systems are shown in Fig.\,\ref{fig:disconnected}. Each order in Eq.\,\eqref{eq:smatexp} has poles at $\epsilon^j$, which cancel out only if all terms of a given order are calculated simultaneously. For example, for $j=2$ both terms of $2\left<S_{\epsilon,1}^{(2)} 
 \right>_C-\left<S_{\epsilon,1}^{(1)} \right>_C^{2}$ must be calculated together to cancel the $1/\epsilon^2$ pole. 
 
The $S$-matrix formalism is not limited to the evaluation of QED energy shifts. It can also be used for the evaluation of radiative corrections to one- \cite{isv2004,alp2009} and two-photon \cite{lss2005} emission probability for example, and line shapes \cite{alps2008}. 

From the definition of the $S$-matrix \eqref{eq:s-matrix} and the evolution operator 
(\ref{eq:evolop}) one obtains for the matrix element of order $j$:
\begin{equation}
 S_{\epsilon,g}^{(j)}=\frac{\left(-ig\right)^{j}}{j!}\int d^{4}x_{j}\ldots \int d^{4}x_{1} 
 e^{-\epsilon \left|t_{j}\right|}\ldots
 e^{-\epsilon \left|t_{1}\right|}T\left[H_{I}\left(x_{j}\right) \ldots 
 H_{I}\left(x_{1}\right)\right].
 \label{eq:sjdef}
\end{equation}
These matrix elements can be expressed in terms
of the electron propagator and photon propagator. The electron propagator is connected to the Dirac bound electron Green's function by
\begin{eqnarray}
 S_{F}\left(x,y\right)&=&\left<0\right| T\left[ 
 \psi\left(x\right)\bar\psi\left(y\right)\right]\left|0\right>\nonumber \\
 &=&\left\{
 \begin{array}{cc}
        \sum_{E_{n}>0} \phi_{n}\left(x\right)\bar\phi_{n}\left(y\right)& 
        t_{x}>t_{y}  \\
        -\sum_{E_{n}<0} \phi_{n}\left(x\right)\bar\phi_{n}\left(y\right) & 
        t_{x}<t_{y}
 \end{array}\right.
 \nonumber \\
 &=&\frac{-i}{2\pi}\int_{C_F}dz
  G\left(\vec x_2,\vec x_1,z(1+i\delta)\right)
\gamma^{0} e^{-iz\left(t_2-t_1\right)}.
 \label{eq:sft}
\end{eqnarray}
The electron Green's function in \eqref{eq:sft} is the solution of \cite{mps1998,iam2017}
\begin{equation}
        \left(-i\vec \alpha \cdot \vec \nabla_{2}+V\left(|\vec x_{2}|\right)+\beta m 
        -z\right) G\left(\vec x_2,\vec x_1,z(1+i\delta)\right)=\delta 
        \left(\vec x_{2}-\vec x_{1}\right)\, .
        \label{eq:dirgreen}
\end{equation}
The energies of the bound states are given by the poles of the Green's function along the real axis.

The contraction of the two photon field operators in \eqref{eq:sjdef} gives
\begin{equation}
\left<0 \right|A_{\mu}\left(x_{2}\right)A_{\nu}\left(x_{1}\right)\left|0\right>=
g_{\mu\nu}D_F\left(x_{2}-x_{1}\right)
\end{equation}
where
\begin{eqnarray}
D_F\left(x_{2}-x_{1}\right)&=&
-\frac{i}{\left(2 \pi\right)^4}\int d^{4}q \frac{e^{-iq\cdot\left(x_{2}-x_{1}\right)}}{q^2+i\delta}
\nonumber \\
&=&\frac{1}{\left(2 \pi i\right)}\int_{-\infty}^{+\infty} dq_0 H\left(\vec x_{2}-\vec x_{1},q_0\right)
e^{-iq_0\left(t_{2}-t_{1}\right)}
\label{eq:photprop}
\end{eqnarray}
In Eq. \eqref{eq:photprop}, $H\left(\vec x_{2}-\vec x_{1},q_0\right)$ is
the photon Green's function, given by
\begin{eqnarray}
H\left(\vec x_{2}-\vec x_{1},q_0\right)&=&-\frac{e^{-bx_{21}}}{4 \pi x_{21}} \nonumber \\
x_{21}=\left| \vec x_{2}-\vec x_{1}\right|,&&   b=-i\left( q_0^2+i\delta\right)^{\frac{1}{2}},\,  \Re(b)>0.
\label{eq:photgreen}
\end{eqnarray}

As noted by Dyson, the expansion in power of $\alpha$ of Eq. \eqref{eq:gmlt} has a radius of convergence equal to zero \cite{dyson1952}. The series in $\alpha$ is thus only an asymptotic series that  diverges for $n\ge 1/\alpha$. Thanks to the small value of $\alpha$, this is not an issue unlike for the strong interactions.

The first-order contribution in Eq. \eqref{eq:smatexp}, the mass renormalization term, can be written as:
\begin{equation}
 \Delta E_{n}^{(1)}=\lim_{\epsilon\rightarrow 0}\frac{1}{2} i\epsilon \left<S_{\epsilon,1}^{(1)} \right>_{c}=
 -\delta m
 \int d\vec x
  \phi^{\dag}_{n}\left(\vec x\right)\gamma^{0}\phi_{n}\left(\vec x\right)\, .
 \label{eq:1stshift1}
\end{equation}

The three possible second-order connected diagrams are shown in Figs. \ref{fig:feynman_ord_alpha} (first order, one-electron QED corrections) and \ref{fig:feynman_ee} (electron-electron interaction). They originate from the second-order term in Eq.~\eqref{eq:smatexp} which can be explicitly written  (in natural units) as
\begin{eqnarray}
\label{eq:eeinter}
 \left<S_{\epsilon,1}^{(2a)} \right>_{c}&=& \frac{1}{\left(4 \pi i\right)}\int_{-\infty}^{+\infty} dq_0
 \int d^{4}x_{2} \int d^{4}x_{1}
 e^{-\epsilon \left(\left|t_{2}\right|+\left|t_{1}\right|\right)}
e^{-iq_0\left(t_{2}-t_{1}\right)}\frac{e^{-bx_{21}}}{4 \pi x_{21}}\nonumber \\
&& \bigg\{ \sum_{m_2 n_2 m_1 n_1}  e^{i\left(E_{n_{2}}-E_{m_{2}}\right)t_{2}} 
 e^{i\left(E_{n_{1}}-E_{m_{1}}\right)t_{1}}  \nonumber \\
&&\times  \phi^{\dag}_{n_{2}}\left(\vec x_{2}\right)\gamma^{0}\gamma^{\mu}\phi_{m_{2}}\left(\vec x_{2}\right) 
\phi^{\dag}_{n_{1}}\left(\vec x_{1}\right)\gamma^{0}\gamma^{\nu}\phi_{m_{1}}\left(\vec   x_{1}\right) \nonumber \\
  &&\times \left< N_{q};0 \right|:a^{\dag}_{n_{2}}a_{m_{2}} a^{\dag}_{n_{1}}
     a_{m_{1}}:\left| N_{q};0\right> 
  \nonumber \\
  &&
 -2Tr\left[\gamma^{\mu}\frac{-i}{2\pi}\int_{-\infty}^{+\infty}dz G\left(\vec x_2,\vec x_2,z(1+i\delta)\right)
\gamma^{0} \right]  \nonumber \\
&&\times\sum_{n \,m}e^{i\left(E_{n}-E_{m}\right)t_{1}}
\phi^{\dag}_{n}\left(\vec x_{1}\right)\gamma^{0}\gamma^{\nu}\phi_{m}\left(\vec 
  x_{1}\right) \left< N_{q};0 \right| a^{\dag}_{n}a_{m} \left| N_{q};0\right>
   \nonumber \\
  &&+\frac{-i}{\pi}\int_{-\infty}^{+\infty}dz\sum_{n\, m}e^{i\left(E_{n}t_{2}-E_{m}t_{1}-iz\left(t_2-t_1\right)\right)}\nonumber  \\ 
&&\times \phi^{\dag}_{n}\left(\vec x_{2}\right)\gamma^{0}\gamma^{\mu}
G\left(\vec x_2,\vec x_1,z(1+i\delta)\right)\gamma^{0}\gamma^{\nu}\phi_{m}\left(\vec x_{1}\right) \nonumber \\
  &&\times \left< N_{q};0 \right| a^{\dag}_{n}a_{m} \left| N_{q};0\right> \bigg\}.
\end{eqnarray}
Those diagrams are of the order of $\alpha/\pi$ since they have two vertices. 

\subsection{Two-times Green's function method}
\label{subsec:two-times}

This method is based on the generalisation of the Green's function \eqref{eq:sft} to a system of 
$N$ electrons \cite{sha2002,art2017}. The $2N-$times Green's function is defined as
\begin{equation}
\label{eq:ngreen}
G(x_1'\cdots x_N';x_1\cdots x_N)=\left< 0|T\left[\psi(x_1')\cdots \psi(x_N')\bar\psi(x_1)\cdots \bar\psi(x_N)\right]|0\right>\,.
\end{equation}
It can be expressed as
\begin{eqnarray}
\label{eq:ngreen-exp2}
\lefteqn{G(x_{1}^{\prime},\dots x_{N}^{\prime};x_{1},\dots x_{N})
\;\;\;\;\;\;\;\;\;} \\
&=&\frac{
\langle 0|T\left[\psi_{\rm in}(x_{1}^{\prime})\cdots\psi_{\rm in}(x_{N}^{\prime})
\overline{\psi
}_{\rm in}
(x_{N})\cdots\overline{\psi}_{\rm in}(x_{1})\right]
\exp{\{-i\int d^{4}z \; H_{I}
(z)\}}
|0\rangle}
{\langle 0|T\left[\exp{\{-i\int d^{4}z \; H_{I}
(z)\}}
\right]|0\rangle}\nonumber\\
&=&\Bigl \{ \sum_{m=0}^{\infty}\frac{(-i)^m}{m!}\int
d^4y_1\cdots d^4y_m \;
\langle 0|\left[T\psi_{\rm in}(x_{1}^{\prime})\cdots\psi_{\rm in}(x_{N}^{\prime})
\overline{\psi
}_{\rm in}
(x_{N})\cdots\overline{\psi}_{\rm in}(x_{1})\right.\nonumber\\
&\times& \left.
H_{I}(y_1)\cdots
H_{I}(y_m)\right]|0\rangle\Bigr\}
\Bigl \{ \sum_{l=0}^{\infty}\frac{(-i)^l}{l!}\int
d^4z_1\cdots d^4z_l \;
\langle 0|T\left[ H_{I}(z_1)\cdots
H_{I}(z_l)\right]|0\rangle\Bigr\}^{-1}.
\nonumber
\end{eqnarray}

Two-times Green's function method starts by keeping only two times in Eq. \eqref{eq:ngreen-exp2}, setting $t_1\equiv t_2 \cdots \equiv t_N \equiv t$ and $t^{\prime}_1\equiv t^{\prime}_2 \cdots \equiv t^{\prime}_N \equiv t^{\prime}$. This operation does not lead to any loss of information. 
For an isolated level $a$ of an $N$-electron atom, with an unperturbed energy $E_a^{(0)}$, the energy shift is given by 
\begin{equation}
    \label{eq:e-shift-2t}
    \Delta E_a=\frac{\frac{1}{2\pi i}\oint_{\Gamma} dE \left(E-E_a^{(0)}\right) \Delta\mathcal{G}_{aa}(E)}{1+\frac{1}{2\pi i}\oint_{\Gamma} dE\Delta \mathcal{G}_{aa}(E)},
\end{equation}
where $\mathcal{G}_{aa}(E)$ is the mean value for state $a$ of the Fourier transform of the two-times Green's function \eqref{eq:ngreen-exp2}. A perturbation expansion of $\mathcal{G}_{aa}(E)$ in powers of the fine structure constant $\alpha$ generates results similar to those shown in section \ref{subsec:s-matrix}.

In the case of two quasi-degenerate levels, one can use the 4-times Green's function in a similar manner. 
\begin{equation} 
\label{eq:4-green}
G(x_{1}^{\prime} x_{2}^{\prime};x_{1} x_{2})
=\frac{
\langle 0|T\left[\psi_{\rm in}(x_{1}^{\prime})\psi_{\rm in}(x_{2}^{\prime})
\overline{\psi}_{\rm in}
(x_{2})\overline{\psi}_{\rm in}(x_{1})\right]
\exp{\{-i\int d^{4}z \; H_{I}
(z)\}}
|0\rangle}
{\langle 0|T\left[\exp{\{-i\int d^{4}z \; H_{I}
(z)\}}
\right]|0\rangle}.
\end{equation}
In this case, the perturbation expansion is realized on the subspace containing the two quasi-degenerate levels. This procedure can formally be extended to any number of quasi-degenerate states. 

\subsection{Covariant evolution-operator procedure}
\label{subsec:evol-op}
The covariant evolution-operator method has been developed in Refs.~\cite{lsa2004,lsh2006,lsh2011,hsl2015,lai2017}. 
The method also starts from the evolution operator, and applies Relativistic Many-Body Perturbation Theory methods (RMBPT). It is closely related to the two-times Green's function method discussed in the previous subsection.  In contrast to the $S$-matrix formalism, which cannot handle quasi-degenerate states due to the energy-conservation condition caused by the integration over all times, the covariant evolution operator procedure has been successfully applied to quasi-degenerate states. The method is based on the fact that at $t=0$, the Green's operator is equivalent to the RMBPT wave operator, which is obtained as solution of a generalized Bloch equation. Additionally, in the standard evolution operator, time runs only in the forward direction and is therefore not relativistically covariant. By allowing the time to evolve forwards as well as backward, the relativistic covariance is restored. This method allows to evaluate non-QED many body effects to high-order and to take into account first and second order QED diagrams as well, at least in simple systems.

\subsection{Calculation of QED corrections}
\label{subsec:qed-res}

\begin{figure}[tb]
\vspace{-10pt}
\centering
\begin{tabular}{@{\extracolsep{8mm}}lll}
\\
\\
(a)\quad\quad
\begin{fmffile}{self_energy_screen1}
	\fmfframe(5,7)(0,7){
	\begin{fmfgraph*}(80,120)
		\fmfstraight
		\fmfbottom{in,i2,i3}
		\fmftop{out,o2,o3}
		\fmf{dbl_plain,tension=7}{in,v1}
		\fmfdot{v1}
		\fmf{dbl_plain,tension=5,label=$S_F$,label.side=left}{v1,u1,v2}
		\fmf{photon,left,label=$D_F$,label.angle=180}{v1,v2}
		\fmfdot{v2}
		\fmf{dbl_plain,tension=7}{v2,out}
		\fmf{dbl_plain,tension=7}{i3,u3,o3}
		\fmfdot{u1}
		\fmfdot{u3}
		\fmffreeze
		\fmf{photon,label=$D_F$,label.angle=0}{u1,u3}
		\fmfv{label=$\Psi_{n\ell j}$,label.angle=-90}{in}
		\fmfv{label=$\Psi_{n'l'j'}$,label.angle=-90}{i3}
		\fmfv{label=$e\gamma^\nu$,label.angle=180}{v1}
		\fmfv{label=$e\gamma^\mu$,label.angle=180}{v2}
		\fmfv{label=$e\gamma^\mu$,label.angle=180}{u1}
		\fmfv{label=$e\gamma^\mu$,label.angle=0}{u3}
		\fmfv{label=$\bar{\Psi}_{n\ell j}$,label.angle=90}{out}
		\fmfv{label=$\bar\Psi_{n'l'j'}$,label.angle=90}{o3}
	\end{fmfgraph*}
	}
\end{fmffile}
&
(b)\quad\quad
\begin{fmffile}{self_energy_screen2}
	\fmfframe(5,7)(0,7){
	\begin{fmfgraph*}(80,120)
		\fmfstraight
		\fmfbottom{in,i2,i3}
		\fmftop{out,o2,o3}
		\fmf{dbl_plain,tension=7}{in,u1}
		\fmf{dbl_plain,tension=7,label=$S_F$,label.side=left}{u1,v1}
		\fmfdot{v1}
		\fmf{dbl_plain,tension=7,label=$S_F$,label.side=left}{v1,v2}
		\fmf{photon,left,label=$D_F$,label.angle=180}{v1,v2}
		\fmfdot{v2}
		\fmf{dbl_plain,tension=7}{v2,out}
		\fmf{dbl_plain,tension=7}{i3,u31}
		\fmf{dbl_plain,tension=7}{u31,v31}
		\fmf{dbl_plain,tension=7}{v31,v32}
		\fmf{dbl_plain,tension=7}{v32,o3}
		\fmfdot{u1}
		\fmfdot{u31}
		\fmffreeze
		\fmf{photon,label=$D_F$,label.angle=0}{u1,u31}
		\fmfv{label=$\Psi_{n\ell j}$,label.angle=-90}{in}
		\fmfv{label=$\Psi_{n'l'j'}$,label.angle=-90}{i3}
		\fmfv{label=$e\gamma^\nu$,label.angle=180}{v1}
		\fmfv{label=$e\gamma^\mu$,label.angle=180}{v2}
		\fmfv{label=$e\gamma^\mu$,label.angle=180}{u1}
		\fmfv{label=$e\gamma^\mu$,label.angle=0}{u31}
		\fmfv{label=$\bar{\Psi}_{n\ell j}$,label.angle=90}{out}
		\fmfv{label=$\bar\Psi_{n'l'j'}$,label.angle=90}{o3}
	\end{fmfgraph*}
	}
\end{fmffile}
\end{tabular}
\caption{Self-energy screening diagrams of order $\left(\alpha/\pi\right)^2$.The other notations are defined in the legend of Fig. \protect \ref{fig:feynman_ord_alpha}.}
\label{fig:feynman_se_screen}
\end{figure}
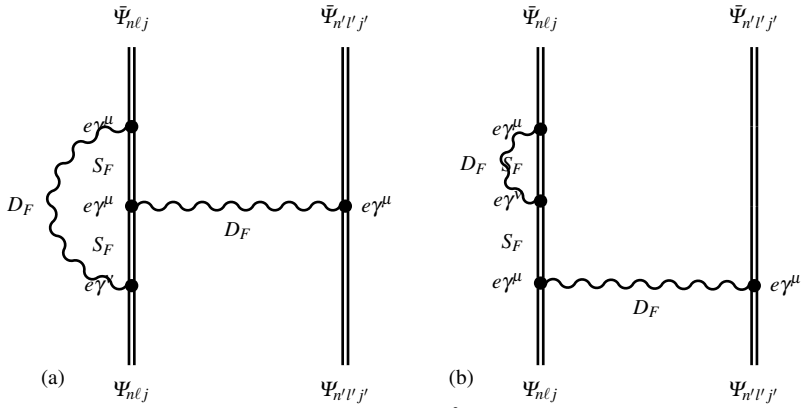

There are several kinds of QED corrections that need to be evaluated for computing transition energies in superheavy elements. In the first category, there are one-electron corrections like the self-energy and the vacuum polarization shown in Fig. \ref{fig:feynman_ord_alpha}. These corrections concern all atoms.  The Feynman diagram for the electron-electron interaction is shown in Fig.  \ref{fig:feynman_ee}. It contains the Breit interaction and all-order retardation corrections. This diagram can be iterated to provide higher-order corrections to the electron-electron interaction as shown in Fig. \ref{fig:feynman_lad_cross}. The latter diagrams and similar ones with more photons provide QED corrections to the correlation energy. The full ladder diagram in Fig. \ref{fig:feynman_lad_cross}(a) contains both the correlation contribution present in many-body theories like RMBPT, MCDF or RCI, and pure QED corrections, which involve positrons. The crossed diagrams in Fig. \ref{fig:feynman_lad_cross}(b) provides pure QED corrections not included in many-body calculations. 
A last category of Feynman diagrams contains electron-electron interaction corrections to one-electron correction, like self-energy screening presented in Fig. \ref{fig:feynman_se_screen}.
These two last categories concern atoms with at least two electrons. 

\subsubsection{One-electron radiative corrections}
\label{subsubsec:one-elec-qed}

The energy shift due to one-electron radiative corrections of order $i$, corresponding to an ensemble of diagrams with $2 i$ vertices, is formally of order $(\alpha/\pi)^i m_ec^2$, but after renormalization it can be written as:
\begin{equation}
    \Delta E_{(n,\kappa)}^{(i)}=\left(\frac{\alpha}{\pi}\right)^i \frac{\left(Z\alpha\right)^4}{n^3} F_{n,\kappa}^{(i)}\left(Z\alpha\right) m_ec^2,
    \label{eq:radiat-behav}
\end{equation}
where $F_{n,\kappa}^{(i)}(Z\alpha)$ is a slowly varying function of $Z$ for a level of quantum numbers $(n,\kappa)$. 
For low $Z$, one can write an expansion of $F_{n,\kappa}^{(i)}(Z\alpha)$ as an expansion in powers of $Z\alpha$ and $\log\left((Z\alpha)^{-2}\right)$. The lower-order coefficients of this expansion can be found in Refs. \cite{mtn2012,tmnt2021}. As shown in \cite{moh1974}, this expansion is not convergent at medium to high-$Z$. For superheavy elements, we will thus only consider results evaluated to all orders in $Z\alpha$.

\begin{figure}[tb]
\centering
        \includegraphics[width=0.67\columnwidth]{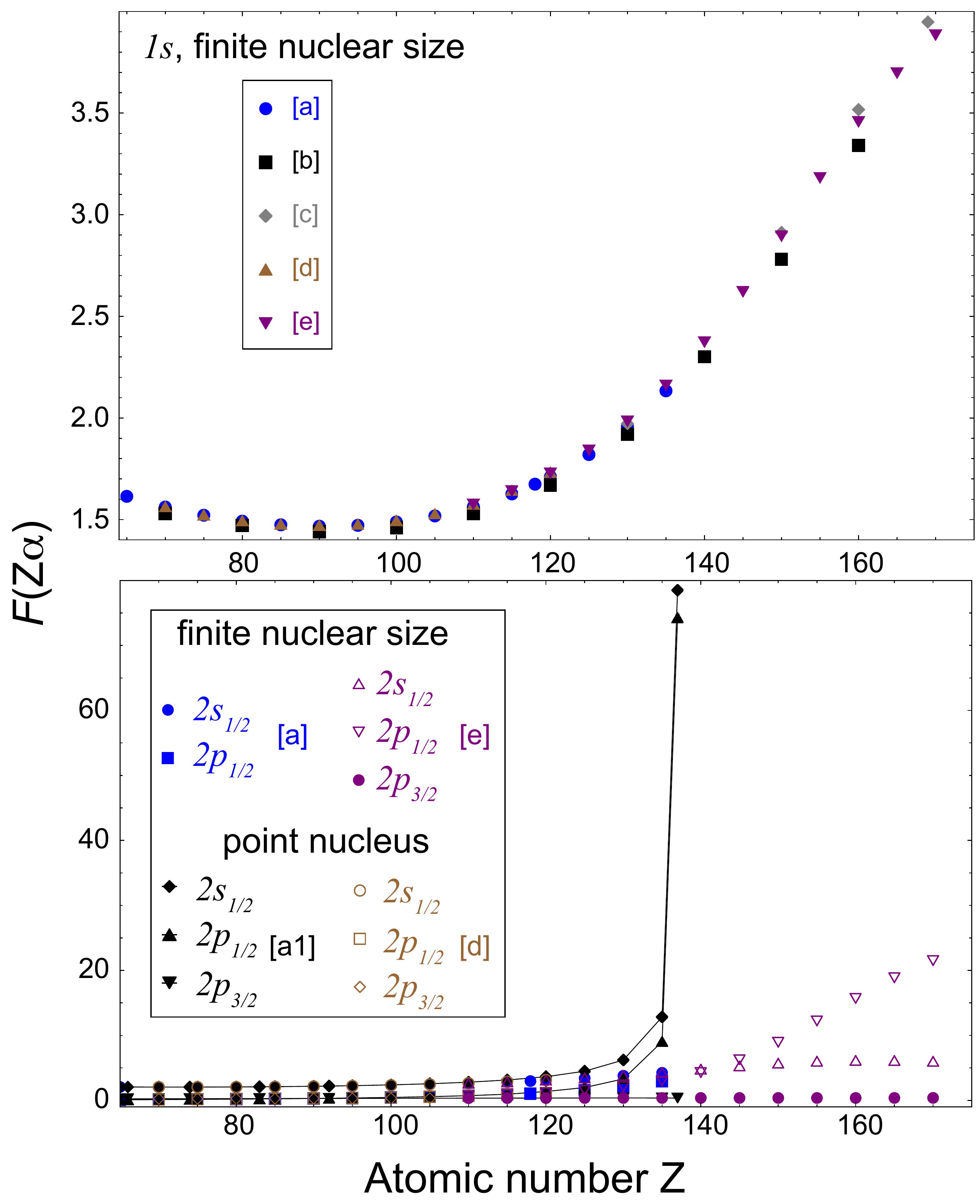} \\
\caption{Values of the $F^{(1)}(Z\alpha)$ function in the high-$Z$ and supercritical region. Top: comparison between finite size values for $1s_{1/2}$. Bottom: comparison between finite-size and point nucleus values for the $n=2$ shell. References: [a]=\cite{iam2022}, [a1]=\cite{ijm2022}, [b]=\cite{caj1976}, [c]=\cite{ssmg1982}, [d]=\cite{ShabaevTupitsyn2013}, [e]=\cite{mgst2022}. }
\label{fig:FZA_SE_1s_2s2p}
\end{figure}

We now discuss the evaluation of the self-energy diagram \ref{fig:feynman_ord_alpha}(b).
The  $1s_{1/2}$ self-energy has been evaluated to all-orders for $5\leq Z \leq 120$ for point nucleus \cite{moh1974a,caj1976,moh1992,iam1998,ShabaevTupitsyn2013}. The finite nuclear-size correction is important for high-$Z$ and small values of the principal quantum number and for $s$ and $p$ states. It is negligible for larger values of $|\kappa|$.  The self-energy for the $1s_{1/2}$ state has been  evaluated in \cite{caj1976,mas1993,bmps1998,mps1998}. In Ref.~\cite{caj1976}, it was evaluated up to $Z=160$ and in Ref.~\cite{ssmg1982} up to $Z=170$. An extension of the evaluation of $F_{1s}^{(1)}(Z\alpha)$ for $i=1$  for point nuclei up to $Z=137$ has been performed recently \cite{ijm2022} and for uniformly charged  nuclei up to $Z=135$ \cite{iam2022} with radii in the range \SIrange{1.5}{7.3}{fm}. The work from Ref. \cite{ShabaevTupitsyn2013} has been extended recently  to $Z=170$ \cite{mgst2022}. In both works, the self-energy for a given $Z$ value is calculated for a specific nuclear size, using the Fermi model. A comparison between the different values of $F_{1s}^{(1)}(Z\alpha)$ for $i=1$ and finite nuclear size is shown in Fig. \ref{fig:FZA_SE_1s_2s2p}. All calculations are in good agreement with each other, that is within the differences of the nuclear model and size applied.

In the case of excited states, the self-energy has been evaluated  for $ns$, $np_{1/2}$, $np_{3/2}$ and $nd_{3/2}$ states for $5\leq Z \leq 110$ and $2\leq n \leq 5$ in Refs. \cite{moh1974,moh1982,mak1992,moh1992,iam1998}. The 
point-nucleus self-energy of $ns$, $np$ and $nd$ states up to $n=5$ can also be found in Ref. \cite{ShabaevTupitsyn2013} for $10 \leq Z \leq 120$. Reference \cite{lim2001}  contains the values of $F_{n,\kappa}^{(1)}(Z\alpha)$ for $nd_{3/2}$ to $ng_{9/2}$ up to $n=5$ for a point nucleus.  The finite size of correction for $2s$ states and $2p_{1/2}$ states can be found in \cite{mas1993,bmps1998,mps1998} for $26\leq Z \leq 100$, for $10 \leq Z \leq 120$ in \cite{ShabaevTupitsyn2013} and for $100 \leq Z \leq 170$ in \cite{mgst2022}. The comparison between different theoretical values with and without finite size correction for $2s$, $2p_{1/2}$ and $2p_{3/2}$ states is shown in Fig. \ref{fig:FZA_SE_1s_2s2p}. The divergence of the point-nucleus values when $Z\to \alpha^{-1}\approx 137$ for states with $|\kappa|=1$ is visible for both $2s$ and $2p_{1/2}$ states. It is in fact even more pronounced than for the $1s$ state.

\begin{figure}[tb]
\centering
        \includegraphics[width=.8\columnwidth]{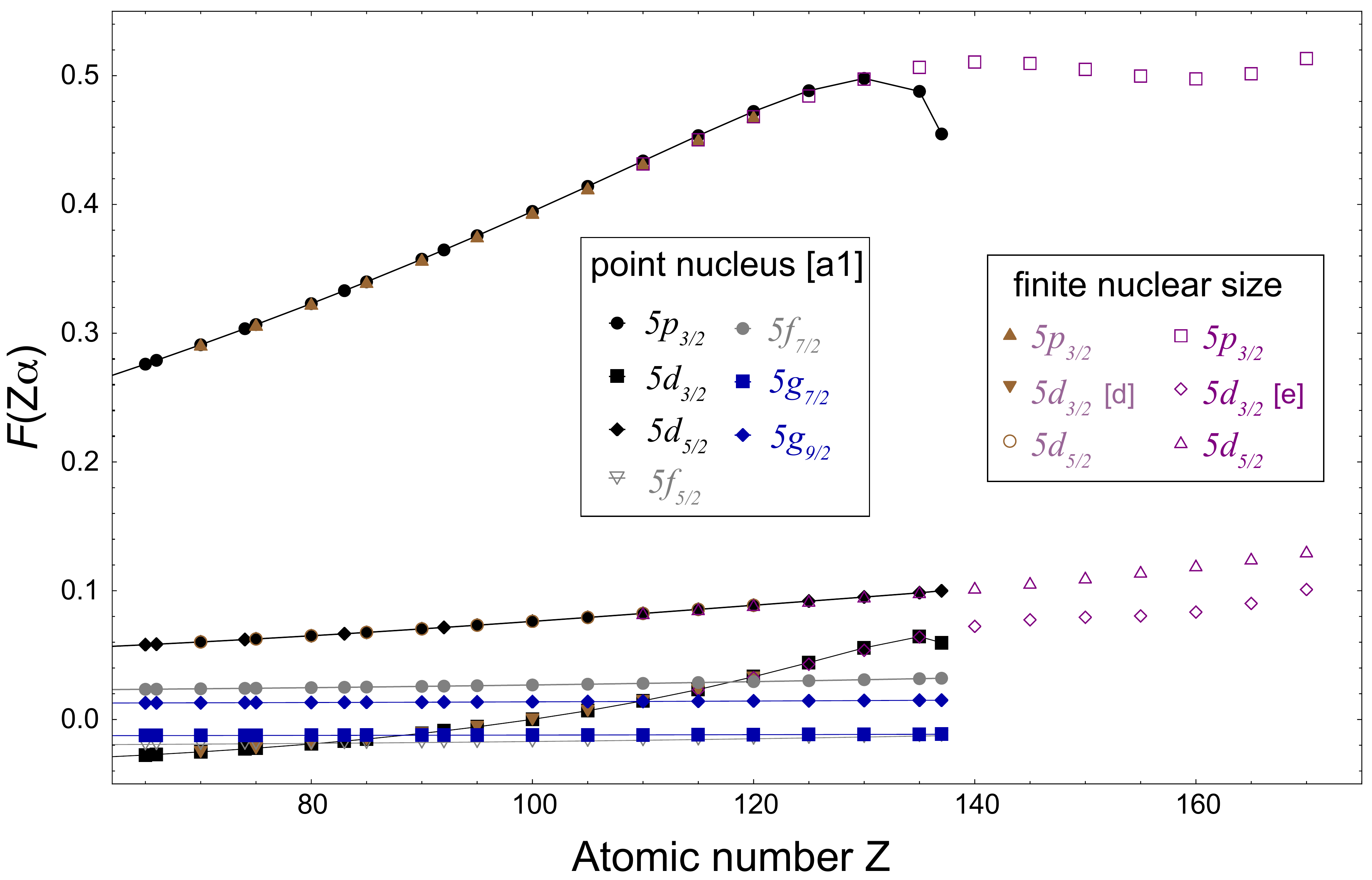}
\caption{Similar as in Fig.~\ref{fig:FZA_SE_1s_2s2p} but for
$F_{5p_{3/2}}^{(1)}(Z\alpha)$, $F_{5d_{3/2}}^{(1)}(Z\alpha)$, $F_{5d_{5/2}}^{(1)}(Z\alpha)$, $F_{5f_{5/2}}^{(1)}(Z\alpha)$, $F_{5f_{7/2}}^{(1)}(Z\alpha)$, $F_{5g_{7/2}}^{(1)}(Z\alpha)$, and  $F_{5g_{9/2}}^{(1)}(Z\alpha)$ functions in the high-$Z$ and supercritical region.}
\label{fig:FZA_SE_up5g}
\end{figure}

For larger values of $n$ and $Z>120$ there are no published results for $F_{n,\kappa}^{(1)}(Z\alpha)$. A large scale effort has been recently undertaken to provide values of $F_{n,\kappa}^{(1)}(Z\alpha)$ to cover the range of interest for superheavy elements. The values of $F_{n,\kappa}^{(1)}(Z\alpha)$ with all possible $\kappa$ for all $1 \leq n \leq 10$ and $Z$ up to \num{137} have been evaluated for point nuclei \cite{ijm2022}. The point nucleus values for all possible $|\kappa|>1$ for $n=5$  are plotted in Fig. \ref{fig:FZA_SE_up5g}, together with values from Refs. \cite{ShabaevTupitsyn2013,mgst2022} for $p$ and $d$ states.
The functions $F^{(1)}_{n,\kappa}(Z\alpha,R)$ including finite nuclear size correction for $3 \leq n \leq 6$ and $| \kappa | =1$ ($s$ and $p_{1/2}$ states) have been evaluated for $Z$ up to \num{135}, with values of $R$, the mean spherical radius in the \SIrange{1.5}{7.3}{fm}  \cite{iam2022}.

The vacuum polarization correction of order one in $(\alpha/\pi)$, presented in Fig.~ \ref{fig:feynman_ord_alpha}(a) can be evaluated with good enough accuracy by using an expansion in power of $Z\alpha$. The potential of order $\alpha(Z\alpha)$, with only one interaction with the nucleus is called the the Uehling potential $V_\text{U}$. The next term in the expansion, of order $\alpha(Z\alpha)^3$ is called the Wichmann-Kroll potential $V_\text{WK}$. All orders calculations of the vacuum polarization have been performed in \cite{sam1988,Persson1993}.
The Uehling potential  \cite{Uehling1935} is evaluated as
\begin{equation}
\label{eq:Uehling}
V_\text{U}(r) = -\frac{2\alpha}{3\pi} \frac{Z}{r}
\int_{[1,\infty)} \text{exp}\left( -2\alpha^{-1} \xi r \right) \left( 1 + \frac{1}{2\xi^2}\right) \frac{\sqrt{\xi^2 - 1}}{\xi^2} ~d\xi
\end{equation}
if one treats the nucleus as a point charge. An analytical formula for the Uehling potential in terms of modified Bessel functions has been provided in Ref.~\cite{Frolov2012}. The expression in \eqref{eq:Uehling} can be extended to a finite nuclear charge distribution \cite{FullertonRinker1976,kla1977}. The Uehling potential can be added to the Dirac equation potential, providing an easy way to include the loop-after-loop vacuum polarization correction to all orders \cite{ind2013}.

The Wichmann-Kroll contribution $V_\text{WK}$ to the vacuum polarization is of order $\alpha(Z\alpha)^3$; it  can be written approximately for $r \to 0$ as \cite{WichmannKroll1956},
\begin{align}
V_\text{WK}(r)\approx &\frac{\alpha(Z\alpha)^3}{\pi}\left[\left(-\frac{3}{2}\zeta(3)+\frac{\pi^2}{6}-\frac{7}{9}\right)\frac{1}{r}+2\pi\zeta(3)\right.\notag\\
&\left.-\frac{\pi^3}{4}+\left(-6\zeta(3)+\frac{\pi^4}{16}-\frac{\pi^2}{6}\right)r+O(r^2)\right]
\end{align}
where $\zeta(n)$ is the Riemann zeta function. For more details see for example Refs.\,\cite{Persson1993,SapirsteinCheng2003}. An efficient numerical method to evaluate $V_\text{WK}$  without low-$r$ expansion is given in \cite{hua1976}. Additional terms on this expansion of order $\alpha(Z\alpha)^5$ and $\alpha(Z\alpha)^7$, corresponding to 5 and 7 interactions with the nucleus in the vacuum-polarization loop, are approximately known.
They have been used in muonic atoms for many years \cite{WichmannKroll1956,bam1978,bar1982,pbao2021}. Numerical methods to evaluate them can be found  in Ref. \cite{hua1976}. 

One should also add next-order terms with $i=2$ in \eqref{eq:radiat-behav}. These terms are  of the order of $(\alpha/\pi)\approx 2\times 10^{-3}$ compared to the two-vertex terms (see Fig. \ref{fig:lambshift}) but the leading coefficients in their $Z\alpha$ expansion can be large.
These  terms represent, \eg  two-loop self-energy, two-loop vacuum-polarization corrections, and  mixed self-energy vacuum-polarization terms. The corresponding Feynman diagrams  are shown for example in Refs.~\cite{yis2008,Thierfelder2010,iam2017}. Some of these terms can be easily calculated, such as the K\"allen-Sabry contribution to the vacuum polarization  \cite{Kallen1955} for which a potential is known \cite{FullertonRinker1976}. The two-loop self-energy terms have been evaluated for one- \cite{yis2003,yis2003a,yis2005,yis2005a,yer2009,yer2010,yer2018} and three-electron atoms \cite{yis2006} for $30\leq Z \leq 100$, but only for $n=1$ and $n=2$ states. Mixed self-energy vacuum polarization diagrams have been evaluated in \cite{lpsk1993,pllp1996,yis2008}. Whilst these diagrams are important for inner shell electron energies, they are not expected to contribute significantly to the outer-shell energies of superheavy elements compared to correlation effects. The evaluation of the diagrams of order $(\alpha/\pi)^2$,  which are easier to calculate, like the  K\"allen-Sabry term or the loop-after-loop Uehling contribution, can provide the needed order of magnitude to assess the importance of the uncalculated ones on specific cases.

\begin{figure}[tb!]
\centering
\includegraphics[width=0.8\linewidth]{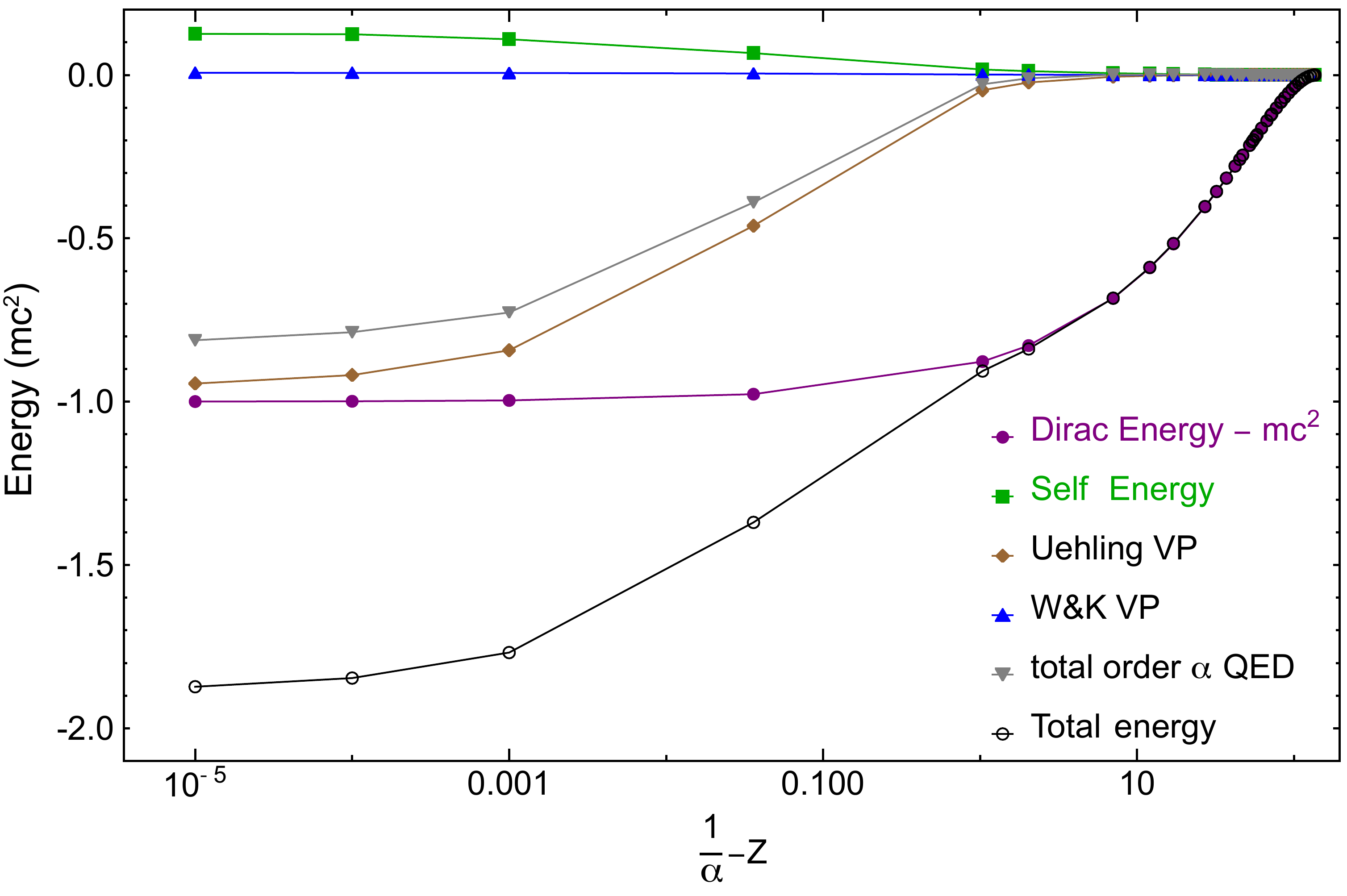}
\caption{\label{fig:comp_near_crit_point} 
Dirac energy, self-energy, and vacuum polarization near $Z_{\mathrm{c}}=1/\alpha$ for a point nucleus. The Uehling and Wichmann and Kroll vacuum polarization contributions have been evaluated with the MDFGME code and the self-energy is from \cite{ijm2022} and evaluated to all order in $Z\alpha$. }
\end{figure}

In Fig.~\ref{fig:comp_near_crit_point} we show the evolution of the  QED contributions of order $\alpha$ and of the Dirac energy for the $1s$ state as a function of $1/\alpha-Z$ with non-integer values of $Z$, to show what happens near the critical $Z_{\mathrm{c}} = 1/\alpha$  in the case of a point nucleus. It shows that the self-energy becomes nearly independent  of $Z$ and that the vacuum polarization becomes the dominant contribution almost one order of magnitude larger than the self-energy. The total energy to that order becomes close to $-m_ec^2$. Although the Wichmann-Kroll contribution is very small, it remains to be checked how the all-order vacuum polarization and the sum of higher-order contributions would behave. The same comparison for finite size nucleus does not show the same effect: the self-energy and vacuum polarization remain of the same size and their values are strongly reduced.

\subsubsection{Two-electron radiative and non-radiative corrections}
\label{subsubsec:two-elec-qed}

Concerning the calculation of atomic spectra of heavy and superheavy elements, the bottleneck in terms of accuracy in such many-electron systems still lies in the treatment of electron correlation  (see discussion in Sec. \ref{sec:elcor}). There are, however, mixed terms between radiative corrections and the electron-electron interaction that need to be considered. The main one is known as self-energy screening (see Fig. \ref{fig:feynman_se_screen}). It has been evaluated by direct calculation of the Feynman diagrams only for $n=1$ and $n=2$ states \cite{pssl1996,iam2001}.
These terms containing self-energy loops cannot be put into the form of an exact potential, and thus cannot be easily generalized to arbitrary atoms. It would therefore be useful to formulate an approximate QED potential that could describe the screened Lamb-shift and other atomic properties sufficiently accurately  and could be successfully used in molecular calculations. One could then introduce a (model) perturbation Hamiltonian to represent the radiative part of QED corrections of the form
\begin{equation}
\Delta \tilde{H}^\text{QED} = V_\text{U} + V_\text{WK} + h_\text{SE}+ h_\text{h.o.t.},
\label{eq:qedHamil}
\end{equation}
where $V_\text{U}$ is the Uehling potential, $V_\text{WK}$ is the Wichmann and Kroll potential, $h_\text{SE}$ the self energy model potential and $h_\text{h.o.t.}$ represents two-loops contributions. The aim of this operator is to include approximate QED corrections to the electron-electron operators, with negligible errors compared to the electron correlation treatment. In addition, the matrix elements of this QED Hamiltonian can be added to the CI matrix or to the Hamiltonian matrix and differential equation in the MCDF procedure. 


The non-radiative part in Eq. \eqref{eq:QEDHamiltonian} is dominated by the electron-electron interaction,  which is obtained by evaluating the Feynman diagram of Fig. \ref{fig:feynman_ee}. It is given in atomic units and in the Coulomb gauge by
\begin{eqnarray}
    V\left(r_{ij},\omega_{ij}\right) & = & \frac{1}{r_{ij} } -\frac{\bm{\alpha}_i \cdot
\bm{\alpha}_j}{r_{ij}} - \frac{\bm{\alpha}_i \cdot
\bm{\alpha}_j}{r_{ij}}
\left(\cos\left(\alpha\omega_{ij}r_{ij}\right)-1\right) \nonumber\\ & + &
\left( \bm{\alpha}_i \cdot \bm{\nabla}_i \right)\left(
\bm{\alpha}_j \cdot \bm{\nabla}_j
\right)\frac{\cos\left(\alpha\omega_{ij}r_{ij}\right)-1} {(\alpha\omega_{ij})^2
r_{ij}}\, ,
\label{eq:eeinteract}
\end{eqnarray}
where  $ \omega _{ij}=E_{i}-E_{j}$ is the energy of the photon exchanged between the two electrons. The $\bm\nabla$ operator acts only on $r_{ij}$ and not on the following wave function.
The Breit operator \cite{Breit1929,Breit1930,Breit1932} in Eq.~\eqref{eq:Breit} corresponds to the expansion of Eq.~\eqref{eq:eeinteract} in powers of $\alpha=1/c$ up to the second order. It is then independent of $ \omega _{ij}$. The frequency-dependent part is called higher-order retardation.
This finite frequency contribution becomes important at high nuclear charges \cite{Thierfelder2010}. The frequency dependent Breit interaction has been explored in detail in many works \cite{maj1971,Mittleman1981,hag1984,gid1987,iad1990,ind1995}.  The main difficulty lies in the definition of $ \omega _{ij}$ in CI or MCDF calculations, where the energy of an individual orbitals is not physical and can also reach very negative values, much lower than $-m_ec^2$ \cite{gid1987,iad1990,ind1995}. The gauge dependence of the resulting energy shift has been discussed in detail in Refs. \cite{Gorceix1988,lin1990}. 

The second-order diagrams of Fig. \ref{fig:feynman_lad_cross} have been evaluated in Refs. \cite{bmjs1993,lpsl1995,mas2000,asl2002} for the ground state and $n=2$ excited states of two-electron atoms. They contains specific QED corrections beyond what can be obtained by many-body treatment of the  interaction in Eq. \eqref{eq:eeinteract} with the necessary projection operators. These corrections contains the positron part of the ladder diagram \ref{fig:feynman_lad_cross}~(a) and the contribution from the cross-ladder diagram \ref{fig:feynman_lad_cross}~(b).

\subsubsection{Effective QED Hamiltonians}
\label{subsubsec:eff-qed-H}

To bring the self-energy term into a useful effective Hamiltonian form, $h_\text{SE}$, is the most challenging part as this operator is inherently non-local. Nevertheless, many attempts were made to estimate the self-energy shift in atomic spectra by approximations. Earlier ones  were summarized in \cite{PyykkoeZhao2003}. Approximations based on effective $Z$ values to account for electron screening were introduced in the early versions of GRASP \cite{dgjp1989}. The Welton approximation \cite{wel1948} was introduced in the MDFGME code in 1987 \cite{igd1987} for $s$-states and generalized to $\ell\geq 0$ in \cite{iad1990}. Effective operators directly based on BSQED have been introduced more recently \cite{ShabaevTupitsyn2013,mgst2022}. 

In Ref.~\cite{PyykkoeZhao2003},  a local self-energy potential was introduced  in a simple Gaussian form
\begin{equation}
h_\text{SE}(r)=\left( b_0+b_1Z +b_2Z^2\right) \text{exp}\left\{ -(\beta_0 +\beta_1Z +\beta_2Z^2)r^2 \right\},
\end{equation}
which serves as a rough estimate. Here $b_i$ and $\beta_i$ are adjustable parameters and the Gaussian is  located close to the nucleus.

A far more accurate expression for an effective self-energy Hamiltonian has been proposed in Ref.\,\cite{Flambaum2005} to be
\begin{align}
h_\text{SE}(r)=\Phi_\text{mag}(r)+\Phi_\text{el}(r)+\Phi_\text{low}(r),
\end{align}
with the magnetic form factor
\begin{align}
\Phi_\text{mag}(r)=\frac{\alpha}{4\pi m}i \mathbf{\gamma}\cdot\nabla \left[\phi(r)\left(\int_1^\infty dt \frac{e^{2trm}}{t^2\sqrt{t^2-1}}-1 \right)\right],
\label{magint}
\end{align}
where $\phi(r)$ is the electric potential of the nucleus. The last two terms are contributions 
from the electric form factor decomposed into a high- and a low-frequency part,
\begin{align}
\Phi_\text{el}(r)=& A(Z) \frac{\alpha}{\pi} \phi(r)\int_1^\infty dt \frac{e^{-2trm}}{\sqrt{t^2-1}}\left[\left(1-\frac{1}{2t^2}\right)\right.\notag\\
&\left.\left(\log(t^2-1)+4\log\left(\frac{1}{Z\alpha}+\frac{1}{2}\right)\right)-\frac{3}{2}+\frac{1}{t^2}\right]
\label{elint}
\end{align}
The (long-range) low-frequency contribution is given by 
\begin{align}
\Phi_\text{low}(r)=-B(Z)Z^4\alpha^5m e^{-Zr/a_\text{B}},
\label{lowint}
\end{align}
where $B(Z)=0.074+0.35Z\alpha$ is a coefficient adjusted to reproduce the radiative shifts for 
the high Coulomb $p$-levels \cite{Flambaum2005}, and $a_B$ is the Bohr radius.  This expression of the self-energy operator was implemented into the program GRASP \cite{FroeseFischer2019} by simply replacing the Coulomb potential $-Z/r$ by its extension to the finite nucleus case \cite{Thierfelder2010}.  Later, this has been correctly folded into the self-energy potential leading to more complicated expressions, which slightly improves the self-energy shifts \cite{Berengut2016}. To improve the SE corrections for the $s$-levels, and especially for the $1s_{1/2}$ level, in multi-electron systems the prefactor $A(Z)$ in \eqref{elint} was chosen to be dependent on the principal quantum number $n$ \cite{Thierfelder2010}. More recently, both coefficients $A(Z)$ and $B(Z)$ were refitted and made dependent on the angular momentum $\ell$ \cite{Berengut2016}.  

To go beyond models with adjustable parameters, one can in principle go back to first principle QED and use a spectral decomposition of the self-energy operator
\begin{equation}
h_\text{SE}^\text{nl} = \sum_{i,j} \ket{\psi_i} D_{i,j}^\text{SE} \bra{\psi_j}
\end{equation}
where $\{ \psi_i \}$ represents a complete set of hydrogenic wave functions (including both continua), and the matrix elements $D_{i,j}^\text{SE}$ need to come from accurate self-energy calculations. This has been   explored \cite{Dyall2013} with limited success because of the basis set restrictions imposed and the underlying slow convergence of this sum, which is well known from direct exact QED evaluations \cite{moh1974,moh1974a,mps1998}. Furthermore, the off-diagonal elements $D_{i,j}^\text{SE}$ with $i\ne j$ are crucial and cannot be neglected. To this end, an additional exponential type semi-local operator has been added to reduce the matrix elements $D_{ij}$ in size for the subsequent spectral decomposition \cite{ShabaevTupitsyn2013,Shabaeev2015},
\begin{equation}
h_\text{SE} = h_\text{SE}^\text{sl}+h_\text{SE}^\text{nl},
\end{equation}
with
\begin{equation}
h_\text{SE}^\text{sl}(r) = \sum_\kappa V_\kappa(r) P_\kappa.
\end{equation}
The semi-local operator
\begin{equation}
V_\kappa(r) = A_\kappa e^{-r/\lambda_c}
\end{equation}
differentiates between the different $\kappa$ states through the projection operator $P_\kappa$ (for the definition see Ref. \cite{Shabaeev2015}). Here $\lambda_c$ is the Compton wavelength. For details see Refs.\cite{ShabaevTupitsyn2013,Shabaeev2015}. A recent extension to superheavy elements up to nuclear charge $Z=170$ has been carried out in Ref.\,\cite{mgst2022}. This scheme gives very accurate results for the self-energy. One wonders if a semilocal ansatz in the same form of a pseudopotential applied commonly in electronic structure theory could be efficiently used as well \cite{Hangele2012,Hangele2013} for all-electron QED treatments in molecules for example. It would certainly be an improvement to the original local ansatz \cite{PyykkoeZhao2003} and possibly of sufficient accuracy in molecular calculations.

\begin{figure}[tb!]
\centering
\includegraphics[width=0.8\linewidth]{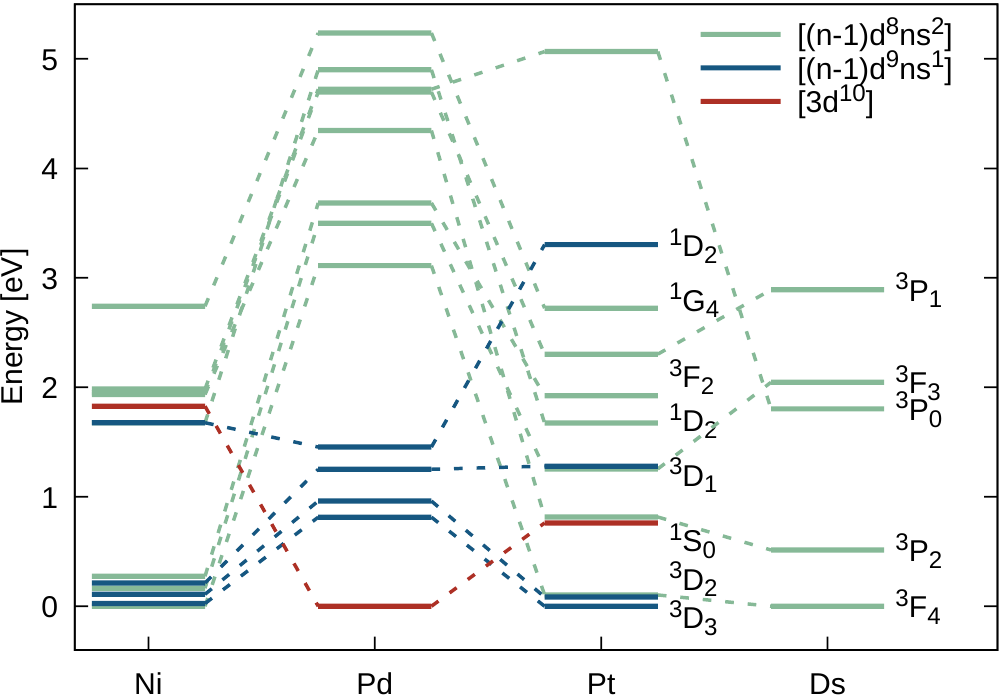}
\caption{\label{fig:Group10} Energy levels for the dominant configurations of the Group 10 elements Ni, Pd, Pt, and Ds. The values for Ni, Pd, and Pt are from the NIST database \cite{NIST-ASD2022}. The Ds levels are from Ref. \cite{lackenby2019Ds}. Different colors are used to distinguish between the three different configurations: green [$(n-1)d^8\;ns^2$], blue [$(n-1)d^9\;ns$]  and black  [$3d^{10}$]. For Pd, there are  intruder states (not shown here) arising from the [$(n-1)d^9\;np$]  configuration (for Pt from the [$(n-1)d^9\;np$] and [$(n-1)d^8\;ns\;np$] configurations), which mix with several of the low energy states shown. Thus, some configuration assignments (especially for the $^3P_0$ level) are approximate at best. For Ds, a dense spectrum arising from the [$6d^7\;7s^2\;7p$] configuration intrudes into the normal spectrum and only few predicted lines of even parity are shown \cite{lackenby2019Ds}. Adapted from Ref. \cite{Schwerdtfeger2020}.}
\end{figure}

\section{Electron Correlation}\label{sec:elcor}
The accurate computational treatment of both static and dynamic electron correlation in atomic open-shell multi-electron systems is a daunting task. Even for the lightest elements such as nickel, a correct description of the many  low-lying states arising from the $3d^8\;4s^2$, $3d^9\;4s$ and  $3d^{10}$ configurations is currently not available. For example, using Gaussian type basis sets (GTOs), Ref.\,\cite{Andersson1992} applied large-scale complete active space second-order perturbation theory (CASPT2) calculations including relativistic corrections for nickel correlating 18 electrons within 14 orbitals. These calculations resulted in the following excitation energies with respect to the $^3D(d^9s$) ground state ($j$-averaged experimental values set in parentheses): $^3F$ \SI{-0.08}{\electronvolt} (\SI{0.03}{\electronvolt}), $^1D$ \SI{0.32}{\electronvolt} (\SI{0.33}{\electronvolt}), $^1S$ \SI{1.77}{\electronvolt} (\SI{1.74}{\electronvolt}). Figure \ref{fig:Group10} shows the energy levels for the Group 10 elements. From this it is clear that the correct prediction of the ground state symmetry is difficult for atoms with dense spectra. This problem will become worse when degenerate high angular momentum states are involved such as in the lanthanides and actinides and for the superheavy elements.

One of the main workhorses in relativistic atomic structure theory is the configuration interaction (CI) method with a predetermined set of CSFs where the radial shapes of the one-electron orbital spinors remain unchanged. In a typical CI procedure, the active virtual and core space are systematically increased and higher angular momentum functions added to test convergence against the final value. The resulting CI wave functions are then used for calculating QED effects, albeit QED matrix elements can be directly added to the CI matrix resulting in correlated QED calculations (this  still needs to be explored for the SHEs). These CI techniques are invaluable for obtaining  accurate properties to, for example, test the standard model \cite{Bieron2009}. However, as the size of a CI calculation scales exponentially with the excitation level (the number of determinants is $N_\text{det}\sim n^mN_\text{v}^m/(m!)^2$ with $n$ being the number of electrons, $N_\text{v}$ the number of virtual orbitals and $m$ the excitation level), the CI method is often combined with many-body perturbation theory for electron correlation (CI+MBPT) to allow for an efficient treatment of important core excitations \cite{Dzuba1996}. Again, because of the large computer time involved one rarely goes beyond second-order MBPT, although calculations for atoms with one valence electron (Cs and Tl for example) have been performed up to third order \cite{bgjs1987,bjs1990}. A mix of MBPT and CC methods has also been used for evaluating electron affinities for Ca and Sr \cite{swl1996}.

A very popular electron correlation method within the quantum chemistry community is coupled cluster (CC) theory originally proposed by Coester and Kuemmel \cite{CoesterKuemmel1960} for nuclear interactions, and subsequently brought into electronic structure theory  \cite{Cizek_1980}. There are several excellent papers, books and reviews on CC applications \cite{Bartlett1991,Kuemmel2003,Bartlett2007,Shavitt2009,bartlett2012,Datta2019,Liu2021,chaudhuri2017many}. In CC theory, the ground state wave function is related to the DHF ground state configuration by an exponential operator containing the cluster operator $T$,
\begin{equation}
\Psi_0 = e^T \Psi_0^{\text{DHF}}\, ,
\end{equation}
with $T=T_1+T_2+\dots$ and $T_n$ are the $n$-particle excitation operators. For example, if one restricts to double excitations only ($T=T_2$, CCSD), the cluster operator is
\begin{equation}
T_2=\sum_{i<j,r<s}t_{ij}^{rs}c_r^\dagger c_s^\dagger c_i c_j \, ,
\end{equation}
where $t_{ij}^{rs}$ are the coupled-cluster amplitudes, determined through a variational procedure, and $c_r^\dagger$ and $c_i$ are electron creation and annihilation operators for single-particle states $r$ (virtual) and $i$ (occupied), respectively. 

This scheme can be extended from the Hilbert space to the Fock space formalism (FSCC) where in addition electrons are removed (ionization) or added (attachment) \cite{Kaldor1991,Visscher2001,eliav2010four,Eliav2015,Oleynichenko2020}. FSCC theory has been successfully applied for atomic properties of heavy and superheavy elements  \cite{Eliav-1994,Kaldor1998x,Eliav2005E119,Sato2015,Borschevsky2015Lv,Borschevsky2013E120}. One may like to chose a multireference wave function instead of $\Psi_0^{\text{DHF}}$ for the starting point in coupled-cluster theory, termed multi-reference coupled-cluster (MRCC) theory \cite{Pal1989,Adamowicz1993,Piecuch2002,Jeziorski2010,eliav2010relativistic,koehn2013,Tang2017,Datta2020}, which, however, has its computational challenges \cite{Evangelista2018}. Because of the steep computational scaling, single-reference coupled-cluster theory is usually restricted to CCSD(T) (the golden standard of quantum chemistry), where the single and double contributions to the coupled-cluster amplitudes are determined variationally, and the triples are obtained by perturbation theory. Most coupled-cluster calculations are using GTOs as the underlying basis set, but the coupled-cluster scheme can easily be adapted to numerical procedures such as FEM using B-splines \cite{Dzuba2007,Tang2017,Tang2020}.

In atomic structure theory a popular approximation to CC theory is the so-called all-order CI method (AOCI) where one keeps only the linear terms in the expansion, \ie $\Psi_0 = (1+T_1+T_2+\dots) \Psi_0^{\text{DHF}}$ \cite{Safronova2008x,Safronova2009,Gharibnejad2011,Safronova2014a}. For core excitations, one can add MBPT (AOCI+MBPT) \cite{Safronova2008x}. This method has successfully been used for predicting spectra of multi-electron systems with large number of electrons \cite{Safronova2014a,Porsev2016}. For an alternative method combining configuration interaction with perturbation theory with a large number of valence electrons see Ref.\,\cite{Berengut2017}. However, high accuracy electronic structure calculations of spectra with uncertainties below  \SI{1e-3}{\electronvolt}  are usually limited to few electron systems \cite{Blundell1990,iad1990,sac2011,Derevianko2008,sac2015}. Such calculations are important to test QED and subsequently the standard model as well \cite{Blundell1990a,Karshenboim2005,Beiersdorfer_2010,Volotka2013,ShabaevQED2018,ind2019}. Despite the computational limitations and challenges to treat relativistic many-electron systems with a high number of electrons, the ionization potential and electron affinity have recently been obtained for gold to \si{\milli\electronvolt} accuracy compared to experiment \cite{Pasteka2017}. Gold is a special case that can still be treated accurately as the ionization and electron attachment involves mainly the valence $6s$ shell. The computational cost comes, however, from a very soft polarizable $5d$ core giving rise to large core and core-valence correlation effects. For these calculations, relativistic CC theory up to pentuple excitations were required, that is single reference DHF-CCSDTQ(P) theory including Breit and lowest-order QED interactions, to achieve almost (but not quite) experimental accuracy \cite{Pasteka2017}.

For systems with two or more electrons in open-shells, as this is the case for most actinide and transactinide elements, the electronic many-body treatment remains a major challenge for the foreseeable future. For example, using MRCI plus relativistic corrections for the ionization potential of neutral uranium, U$[5f^36d^17s^2](^5L_6)\rightarrow$U$^+ [5f^37s^2](^4I_{9/2})$, gave a value of \SI{6.062}{\electronvolt} (at the DHF+Breit level one gets \SI{5.540}{\electronvolt}) \cite{Peterson2015} compared to the experimental value of \SI{6.19405+-0.00006}{\electronvolt} \cite{Coste1982} achieving practically an accuracy in the \SI{0.1}{\electronvolt} region. Such calculations are computationally expensive and often may not be sufficient to predict the correct sequence of states within a window of about \SI{1}{\electronvolt}. For comparison, pseudopotentials which replace the core by an effective Hamiltonian give results accurate to about \SI{0.1}{\electronvolt} for atomic spectra \cite{Dolg2012,SchwerdtfegerPP2011}, and are therefore not always suitable for applications in atomic physics if high accuracy is required. An all-order correlation potential method based on Green's functions has been developed \cite{Dzuba1989x,Dzuba2008x} and applied to many-electron systems into the superheavy element region. Future developments may include density matrix renormalization group (DRMG) methods \cite{Brandejs2020}, or even machine learning algorithms to estimate the electron correlation error compared to experiment. It is clear that for dense spectra, efficient electron correlation methods and algorithms need to be further developed to efficiently study atoms with a high number of electrons (such as the superheavy elements) including QED effects \cite{Lindgren2014,lindgren2016relativistic}. An additional difficulty comes from the very large size reached by the calculation of configuration with two-large angular momenta orbitals like $6d^n \; 5g^m$ for example. The $6d^9 5g^9 \, J=2$ $LSJ$ configuration has \num{1895} $JJ$ configurations and \num{\approx 30,300} determinants, leading to \num{\approx 1.3e7} Coulomb integrals, \num{\approx 3.8e7} magnetic integrals and \num{\approx 3.1e7} retardation integrals. The $6d^6 5g^{12} \, J=0$ $LSJ$ configuration leads to \num{3360} $JJ$ configurations and \num{\approx 256,000} determinants and the number of angular integrals cannot be indexed by a \num{32} bit integer. Such unusual configurations, including for example the $5g^{18}\; J=0$ configuration, may become competing candidates for the ground state of the Og isoelectronic sequence at very large $Z$.

\begin{figure}[tb!]
\centering
\includegraphics[width=0.8\linewidth]{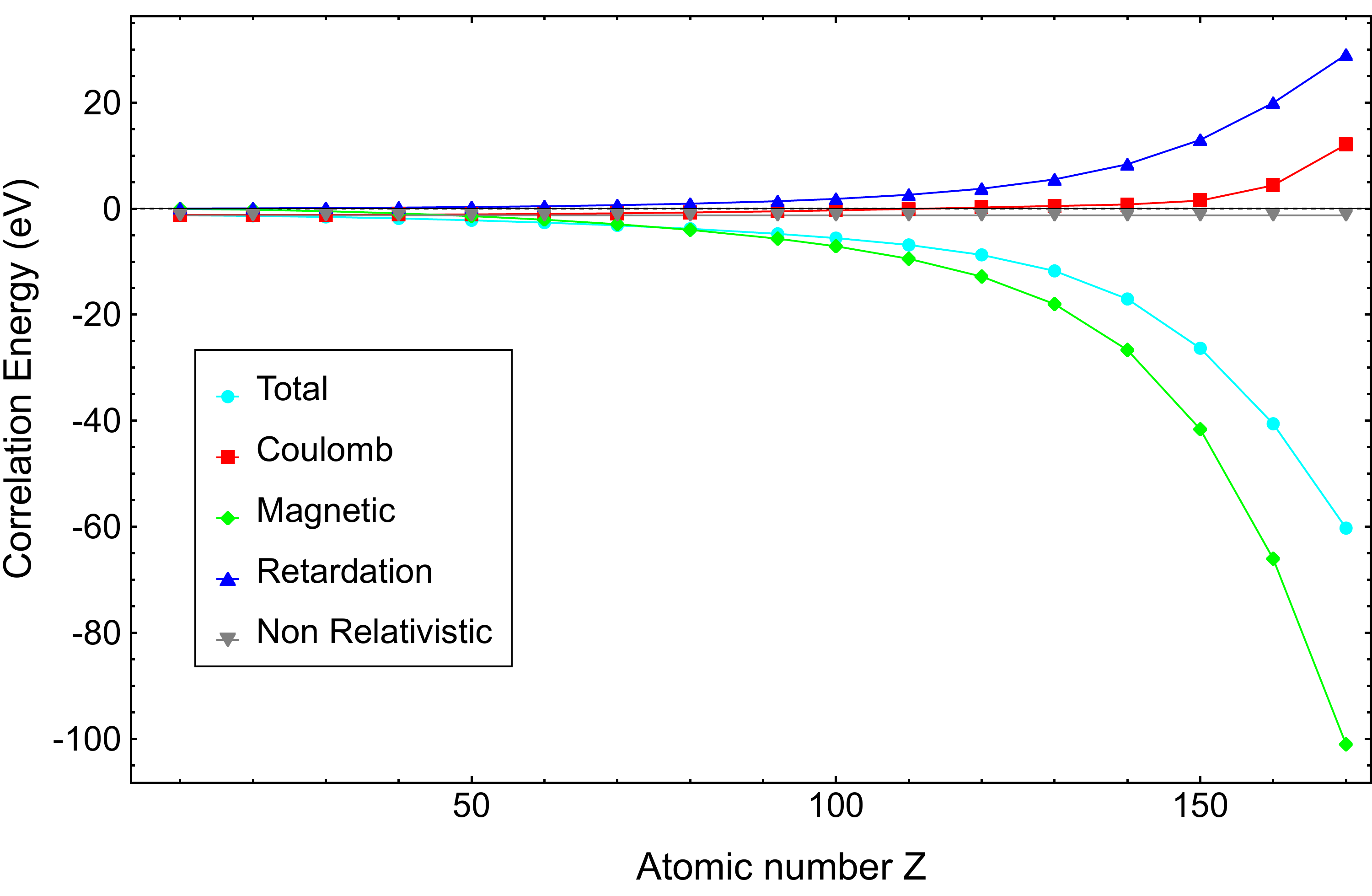}
\caption{\label{fig:he-relativistic-corr} He-like ions contribution to the correlation energy obtained with a fully MCDF calculations using B-spline basis set and the full operator from Eq. \protect \eqref{eq:eeinter}. Orbitals up to $7i$ have been included.}
\end{figure}

Finally, the question arises if one should include the negative energy continuum (NEC) in the electron correlation procedure. For the lighter elements, the gap from the lowest bound state to the negative energy continuum is almost $E=2m_ec^2$, larger than the gap to the positive energy continuum.  The correlation term  estimated perturbatively  turned out to be of the order of $E_\text{cor}(\text{NEC})\sim (Z\alpha)^3$ \cite{saj1996}. For the $Z=50$ He-like ion,  $E_\text{cor}(\text{NEC})$= \SI{0.00471}{\electronvolt} \cite{saj1996}, which implies that this term needs to be included for high precision tests on few electron systems with high nuclear charge. We need to be reminded that in the $Z\rightarrow\infty$ nonrelativistic limit the electron correlation contribution for a He-like atom is \SI{-0.046663254}{{a.u.}} \cite{Loos2010}. For the relativistic case, this  limit is not accurately known \cite{Saue2016}, see also Ref. \cite{Karwowski1991} for a detailed discussion. However, perturbation theory will eventually break down if the occupied level comes close to the negative energy continuum, or even dives into it. Ref.~\cite{Watanabe2007} studied the effect of removing the no-virtual-pair approximation on the correlation energy of the He isoelectronic sequence. They showed that for He-like Ds ($Z=110$) the correlation energy changes from \SI{-2.019}{\electronvolt} to \SI{-1.391}{\electronvolt} due to the NEC inclusion.
The use of projection operators using a B-spline basis set build with direct Dirac-Fock potentials and their effect on the correlation energy were studied in \cite{ind1995} for the ground state of He-like ions. The larger impact on the relativistic correlation energy was shown to be due to the magnetic and retardation interaction. To illustrate the effect near  $Z_c$ we have used the MDFGME code (2022 version) to calculate more precisely the correlation energy for the $1s^2\; ^1S_0$, with a fully relaxed wave function including \num{44} configurations, from $1s^2$ to $7i^2$ and projection operators. The Breit interaction was included in the self-consistent process. The relativistic Coulomb contribution changes sign around $Z=90$. The magnetic contribution is largely dominant, with a value of \SI{-80}{\electronvolt} at $Z=170$, while the retardation contribution going to \SI{25}{\electronvolt} and the Coulomb part to \SI{15}{\electronvolt} as can be seen in Fig. \ref{fig:he-relativistic-corr}.
It should be noted that a phenomenological inclusion of the negative energy continuum is not really consistent as the crossed diagram of Fig. \ref{fig:feynman_ee}, not included in the correlation contribution, is expected to contribute. 

\section{Atomic Structure Calculations of the Superheavy Elements}

\subsection{Dominant Electron Configurations}

Electron configurations are needed for placing the elements into their correct place in the PT and to discuss their chemical behaviour \cite{Schwerdtfeger2020}. Systematic D-HF calculations of total atomic energies of ground state configurations and symmetries (within the $jj$-coupling scheme) for the Li ($Z=3$) to Db ($Z=105$) isoelectronic series up to Og, including QED effects, have been provided  in Ref.~\cite{Rodrigues2004}. Predicted dominant configurations, ionization potentials, electron affinities and dipole polarizabilities for the transactinides, $Z=102-122$, are collected in Table\,\ref{tab:properties1}. Concerning the elements with nuclear charge $Z=105-110$, the Table  does not distinguish between the $j=3/2$ and 5/2 occupations for the $6d$ level. Moreover, whenever spin-orbit coupling becomes large, the assigned (nonrelativistic) $LS$ symmetry has to be taken with some care. For example, the ground-state electron configuration of Fl has a $J=0$ spin and  positive parity, and one expects strong mixing between the $^3P$ and $^1S$ states due to spin-orbit coupling. Furthermore, for a variety of elements, the electronic ground and low lying excited configuration states mix, making it difficult to unambiguously assign a ground-state configuration. 
\begin{table}[tp]
\setlength{\tabcolsep}{0.1cm}
\caption{\label{tab:properties1} Atomic number $Z$ and element symbol E according to IUPAC and predicted ground-state properties:  dominant valence shell configuration and term symbol for the ground electronic state, ionization potential $I_p$ and electron affinity $ E_A $ in \si{\electronvolt}, and dipole polarizability $\alpha_D$ in atomic units. If possible, the most accurate value was taken from the literature. For error estimates and methods used see the cited references.}
\begin{center}
\scriptsize
\begin{tabular}{| c| c| l| c| c| c| c|}
\hline
$Z$ & E & configuration & $I_p$ & $ E_A $ & $\alpha_D$ & Refs.\\
\hline
102 & No    &  [Rn]$5f^{14}7s^2$ & 6.626 & - & 111 & \cite{Borschevsky2007,Chhetri2018,Thierfelder2009pol} \\
103 & Lr 	& [No]$6d^1_\frac{3}{2} (^2D_\frac{3}{2})$			& 4.96	&  -	& 323   & \cite{Safronova2014,Sato2015,Sato2018} \\
104 & Rf 	& [No]$6d_\frac{3}{2}^2 (^3F_2)$ 					& 6.01 	&  -	& 115	& \cite{Safronova2014,eliav2011electronic}  \\
105 & Db	& [No]$6d^3 (^3F_{3/2})$ 							& 6.814 &  1.189& 42.5 	&  \cite{Dzuba2016,Lackenby2018a,arbely2018PhD} \\
106 & Sg 	& [No]$6d^4 (^5D_0)$							    & 7.7 	&  -	& 40.7 	&   \cite{Dzuba2016,Lackenby2019a}  \\
107 & Bh 	& [No]$6d^5 (^6S_{5/2})$ 							& 8.6 	&  -	& 38.4	&  \cite{Dzuba2016,Lackenby2019a}   \\
108 & Hs 	& [No]$6d^6 (^5D_4)$ 							    & 9.5 	&  -	& 36.2	&  \cite{Dzuba2016,Lackenby2019a}   \\
109 & Mt 	& [No]$6d^7 (^4F_{9/2})$ 							& 10.4 	&  -	& 34.2 	&   \cite{Dzuba2016,Lackenby2019a}  \\
110 & Ds 	& [No]$6d^8 (^3F_4)$ 							    & 9.562 & 0.830 & 32.3 	&  \cite{arbely2018PhD}   \\
111 & Rg 	& [No]$6d^4_{3/2}d^5_{5/2} (^2D_\frac{5}{2})$ 	& 11.03 & 1.97  & 30.6	&  \cite{Kaygorodov2022,eliav2011electronic,Dzuba2016}   \\
112 & Cn 	& [No]$6d^{10}  (^1S_0)$ 							& 12.02 & 0 	& 27.64	&  \cite{Kaygorodov2022,Pershina2008,eliav2011electronic} \\
113 & Nh 	& [Cn]$7p_\frac{1}{2}^1 (^2P_\frac{1}{2})$ 			& 7.49	& 0.73 	& 29.85	&  \cite{Kaygorodov2022,Guo2022,Eliav-1996a,Pershina2008x}  \\
114 & Fl 	& [Cn]$7p_\frac{1}{2}^2 (^1S_0)$   					& 8.65	& 0 	& 30.59 &  \cite{Kaygorodov2022,eliav2011electronic,Borschevsky2009,Pershina2008} \\
115 & Mc 	& [Cn]$7p_\frac{1}{2}^2p_\frac{3}{2}^1 (^2P_\frac{3}{2})$	& 5.574	&  0.313& 70.5	&  \cite{Borschevsky2015Lv,Dzuba2016Mc} \\
116 & Lv 	& [Cn]$7p_\frac{1}{2}^2p_\frac{3}{2}^2 (^3P_2)$ 	& 6.855	& 0.776 & -		&  \cite{Borschevsky2015Lv} \\
117 & Ts 	& [Cn]$7p_\frac{1}{2}^2p_\frac{3}{2}^3 (^2P_\frac{3}{2})$	& 7.654	& 1.602 & 76.3	&  \cite{Borschevsky2015Lv,deFarias2017} \\
118 & Og 	& [Cn]$7p^6 (^1S_0)$ 							    & 8.888 & 0.076 & 57.98	& \cite{Kaygorodov2021,Guo2021,Jerabek2018,guo2021ionization} \\
119 & Uue& [Og]$8s^1 (^2S_{1/2})$ 							    & 4.783	& 0.663 & 169.7	& \cite{Landau2001,Eliav2005E119,Borschevsky2013} \\
120 & Ubn& [Og]$8s^2 (^1S_0)$  							        & 5.851	& 0.021 & 162.6	& \cite{Borschevsky2013E120}  \\
121 & Ubu& [Ubn]$8p^1_\frac{1}{2} (^2P_\frac{1}{2})$ 			& 4.447	&  -	& -		&  \cite{eliav2011electronic} \\
122 & Ubb& [Ubn]$7d^1_\frac{3}{2}8p^1_\frac{1}{2} (^1D_2)$		& 5.651	&  -	& -		&  \cite{eliav2011electronic} \\
\hline
\end{tabular}
\end{center}
\end{table}

Table \ref{tab:CSF} lists the ground state linear combination of the CSFs for a selection of elements, obtained within a multi-reference treatment using GRASP \cite{DyaGraJoh89}. For Rf, there is a single dominant configuration, $6d^2_{3/2}$ in contrast to Db, for which  a strong mixing is predicted between the $6d_{3/2}$ and $6d_{5/2}$ levels (the two CSFs with identical configurations have different seniority numbers $\nu$=0 and 2 for the $6d^2_{5/2}$ occupation \cite{grant2007relativistic}).  Also for Ds (2 holes in the $6d$ shell) a substantial mixing between two configurations is expected. Less mixing is predicted between the  strongly spin-orbit separated $7p_{1/2}$ and $7p_{3/2}$ levels in Fl, Mc, and Lv. Hence, the ground states of the $7p$ block elements can be unambiguously assigned to a single dominant configuration.

\begin{table}[tb!]
\caption{\label{tab:CSF} Selected ground state CSF’s obtained within a multi-reference DHF treatment using GRASP.}
\begin{center}
\scriptsize
\begin{tabular}{ c| c| l }
\hline
$Z$ & E & CSF\\
\hline
104 & Rf & $\psi_{J=2}=0.9300\phi(6d^2_{3/2})+0.0350\phi(6d^1_{3/2}d^2_{5/2})-0.1230\phi(6d^2_{5/2})$ \\
105 & Db & $\psi_{J=3/2}=0.7869\phi(6d^3_{3/2})-0.5083\phi(6d^2_{3/2}d^1_{5/2}) - $ \\
& & $0.2510\phi(6d^1_{3/2}d^2_{5/2,0}) + 0.2327\phi_2(6d^1_{3/2}d^2_{5/2,2}) - 0.0727\phi(6d^3_{5/2})$ \\
110 & Ds & $\psi_{J=4}=0.9775\phi(6d^4_{5/2})+0.2110\phi(6d^3_{3/2}d^5_{5/2})$ \\
114 & Fl & $\psi_{J=0}=0.9955\phi(7p^2_{1/2})-0.095\phi(7p^2_{3/2})$ \\
115 & Mc & $\psi_{J=3/2}=0.9932\phi(7p^2_{1/2}p^1_{3/2})+0.099\phi(7p^1_{1/2}p^2_{3/2}) - $ \\
& & $0.061\phi(7p^3_{3/2})$ \\
116 & Lv & $\psi_{J=0}=0.9984\phi(7p^2_{1/2}p^2_{3/2})+0.057\phi(7p^1_{3/2}p^3_{3/2})$ \\
\hline
\end{tabular}
\end{center}
\end{table}

\subsection{Ionization potentials}

Ionization potentials ($I_p$) and electron affinities ($E_A$) are important for discussing the chemical behaviour of an element. For example, Mulliken's (empirical) definition of the (dimensionless) electronegativity $\chi$ \cite{Mulliken1934} of an element takes the sum of both properties in units of \si{\electronvolt}, $\chi= (0.187/\text{eV}) (I_p + E_A)+0.17$. 

\begin{figure}[tb]
  \begin{center}
\includegraphics[width=0.90\textwidth]{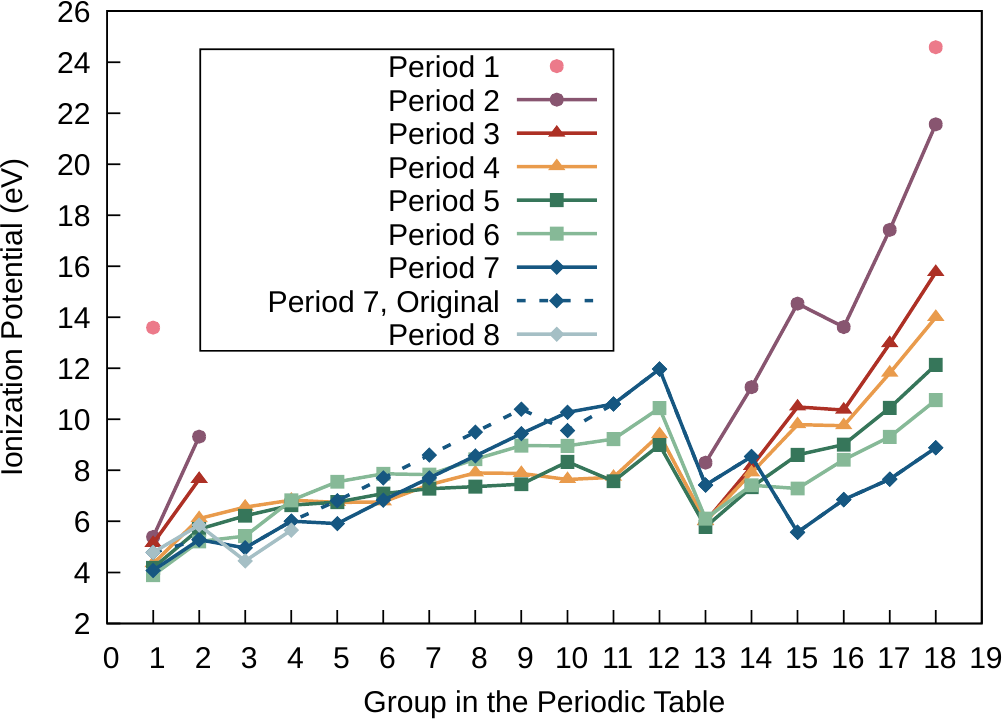}
  \caption{Ionization potentials (in eV) for the $s$-, $d$-, and $p$-block elements along period 4-7 and the four first elements of period 8. Experimental values are from NIST \cite{NIST-ASD2022} and from various authors (see text). For the transactinides, see 
  Table~\ref{tab:properties1}. The Period 7 solid lines correspond to ionization potentials obtained by a linear fit, which are explained in the main text.}
\label{fig:ionizationpotentials}
\end{center}
\end{figure}

The predicted ionization potentials for the transactinides are summarized in Fig.~\ref{fig:ionizationpotentials}.
Only for the elements up to the actinides  ionization potentials have been determined experimentally. The experimental value of  $4.96_{-0.04}^{+0.05}$\,\si{\electronvolt} for Lr is in good agreement with the value of \SI{4.963+-0.015}{\electronvolt} obtained from relativistic coupled cluster calculations, CCSD(T), which include Breit and QED contributions \cite{Sato2015, Sato2018}.

Concerning the accuracy of the data listed in Table \ref{tab:properties1}, the decrease in the ionization potential of Ds compared to Mt and Rg  is a  reminder for the limited accuracy of the electron correlation treatment of open-shell systems, as this decrease is most likely due to the fact that the CIPT (configuration interaction with perturbation theory) method of Refs.~\cite{Dzuba2016,Lackenby2019a} overestimates ionization potentials of the last $6d$-block elements (they give error bars of about \SI{1}{\electronvolt}). Their CIPT values show a steady increase in the ionization potential from Mt (\SI{10.3}{\electronvolt}) to Ds (11.2\,eV) to Rg (\SI{12.2}{\electronvolt}) and Cn (\SI{13.1}{\electronvolt}). We therefore took their lowest listed values (fitting parameter $a$=0 in Table III of Ref.~\cite{Dzuba2016}) to show a more likely trend for the transactinides in Fig.\,\ref{fig:ionizationpotentials}, i.e., \SI{5.91}{\electronvolt} (Db), \SI{6.83}{\electronvolt} (Sg), \SI{7.70}{\electronvolt} (Bh), \SI{8.57}{\electronvolt} (Hs), \SI{9.43}{\electronvolt} (Mt), and \SI{10.3}{\electronvolt} (Ds). The most accurate values for the ionization potentials, electron affinities and polarizabilities (including excited states to confirm the correct ground state symmetry) of a transactinide element come from Fock-space coupled cluster theory \cite{Kaldor1991,Kaldor1998x,Eliav2005E119,Eliav2015,Oleynichenko2020}.

Figure \ref{fig:ionizationpotentials} shows a smooth increase in the ionization potentials for the $6d$ (Period 7) elements  due to an increase in nuclear charge causing the $6d$ shell to become more compact. This is consistent with the lighter $d$-block elements. They do, however, start at a lower ionization potential compared to the lighter systems. This can be explained by the more diffuse and relativistically-expanded $6d$ orbitals. However, starting at Bh, where occupation of the relativistically-destabilized $5d_{5/2}$ shell begins, we observe a higher ionization potential compared to the lower $d$-block elements in the PT, most likely due to a less sufficient screening of the nucleus by the $6d$ orbitals compared to the $5d$ orbitals. A drop in the ionization potential from the Group 12 to the Group 13  is expected as the underlying $ns$-shell and fully occupied $(n-1)d$-shell are more compact compared to the  $np$-shell. Moreover, the $s$-shell undergoes a strong relativistic stabilization. 

\subsection{QED effects}
\begin{figure}[tb]
  \begin{center}
\includegraphics[width=0.9\textwidth]{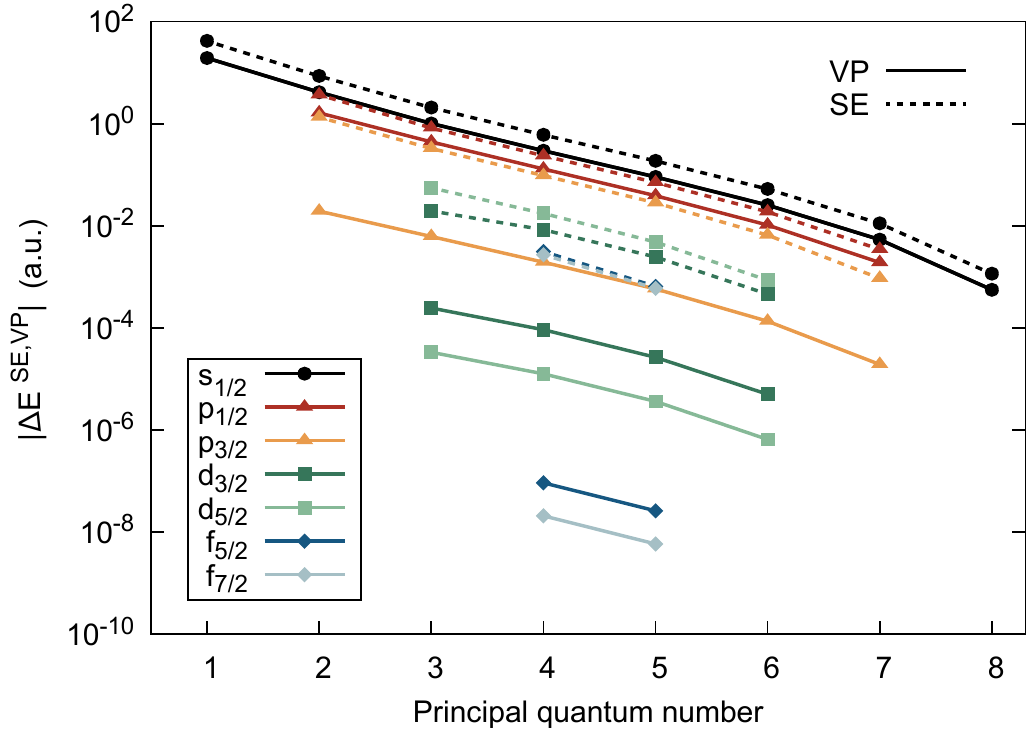}
  \caption{Absolute values for the vacuum polarization $|\Delta E_{n\ell j}^{\rm VP}|$ (solid lines) and self-energy $|\Delta E_{n\ell j}^{\rm SE}|$ (dashed lines) for different $(n\ell j)$ shells (per electron) of Ubn ($Z$=120). The formalism of Ref.~\cite{Flambaum2005} for the self-energy and the Uehling potential with a finite nuclear charge distribution was used \cite{Thierfelder2010}. All shell contributions to the VP have a negative sign, all shell contributions to the SE have a positive sign except for the $4f_{5/2}$ and $5f_{5/2}$ shells.}
  \label{fig:QED1}
  \end{center}
  \end{figure}

Calculated vacuum polarization $\Delta E_{n\ell j}^{\rm VP}$ and self-energy  $\Delta E_{n\ell j}^{\rm SE}$ (absolute value) contributions  to the total electronic energy for element Ubn ($Z=120$) are shown in Fig.\,\ref{fig:QED1}. As both VP and SE operators act in the close vicinity of the nucleus \cite{Schwerdtfeger2015},  the major QED contributions come from the $1s_{1/2}$ and $p_{1/2}$ shells. We also observe a decrease in QED contributions with increasing principal quantum number $n$ for fixed $(\ell j)$, a decrease with increasing angular quantum number $\ell$ for fixed $n$ and increasing total quantum number $j$ for fixed $n\ell$. This can all be explained by the changing density around the nucleus with changing combinations of $(n \ell j)$.
  
As we are interested in valence shell properties, one needs to know if QED-related  changes in shell occupations stem from the valence shell or from relaxation effects of deeper-lying core shells. As shown in Eq.\,\eqref{eq:radiat-behav}, QED effects of the valence shell vary as $1/n^3$. Figures \ref{fig:FZA_SE_1s_2s2p} and \ref{fig:FZA_SE_up5g} show the strong decrease of $F^{(1)}_{(n,\kappa)}(Z\alpha)$ with increasing $|\kappa|$ values for calculations with finite-size correction. This is illustrated in Fig.\,\ref{fig:QED1} for Ubn ($Z=120$). We therefore list in Table\,\ref{tab:ionization} the QED per-shell D-HF contributions to the ionization potentials for some selected atoms, Cn, Nh, and Ubn ($Z=112$, 113 and 120), as well as the QED contributions to the electron affinities of Nh and Uue ($Z=119$). 

\begin{table*}[tb]
\caption{\label{tab:ionization} QED contributions to the ionization potentials of Cn, Nh and Ubn and electron affinities of Nh and Uue obtained from D-HF calculations using GRASP \cite{Thierfelder2010}. For Nh the QED contributions are shown in eV rather than in percentage as cancellation effects in the valence shell lead to a very small total QED contribution to the ionization potential.}
\begin{center}
\scriptsize
\begin{tabular}{ l| c  c  c | c  c  }
\hline
Element & Cn & Nh & Ubn & Nh & Uue \\
\hline
& \multicolumn{3}{c|}{ionization potentials} & \multicolumn{2}{c}{electron affinities} \\
\hline
Transition & $^1S_0(6d^{10}7s^2)$ & $^2P_{1/2}(7s^27p^1_{1/2})$ & $^1S_0(8s^2)$ & $^2P_{1/2}(7p_{1/2}^1)$ & $^2S_{1/2}(8s^1)$   \\
 & $\rightarrow$ $^2D_{5/2}(6d_{3/2}^{4}d_{5/2}^{5}7s^2)$ & $\rightarrow$$^1S_0(7s^2)$ & $\rightarrow$$^2S_{1/2}(8s^1)$ & $\rightarrow$$^1S_0(7p_{1/2}^2)$ & $\rightarrow$$^1S_0(8s^2)$   \\
\hline
QED tot (\si{\electronvolt}) & 0.0227 & -0.0003 & 0.0110  & 0.0021 & -0.0020	\\
QED VP (\si{\electronvolt}) & -0.0150 & 0.0029 & -0.0060 & 0.0005 & 0.0031	\\
QED SE (\si{\electronvolt}) & 0.0377 & -0.0032  & 0.0170 & 0.0026 & -0.0051	 \\
\hline
  &  $7s$   (79.9\%)          & $7p_{1/2}$ (\SI{-0.0109}{\electronvolt})   & $8s$  (78.1\%)		& 7s (356\%) 		    & $8s$ (123 \%)  \\
        &  $6d_{5/2}$ (-17.6\%)     & $7s^2$ (\SI{0.0104}{\electronvolt})        & $7p_{3/2}$ (35.3\%)	& $7p_{1/2}$ (-227\%) 	&$7p_{3/2}$ (-68.9\%) \\
 \multicolumn{1}{c|}{Contributions}  &  $6p_{3/2}$   (22.2\%)     &		                    & 7s (-16.7\%)		    & $6d_{5/2}$ (18.9\%)	&$7s$ (41.6 \%)\\
 \multicolumn{1}{c|}{to total}  &  $6s$  (14.5\%)           &		                    & $6p_{3/2}$(4.9\%)     &$6d_{3/2}$ (11.9\%)	&$6p_{3/2}$ (-7.3\%) \\
 \multicolumn{1}{c|}{QED effect} &  $6d_{5/2}$ (7.9\%)        &		                    & $6s$ (-3.8\%)	        & $6p_{3/2}$  (-20.6\%)	&$6s$ (6.2 \%)\\
                &  $6p_{1/2}$ (2.0\%)       &		                    &		                &$6p_{1/2}$  (-29.7\%)	& \\
                &  $1s$   (-0.7\%)          &		                    &		                & 6s (-19.7\%)		    &\\
\hline
\end{tabular}
\end{center}
\end{table*}

Regarding the valence shell ionization potential, for Cn most of the QED effect comes from the relaxation of the $7s$ shell. The large contribution from $6p_{3/2}$ may come as a surprise, whereas for the $6p_{1/2}$ shell we have cancellation effects between the VP and SE contributions. For the Nh ionization, we see a slightly different picture with a small QED contribution due to an almost perfect cancellation of the $7p_{1/2}$ and $7s^2$ shell contributions. For the Ubn ionization potential, we obtain the largest contribution from the $8s$ shell, but the $7p_{3/2}$ and $7s$ shells also have substantial contributions (our result here deviates from the value given in Ref.~\cite{Thierfelder2010}).
Regarding the individual shell contributions to the electron affinities (defined here as positive
values) of Nh and Uuh, we observe for both elements that the main contribution comes from the shell where the electron occupation is altered, however there are also large contributions of shells where the occupations stay the same.
These results show that relaxation effects are important. For Cn, the dominant contribution does not come from the shell where the electron occupation is altered as QED effects are much larger for relaxing the $s$ shell than removing an electron from the underlying $d$-shell.

\begin{figure}[tb]
  \begin{center}
\includegraphics[width=0.7\columnwidth]{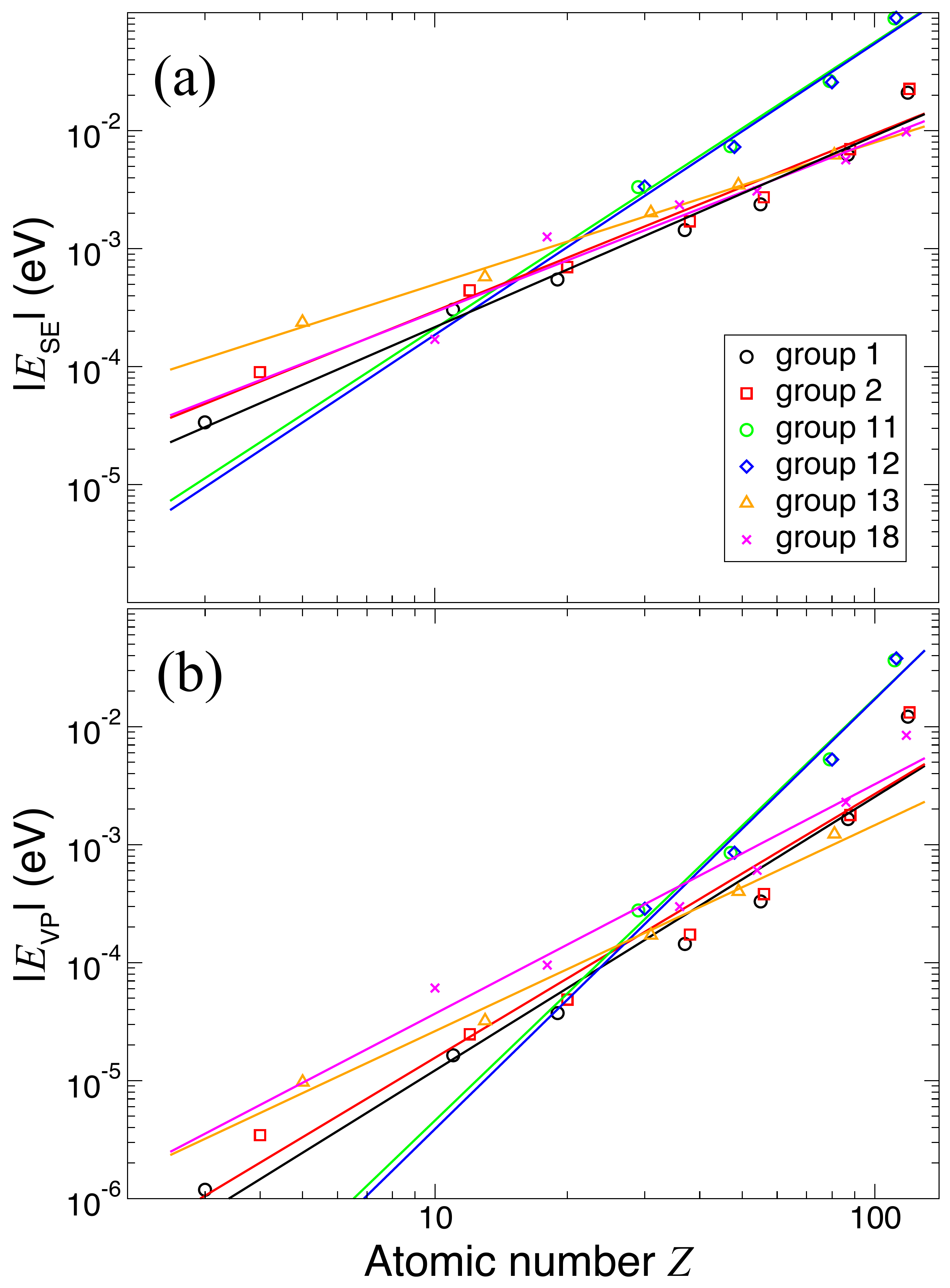}
\caption{\label{fig:SEvzZ} (a) Self-energy (SE) and (b) vacuum polarization (VP) energy contributions 
to the ionization potential against the atomic number $Z$ for different atoms within a Group of the Periodic Table. Data are taken from Ref.~\cite{Thierfelder2010}}
  \end{center}
\end{figure}

We note that Og was predicted to be the first rare gas element with a non-zero electron affinity of \SI{0.080}{\electronvolt}  \cite{Eliav-1996,Lackenby2018,guo2021ionization}, where the Breit interaction contributes with
\SI{-3E-4}{\electronvolt}  and QED with \SI{-3E-3}{\electronvolt}, the latter already comparable in size to the quadruple contributions in a CC treatment \cite{guo2021ionization}.

As shown in Fig.~\ref{fig:SEvzZ}, the VP and SE contributions for the valence shell ionization potentials approximately 
obey a simple power law \cite{Thierfelder2010}
\begin{align}
E(Z)=C Z^\gamma.
\label{polynomialfit}
\end{align}
The exponent $\gamma$ of the VP is roughly 40 to 50\% larger than the one of the SE, and we observe a lower scaling for the valence-$p$ shell compared to the $s$-states.
An logarithmic-scale extrapolation  to high $Z$ shows that the VP and SE curves
cross at $Z=160$ for the Group 11 and $Z=139$ for the Group 1 elements.

For the superheavy elements beyond nuclear charge $Z=120$, the accurate treatment of electron correlation methods still remains the major bottleneck. It could therefore be sufficient to include QED and Breit interactions at a perturbative level, especially if higher-$\ell$ states are involved and shell relaxation effects are small. For example, Ref.\,\cite{ibj2011} performed calculations on a series of transactinides with $Z\ge$140 . For the elements with $Z = 171$, \num{172}, and \num{173} QED effects contribute to the ionization potential by \SI{1.7}{\percent}, \SI{0.1}{\percent}, and \SI{1.2}{\percent}, respectively (the very low value
for $Z = 172$ comes from an almost exact screening of the SE and an exact cancellation of VP in the neutral and singly ionized case \cite{ibj2011}.

The accurate treatment of QED effects is more important when it comes to the prediction of inner shell ionization potentials \cite{Gaston2002,Thierfelder2009a}, especially for high-$Z$ few electron systems \cite{ind2019}, or for $K$-capture rates for neutron deficient nuclei \cite{Bambynek1977}. For the latter, little theoretical work in this direction has been done so far \cite{Pachucki2007}.

\subsection{Atomic static dipole polarizabilities}

Atomic static dipole polarizabilities $\alpha_D$ are very useful quantities for chemical reactions and are, for example, used in the simulation of temperature dependent atom-at-a-time experiments of transactinides absorbed on surfaces such as gold or quartz
\cite{Pershina2008,Pershina2008,Pershina2008x,Pershina2008y,Pershina2009,Turler-Pershina-2013,Pershina2016x,Pershina-2018,Trombach2019}. Dipole polarizabilities are proportional to the inverse of the ionization potential, $\alpha_D\sim I_p^{-1}$ \cite{chandrakumar2004relationship}, which can be argued from the sum-over-states formula. Empirically one approximately finds  $\alpha_D= 6.67 I_p^{-2}$\,a.u.  (with RMSD of \num{0.93} a.u.).

Calculated dipole polarizabilities are listed in Table\,\ref{tab:properties1}. They have recently been reviewed for the known elements in the PT \cite{Nagle2019}, where periodic trends were also discussed. Note that we only discuss here the scalar component of the polarizability tensor as shown in Table\,\ref{tab:properties1}. If strong spin-orbit coupling is involved, one requires knowledge of the $M_J$-resolved components for the Stark effect in open-shell atoms \cite{bonin1997electric}.

The start of a new shell occupation usually increases the polarizability; this is seen for Lr, Nh, Mc and element 119. The rather large DHF+CI+Breit+QED polarizability for Lr comes with large estimated uncertainties \cite{Safronova2014,Nagle2019}. Nevertheless, we see a decreasing trend in dipole polarizabilities across the transactinide series, and perhaps the spin-orbit splitting is not large enough to explain an increase in $\alpha_D$ from Sg to Bh upon occupation or the $6d_{5/2}$ shell. 

\subsection{Electron Localization Function}

The concept of the electron localization function (ELF) was introduced in Ref.~\cite{Becke1990}, and later applied to atoms, molecules and the solid state \cite{Savin1994,Savin1997}. More recently, the concept was extended to nucleon localization functions (NLF) in nuclear structure theory \cite{Reinhard2011,Zhang2016,TongLi2020}. The ELF provides a measure of finding an electron in the vicinity of another same-spin electron located at a given position $\vec{r}$:
\begin{equation}
D_\sigma(\vec{r})=\left[1+\left(\frac{\tau_{\sigma}(\vec{r})\rho_{\sigma}(\vec{r})-\frac{1}{4}|\vec{\nabla}\rho_{\sigma}(\vec{r})|^2}{\rho_{\sigma}(\vec{r})\tau^\mathrm{TF}_{\sigma}(\vec{r})}\right)^2\right]^{-1},
\end{equation}
where spin $\sigma$ is  $\uparrow$ or  $\downarrow$, $\rho_{\sigma}$ is the electron spin density, $\tau_{\sigma}$ is the kinetic energy density, $\vec{\nabla}\rho_{\sigma}$ is the electron density gradient, and $\tau^\mathrm{TF}_{\sigma}$ is the Thomas-Fermi kinetic energy (of the uniform electron gas). The ELF can be seen as a tool to identify regions where electrons are localized or delocalized (for a review see for example Ref.~\cite{fuentealba2007}). The ELF assumes values between 0 and 1, where a value close to 1 indicates that the probability of finding two same-spin electrons close to each
other is very low (high level of electron localization) and the ELF value of \num{0.5} corresponding to the limit of a hypothetical uniform Fermi gas of the same density (high level of electron delocalization). ELFs are therefore ideal to show the changes in electron localization due to the influence by other atoms or by relativistic effects. The discussion of ELF and NFL in superheavy nuclei can be found in Ref.
~\cite{Jerabek2018}

\begin{figure}[tb]
\centering
\includegraphics[width=1.0\columnwidth]{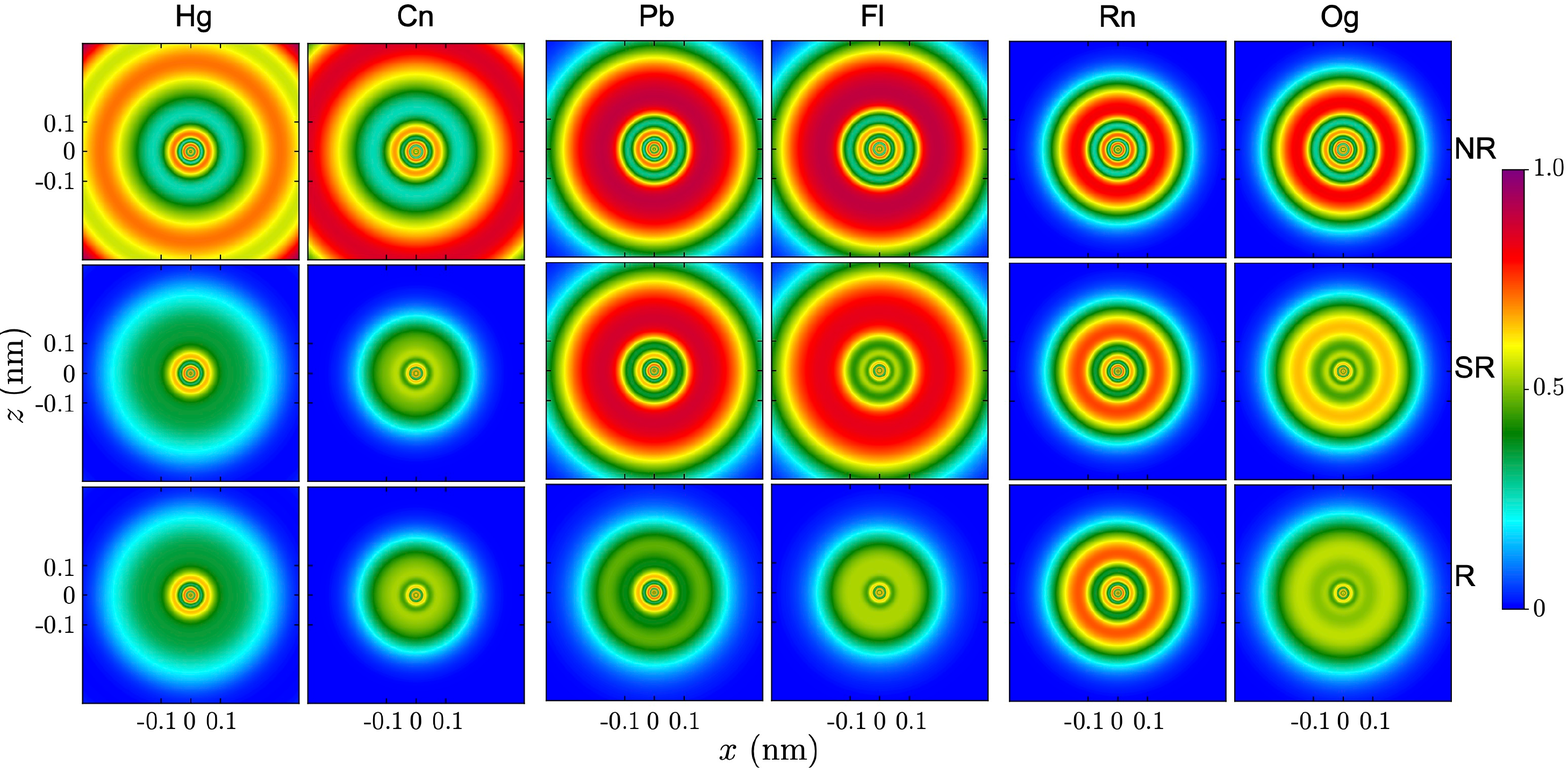}
\caption{\label{pic:ELF} ELFs $D_\sigma(r)$ from nonrelativistic (NR, top), scalar relativistic (SR, middle) and fully relativistic (R, bottom) DFT calculations for Hg and Cn (left), Pb and Fl (center), and Rn and Og (right). Adopted from Ref.\,\cite{Florez2022}.}
\end{figure}

To demonstrate the importance of relativistic effects for the superheavy elements we show in Fig. \ref{pic:ELF} the ELFs for Cn, Fl, and Og in comparison with their lighter congeners. The apparent delocalization is due to scalar relativistic effects in the case of Cn, and due to spin-orbit effects for both Fl and Og. The appreciable relativistic effects are already notable for Hg. It has been  argued that the large spin-orbit splitting in the $7p$  shell of Og with \SI{10.13}{\electronvolt} is responsible for a uniform-gas-like behavior in the valence region \cite{Jerabek2018}; see also discussion in Ref.~\cite{Kaygorodov2021}.

\subsection{Examples of relativistic effects on the chemistry of SHE}

\begin{figure}[tb]
  \begin{center}
\includegraphics[width=0.77\textwidth]{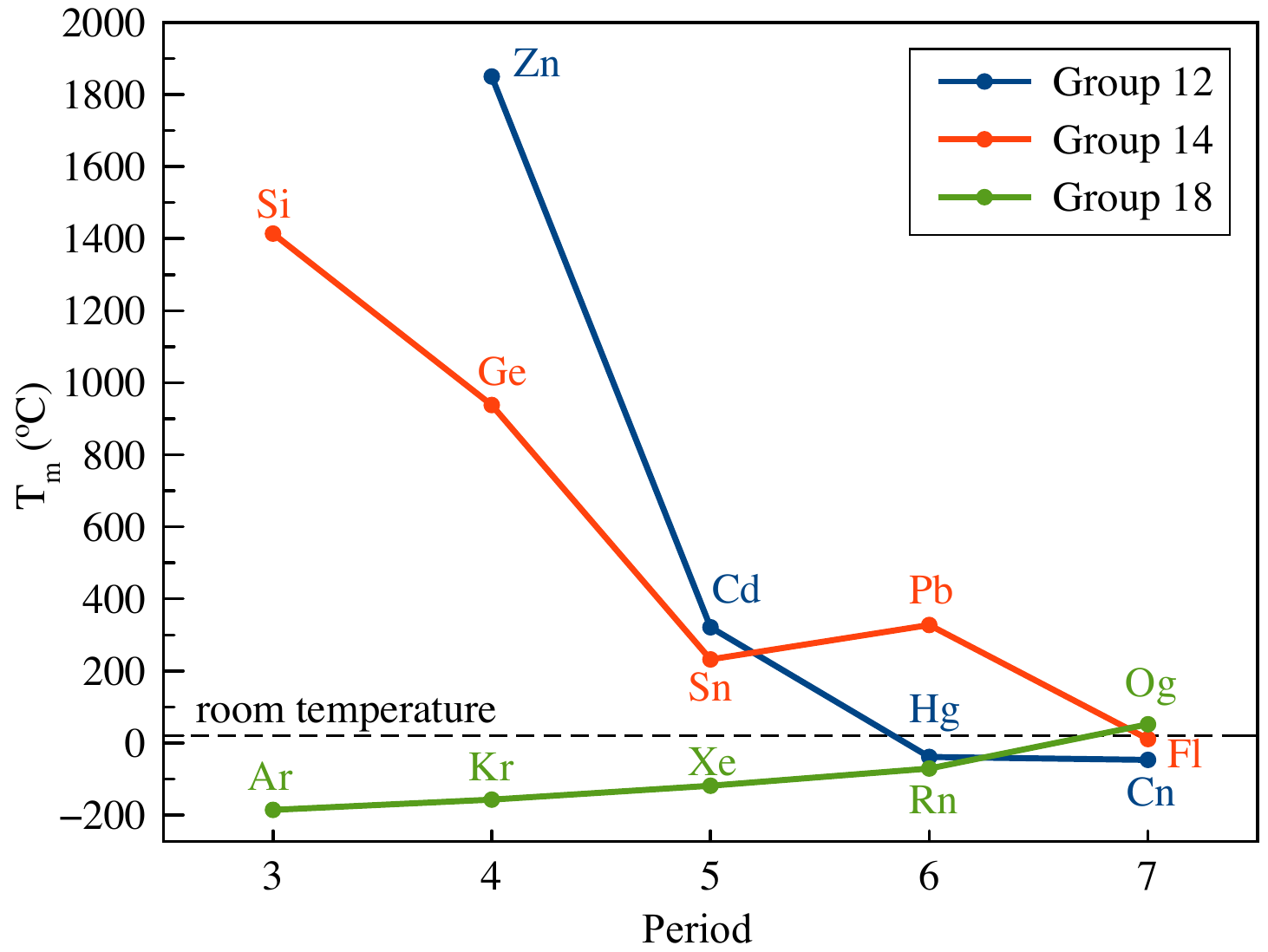}
  \caption{Melting points for the Group 12, 14 and 18 elements of the periodic table. Experimental values are from \cite{CRC2016}. For Cn, Fl and Og the melting points are taken from computational simulations  \cite{Mewes-2019,Mewes2021,Florez2022}.}
\label{fig:Melting}
\end{center}
\end{figure}

Relativistic effects have a profound influence on the chemistry of the transactinides.
For example, due to the large relativistic $7s$ stabilization, Cn is expected to behave like a rare gas with a low predicted melting and boiling point of  \SI{-18+-30}{\celsius} and \SI{77+-10}{\celsius}, respectively (at the nonrelativistic level ca. \SI{370}{\celsius} and \SI{970}{\celsius} respectively) \cite{Mewes-2019,Mewes2021}. 
In contrast, relativistic effects can cause the Group-18 element  Og to be a semi-conductor and a solid at room temperature (RT) with an electron localization function resembling that of a Thomas-Fermi gas \cite{Jerabek2018,Mewes2019,Smits2020a}. The trends in melting points for the Group 12, 14 and 18 of the periodic table are summarized in Fig.\,\ref{fig:Melting}. Each of the predicted melting points for the three superheavy elements are very close to room temperature and are highly influenced by relativistic effects.

As another example, we take the closed $7p^2_{1/2}$ shell of the Fl atom (Period 7, Group 14), for which the relativistic effects are predicted to cause  a large $7p_{1/2}-7p_{3/2}$ spin-orbit splitting. This explains the maximum in the ionization potential in Fig.\,\ref{fig:ionizationpotentials} and its expected low melting temperature \cite{Mewes2021}.
Furthermore, the $d$-block transactinides are expected to always ionize out of the $6d$ shell as the fully occupied $7s$ shell undergoes a very strong relativistic energetic stabilization and contraction. This was pointed out early on for Rg, which adopts the $d^9s^2$ configuration rather than the usual $d^{10}s$ arrangement \cite{Fricke1977DS,Eliav-1994}. 
For further reading, see Refs.~\cite{hoffman2008t,Turler-Pershina-2013,Pershina2014x,pershina2019relativity}.

\section{General Considerations}
\begin{figure}[tb]
  \begin{center}
\includegraphics[width=.8\textwidth]{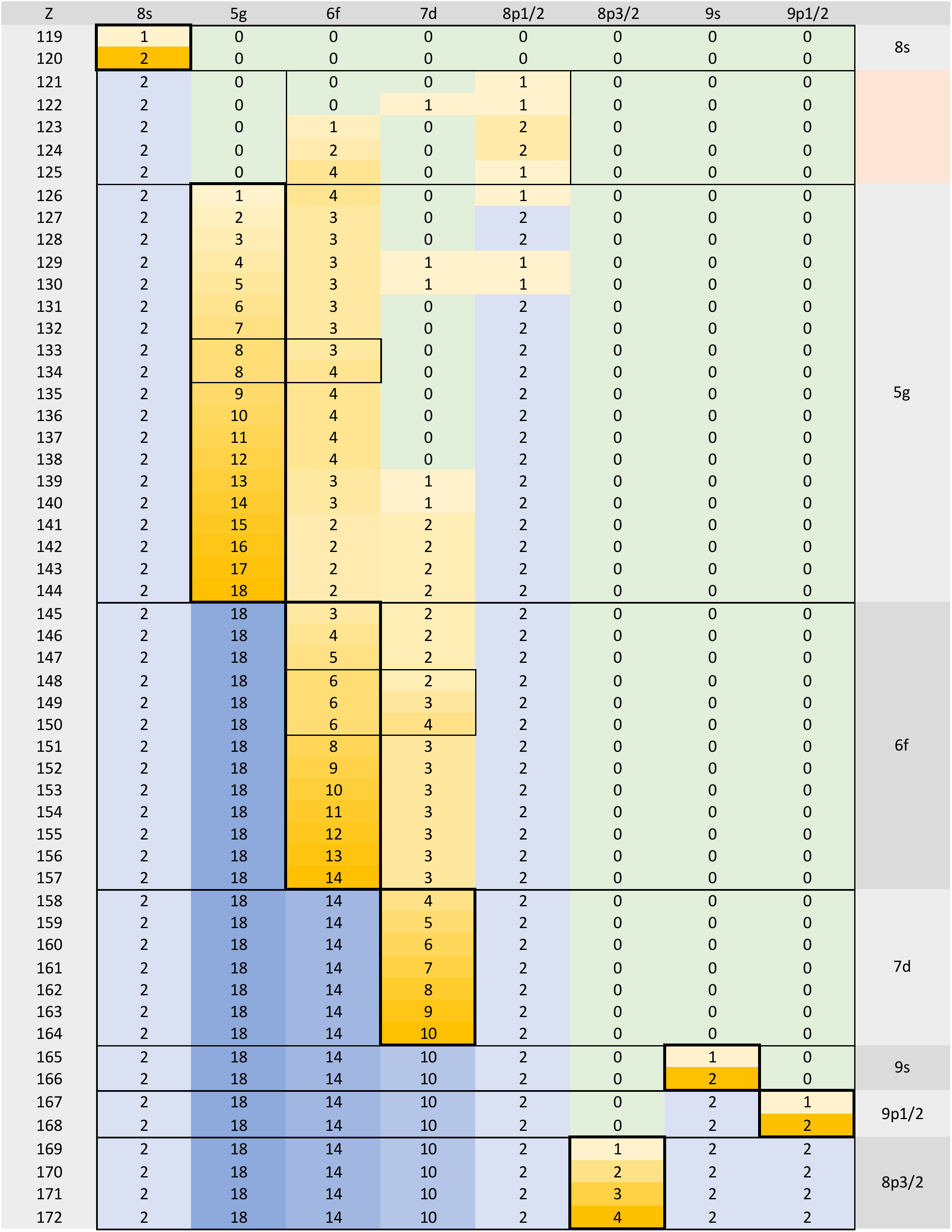}
  \caption{Predicted ground state configurations for the elements with atomic numbers Z= 1 Empty orbitals in green, filled orbitals in blue and the partly filled orbitals in yellow. 19-172  \cite{Nefedov2006}. This diagram visualizes the difficulty with placing the SHE elements in the PT. Whereas some elements have a definite place, a handful of elements, such as $Z = 121-125$, cannot be uniquely placed. The elements $Z = 133$ and $Z = 134$ share the same place, just as $Z = 148$, 149 and 150. Also the $5g$ and $6f$ elements  contain higher-order occupations of  $6f$ and $7d$ state.}
  \label{fig:SHE_PT}
 \end{center}
 \end{figure}

\subsection{Placing new elements on the periodic table}
The correct placement of the elements in the PT (see Fig.\,\ref{fig:PT}) has been a matter of intense debate  \cite{Scerri2012periodic,scerri2013cracks,Pyykkoe2019}. Besides the evident ordering of elements according to their atomic number $Z$, the additional two principles for the placement of atoms into the PT are the Pauli principle, derived from the spin-statistics theorem, and the Aufbau principle, derived from mean-field theory \cite{Schwerdtfeger2020,SchwarzPT2022}.  The Pauli principle dictates that only one electron can be put into a (spin-)orbital, or in the relativistic case into an orbital with quantum numbers $(n,\ell, j, m_j)$. The electrons then fill the orbitals in the order of increasing orbital energy, leading to the  Aufbau principle, where the leading configuration for each element can be obtained from either Kohn-Sham or from MCSCF theory \cite{Schwerdtfeger2020}. By formatting the PT in two dimension, the configurations of valence electrons are repeated within a row with increased principal quantum numbers. Since, in the nonrelativistic approach, the valence electrons dictate the angular distribution of the wave functions, there is a periodicity of the chemical properties \cite{pyykko2016periodic, Schwarz2019, Schwerdtfeger2020}.  Relativistic effects, on the other hand, can substantially alter the behavior of an element. This is especially observed for the main group and late transition elements \cite{Pyykko-1979, Pyykko-1988, pyykko-2012relativistic, Pyykko2012PT, Turler-Pershina-2013, Pyper2020, Schwerdtfeger2020}. 
These relativistic effects as well as electron correlation contributions lead to exceptions to the Aufbau principle, which cause difficulties with the element placement. Electron configurations can alter within a group of the PT as well, and the Group 10 elements serve as a perfect example. Here one has the dominant electron configuration $3d^84s^2$ for Ni, $4d^{10}$ for Pd,  $5d^96s^1$ for Pt, and $6d^87s^2$ for Ds. Yet, inspection of the excited states reveals that apart from the principal quantum number, all three configurations $(n-1)d^8ns^2, (n-1)d^9ns^1$ and $(n-1)d^{10}$ are close in energy \cite{Schwerdtfeger2020}. Another difficulty is the correct placement (starting and ending point) of the $f$-block elements \cite{scerri2013trouble,scerri2018}, see Fig. \ref{fig:SHE_PT}. This makes the PT a somehow fuzzy concept in the superheavy region.

\subsection{Dominant ground state configurations predicted by electronic structure calculations}
A number of authors have tried to predict the dominant electron configuration(s) beyond the element with nuclear charge $Z=118$ \cite{Fricke1971, fricke1975, fricke1976chemical, Fricke1977DS, Umemoto1996, Nefedov2006, hoffman2008t, pyykko2011PT}. Early attempts were made by using D-HF-Slater calculations  for the range $Z=118-131$ \cite{MannCromer1969,Mann1970,Fricke1971}.  However, these mean-field studies did not explicitly include static and dynamic electron correlations. Since the total energies of different configurations differ little from one another in many cases, inclusion of configuration interaction can lead to a change of the ground-state symmetry and configuration. To improve the early results,  multi-configuration Dirac–Fock calculations have been carried out \cite{Nefedov2006} with inclusion of a total angular momentum coupling scheme (see Sec. \ref{sec:MCDHF}) and the Breit correction. These results were  extended to chemically plausible ions \cite{pyykko2011PT}, to differentiate between free ions and ions in chemical compounds. The resulting PT is shown in Fig.\,\ref{fig:PT}. In this work,  the starting point of the $5g$ elements  has been placed at $Z=121$ (in contrast to Ref.~\cite{Nefedov2006} that places the starting point at $Z=$126). Because of the high density of states in the region beyond $Z=120$, it becomes quite difficult to pick the dominant configuration and ground state symmetry. For example, Ref.~\cite{ibj2011} showed from average level (AL) calculations that for the element with $Z=140$ the ground state configuration is $8s^28p^27d6f^35g^{14}$  in agreement with Ref.~\cite{Nefedov2006}, but this could change if the state of specific symmetry is optimized and more configurations are included. This shows that more work is required to determine the  ground state configurations and symmetries, as well as associated chemical behaviour \cite{pyykko2011PT}. Possible candidates for configurations for the elements up to $Z=172$, taken from Ref.~\cite{Nefedov2006}, are visualized in Fig.\,\ref{fig:SHE_PT}.  Needless to say that for the superheavy elements  relativistic effects are crucial and need to be correctly accounted for as they can lead to a noticeable violation of simple regularities such as for the Group 11 elements where Rg adopts a configuration with a hole in the $6d_{5/2}$ shell instead of the $7s$ shell \cite{Eliav-1994}.

The third pillar (beside the atomic number and electron configuration) for building the PT comes from chemical properties, similar as how the plausible chemical ions were taken into consideration for the placement of atoms in Ref.~\cite{pyykko2011PT}. We have learned in the past few decades that relativistic effects can alter chemical properties substantially \cite{Pyykko-1979,Pyykko-1988}, especially in the superheavy element region \cite{Turler-Pershina-2013,giuliani2018,Schwerdtfeger2020}. Regarding the chemical properties of the SHE beyond $Z=120$, there is no information available except from the computational studies. For example, it has been shown \cite{Dognon2017} that the $5g$-electrons for the elements with atomic numbers $Z=125-129$ are core-like and only act as spectators, similar to the $4f$ electrons in the lanthanides. It is clear that more work needs to be done  to study the chemical properties in the $Z>120$ region, which will help designing future  atom-at-a-time chemistry experiments \cite{eichler2019periodic}.

\subsection{Periodic Table - How far can we go?}
Oganesson was the heaviest element and nihonium the last element to be added officially into the PT \cite{Karol2016a}. Thus the 7th period of elements is now complete. The question arises if one can go much further in the atomic number. From an electronic point of view, there is no limitation to  $Z$. While the correct description of multi-electron systems with $Z>Z_{\mathrm{c}}\approx 170$ is still a difficult problem, and the inclusion of  Gamow states for multi-electron systems  needs to be addressed, the real limitation to the PT comes from the nuclear stability \cite{Nazarewicz2002,giuliani2018}. In the transactinide region, $Z > 103$, in early days known as the sea of instability \cite{seaborg1969}, the half-life of the elements varies between hours $\left(^{266}_{103}\text{Lr}\right)$ and seconds $\left(^{294}_{118}\text{Og}\right)$ or below. Although there is a small predicted region of increased stability between $Z$=114-126 and $N=184$ with predicted lifetimes of hours or even days \cite{Oganessian2015a,Duellmann2018,giuliani2018,chapman2020}, it is currently not clear how far the PT can be extended from the nuclear point of view.  Indeed, the IUPAC defines an element to exist if its lifetime is longer than $T_{\rm el}$\SI{\approx 1E-14}{\second}, which is the time it takes for electron cloud to form around the nucleus.
This means that for atomic nuclei living shorter than $T_{\rm el}$
it makes no sense to talk about atoms and chemistry \cite{Nazarewicz2002,giuliani2018}.  

\begin{figure}[tb]
 \begin{center}
  \includegraphics[width=1.0\columnwidth]{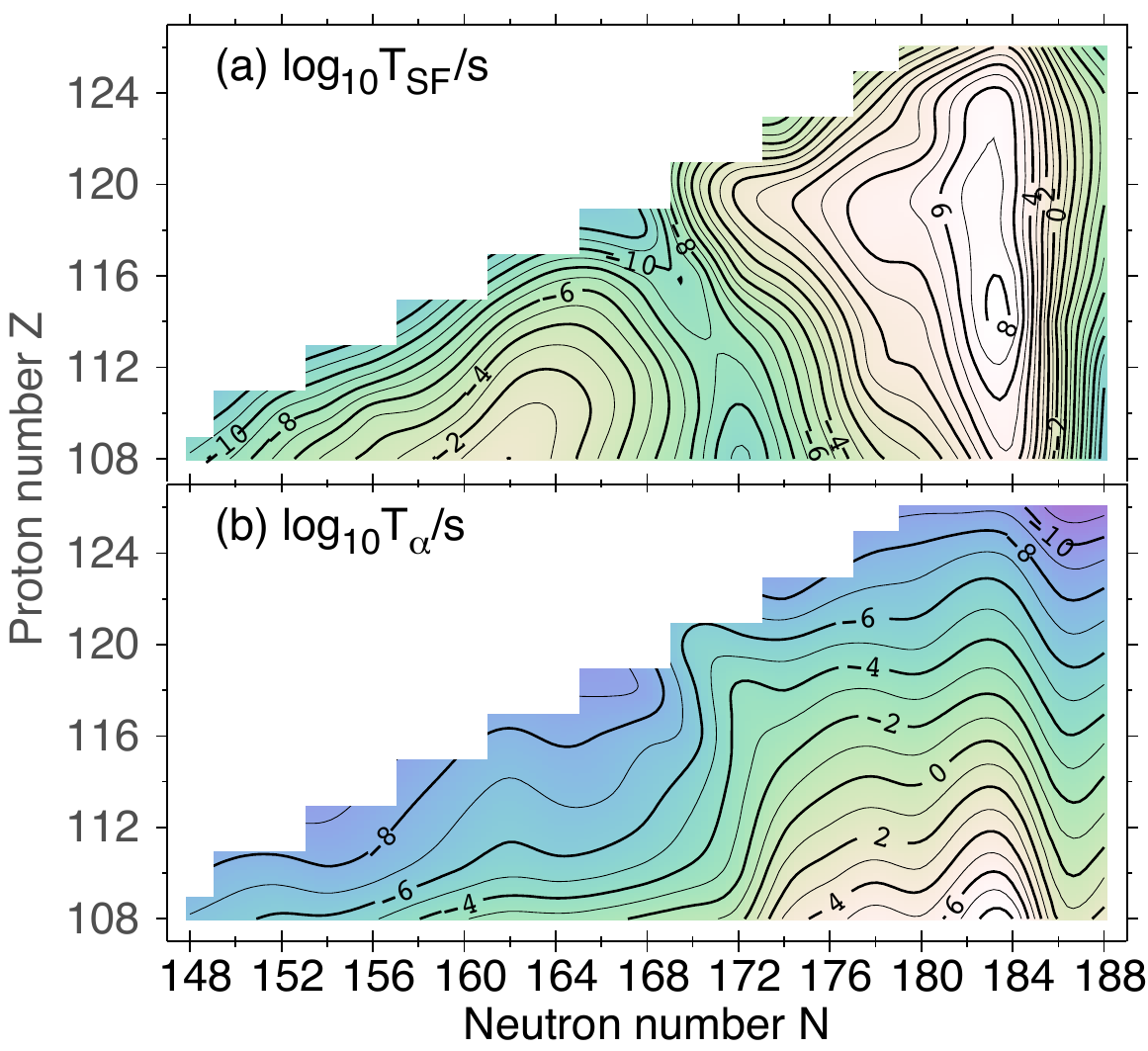}
  \caption{\label{fig:lifetimes}  Summary of theoretical predictions \cite{Staszczak2013} for decay modes of superheavy  nuclei obtained with nuclear DFT: (a)  SF half-lives; (b) $\alpha$-decay  half-lives.}
   \end{center}
 \end{figure}

The lifetimes of most known superheavy nuclei are governed by the
competition between $\alpha$-decay and spontaneous fission (SF). 
The  corresponding lifetimes predicted by a particular DFT model \cite{Staszczak2013}  are shown in Fig.~\ref{fig:lifetimes}. For a survey of various predictions of $\alpha$-decay and SF lifetimes, see Refs. \cite{Heenen2015} and \cite{Baran2015}, respectively.
The shortest SF half-lives,
reaching down to \SI{E-10}{\second}, are predicted for nuclei from a narrow corridor formed by $^{280}$Hs, $^{284}$Fl, and $^{284}$Og. This corridor of fission instability separates the regions of superheavy nuclei synthesized in hot- and cold-fusion
reactions. 
Moving towards more neutron-rich nuclei beyond $N=184$, dramatic decrease of SF lifetimes, below \SI{e-15}{\second} is expected, see Fig.~\ref{fig:lifetimes}(b) and  Ref.~\cite{Giuliani2017}.

It is to be noted that predictions of nuclear models in the region of superheavy nuclei
are very sensitive to both input (forces, functionals) and theoretical framework used.
Consequently, theoretical lifetime estimates, especially for SF,  often differ by many
orders of magnitude \cite{Baran2015}.
The heaviest nuclei synthesized so far are all
proton-rich; hence, they can in principle decay by means of electron
capture or $\beta^+$/EC process. So far, no such decay modes have
been observed in the upper superheavy region, and this indicates
that they cannot compete with the dominant $\alpha$ -decay and SF modes. Indeed, according to
theory $\beta^+$/EC  lifetimes shorter than \SI{1}{\second} are expected in nuclei that
lie rather far from the current superheavy region \cite{Heenen2015}.

Experiments to synthesize new superheavy nuclei and elements beyond Og are  underway \cite{Khuyagbaatar2020,Oganessian2022Mc}. If discovered, these systems will be crucial for    testing  many-body nuclear structure theories \cite{giuliani2018}. To explore their chemistry will  be very challenging, however.

\section{Conclusions}

Atomic structure theory developed enormously over the past decades to the extend where QED can be tested to high precision for few-electron systems. However, for many-electron systems the accurate description of both QED and electron correlation effects remains a major challenge \cite{lindgren2016relativistic}. But even here progress has been made for elements with large atomic numbers \cite{ibj2011}. While the negative energy continuum is required for QED,  is creates a formidable conceptual and computational challenge, especially when bound states approach the negative energy continuum threshold. The correct description of diving (Gamow) states within a multi-electron formalism including QED effects still needs to be explored. It is clear that effects associated with the negative energy continuum distinguishes the Dirac from the Schr\"{o}dinger equation in both mathematical and in physical terms \cite{thaller1992}. Once these problems are solved, there is no limitation to the treatment of atoms beyond the critical nuclear charge. The PT, seen as the foundation for chemistry, is therefore \textit{not} limited to a certain nuclear charge region, but limited by nuclear stability \cite{giuliani2018}. The future looks bright for superheavy element synthesis and associated chemistry experiments, which require the support of accurate electronic  and nuclear  structure theory.

\section{Appendix A: The Self-Adjointness of the Dirac-Coulomb Hamiltonian}\label{sec:appA}
 
In the two appendices we address some of the more mathematical features of the Dirac equation, the self-adjointness problem and (in the next section) the rigged Hilbert space formalism for Gamow states. Both aspects often lead to some misunderstandings in the community and are therefore discussed briefly here.

To explain the non-self-adjointness at critical charge in correct mathematical terms, we require the $L_2$-norm of the derivative $||\frac{d}{dr}\phi||_2$ to exist (or the gradient norm for the three-dimensional case) for the eigensolutions as the Dirac equation is a first-order differential equation (see discussion of Sobolev spaces further below). The radial solutions for a point charge nucleus $\phi(r)$ have the general form $\phi(r)=a_{n\kappa}(2Zr)^\gamma e^{-Zr}f^{P,Q}_{n\kappa,Z}(r)$ with the exponent $\gamma=\pm\sqrt{\kappa^2-(Z\alpha})^2$ and $f^{P,Q}_{n\kappa,Z}(r)$ containing expressions of $P_{n\kappa}(r)$ and $Q_{n\kappa}(r)$ in terms of confluent hypergeometric functions \cite{Gordon1928}. Unlike for the Schr{\"o}dinger equation, $\gamma$ is a non-integer and derivatives lead to negative $r$-exponents. As a result, both $||\frac{d}{dr}\phi||_2$ and  $||H_{\rm D}\phi||_2$ become infinite if $\gamma\le \frac{1}{2}$ leading to $Z_{c_1}\!\alpha=\sqrt{3}/2$ as discussed in Sec.~\ref{EPDE}. In contrast, for the nonrelativistic  radial Schr{\"o}dinger equation, all derivative norms exist, \ie $||\frac{d^n}{dr^n}\phi_{\rm NR}||_2<\infty$, as only integers appear in the $r$-exponent. This leads often to misunderstandings as the $L_2$ Hilbert space only requires the norm $||\phi||_2$ to exist, but as soon as we introduce an unbound differential operator such as $H_D$ we have to deal with the domain of such an operator and the existence of certain expectation values and derivative norms. For a more detailed analysis using the Weyl's limit point - limit circle theorem, generalized to the Dirac equation by Weidmann \cite{Weidmann1982}, the reader is referred to a recent paper by Gallone \cite{Gallone2017}. The various norm existences for different nuclear charge regions are summarized in Table~\ref{tab:norms}.
\begin{table}[tb]
\caption{\label{tab:norms} Norm and expectation value existences for different ranges of nuclear charges $Z$ and parameter $\pm \gamma(Z)=\pm \sqrt{1-(Z\alpha)^2}$ appearing in the radial 1s function of hydrogenic atoms with potential $V(r)=-Z/r$. The symbol S stands for the Sobolev norm ($S=1$ for Dirac and 2 for Schr\"odinger) which requires the gradient norm to exist for the Dirac-Coulomb operator (first and second derivatives for the Schr\"{o}dinger case).}
\begin{center}
\begin{tabular}{|c|c|c|c|c|c|}
\hline
System & range & $\norm{\phi}_2$ & $ \bra{\phi}H \ket{\phi} $ & $\norm{\phi}_2^{\rm (S)}$ & $\norm{H\phi}_2$\\
\hline
Schr\"{o}dinger & $Z>0$ & yes & yes & yes & yes \\
Dirac $+\gamma(Z)$ & $0<Z\alpha<\sqrt{3}/2$     & yes  & yes & yes & yes \\
Dirac $+\gamma(Z)$ & $\sqrt{3}/2\le Z\alpha<1$  & yes & yes & no  & no \\
Dirac $-\gamma(Z)$ & $0<Z\alpha<\sqrt{3}/2$      & no   & no  & no  & no \\
Dirac $-\gamma(Z)$ & $\sqrt{3}/2\le Z\alpha<1$  & yes  & no  & no  & no \\
\hline
\end{tabular}
\end{center}
\end{table}

To rephrase the self-adjointness condition in different terms, a self-adjoint extension of the radial Dirac operator should have the following domain \cite{Weidmann1982,thaller1992}: dom$(H_{\rm D})=\{\phi\in L_2(\mathbb{R}_+)^2 |$ each component of $\phi$ is locally absolutely continuous; $H_{\rm D}\phi \in L_2(\mathbb{R}_+)^2$; deficiency indices $d\{\phi(r=0)\}=(0,0)\}$ (see also Hogreve \cite{Hogreve_2012}), where $L_2(\mathbb{F})^n\equiv L_2(\mathbb{F})\otimes \mathbb{C}^n$ over a field $\mathbb{F}$ ($\mathbb{F}\equiv\mathbb{R}_+$ and $n=2$ for the radial Dirac equation) \cite{thaller1992}. This allows us to select the Sobolev space $\mathcal{W}_{1,2}(\mathbb{R}_+)^2$ \cite{richtmyer1978principles,thaller1992} as the natural domain for the (unbound) Dirac operator (or similarly for a four-component wave function in the three-dimensional case $\mathcal{W}_{1,2}(\mathbb{R}^3)^2$) lying dense in $L_2(\mathbb{R}_+)^2$, such that dom$(H_{\rm D})\subseteq\mathcal{W}_{1,2}(\mathbb{R}^3)^2\subseteq L_2(\mathbb{R}^3)^2$. The domain problem of the Dirac-Coulomb operator has been very recently discussed and critically analyzed by Estaban \cite{Esteban2020}. In the {\it subcritical nuclear charge region} ($\sqrt{3}/2\le Z\alpha\le 1$), $\phi$ is an eigenfunction to a non-self-adjoint Dirac-Coulomb Hamiltonian with real eigenvalues and norm $||\phi||_2<\infty$, but does not belong to dom$(H_{\rm D})$ ($H_{\rm D}$ self-adjoint)! More generally, one looks for the largest subdomain of the Hilbert space that remains invariant under the action of certain powers of required operators (observables) including the Hamiltonian of the system, which is known as the maximal invariant subspace of the algebra generated by these operators \cite{deMadrid2005}.

A note of caution should be added here. If we eliminate the small component and focus on the resulting second-order differential equation, the underlying Sobolev space is now $\mathcal{W}_{2,2}(\mathbb{R}^3)^2$, which makes the conditions more stringent for the norm existence. In any case, we seek for an appropriate self-adjoint extension of $H_D$ \cite{Esteban2007,Arrizabalaga2013,gitman2012self} as physics does not restrict atoms to a maximum critical charge (except for nuclear instability which is an entirely different matter \cite{Nazarewicz_2016Challenges}). The necessary boundary condition for securing the self-adjointness of the Dirac operator at the origin has been discussed in detail by Kuleshov \cite{kuleshov2015vs} and Gitman \cite{Voronov2007,gitman2012self}. The physical realization of this boundary condition is the introduction of a finite nuclear charge density \cite{Pomeranchuk1945}.

Last we mention that mathematically, there are many self-adjoint extensions which can be realized for the Dirac operator. For example, Kato showed that a potential energy matrix of the form $V_{ik}=a/2r+b$ with $a<1$ and $b>0$ makes the Dirac operator essentially self-adjoint on certain domains \cite{kato2013perturbation}. For a more recent discussion on possible self-adjoint extensions we refer to \cite{Arrizabalaga2013}.

\section{Appendix B: The Rigged Hilbert Space Formalism}\label{sec:appB}

If we consider the spectrum of the Dirac-Coulomb operator, $\sigma(H_D)$, we need to include the discrete (d) and both the positive (+) and negative ($-$) continuum states (c) (cf. Figure \ref{fig:DS}), \ie $\sigma(H_D)=\{\phi^\text{d}\}\cup\{\phi_{+}^\text{c}\}\cup\{\phi_{-}^\text{c}\}$. The continuum states are important in scattering (resonance) theory and are essential for the quantum mechanical completeness relation. It is well known, however, that unlike the discrete states, the continuum states lead to domain problems for unbound operators, \ie they are not normalizable and therefore do not belong to the quantum mechanically relevant $L_2$ space. As a result, von Neumann's original Hilbert space formalism requires an extension to include such (generalized) functions. Continuum states are properly defined within an extended or rigged Hilbert space ($\mathcal{G}$) formalism  \cite{maurin1966,bohm1978,Bohm1997,perelomov1998quantum,Gadella2003,antoine2009rigged,Antoine2021} originally proposed for the quantum mechanical framework by Roberts, Antoine and Bohm \cite{Roberts1966,Roberts1966a,bohm1967rigged,Antoine1969,Antoine1969b}.\footnote{The term rigged Hilbert space is misleading as $\mathcal{G}$ is not a Hilbert space per se, but is generated from a Hilbert space $\mathcal{H}$.} In fact, rigged Hilbert spaces are the structures required for both the discrete orthonormal and continuous bases to coexist \cite{Trapani2019}. Their existence for self-adjoint operators on separable Hilbert spaces is guaranteed by the Gelfand-Maurin nuclear spectral theorem.

In strict mathematical terms, the rigged Hilbert space (RHS) $\mathcal{G}$ is defined as a triple of topological vector spaces $\mathcal{G}=(\Phi,\mathcal{H},\Phi^\times)$, called the Gelfand triple \cite{gelfand1964}, generated by an infinite dimensional separable Hilbert space $\mathcal{H}$ such that the denseness relation is $\Phi\subseteq\mathcal{H}\subseteq\Phi^\times$. Here, $\Phi$ is a (complete) nuclear Fr\'echet space, also called test-function space (not necessarily a Hilbert space, which enables to use the nuclear spectral theorem of Gelfand and Maurin \cite{gelfand1964,maurin1966}), and $\Phi^\times$ is the topological dual (or topological conjugate) of $\Phi$, that is the complete space of continuous anti-linear functionals on $\Phi$ (also called distribution or Schwartz space). The RHS structure includes an inductive limit of a sequence of topological spaces $\Phi^{(n)}$ in which the topologies get rapidly coarser with increasing $n$ \cite{blanchard2015,Antoine2021}, \eg we might think of a series of Sobolev spaces $\mathcal{W}_{k+1,2}\subseteq\mathcal{W}_{k,2}$, with $\mathcal{W}_{1,2}\subseteq L_2$. It is clear from this example that these subspaces have different norms (or semi-norms for the more general nuclear spaces  \cite{Antoine2021}).

The RHS formalism provides a correct mathematical foundation to Dirac's original bra and ket notation \cite{dirac1981book,Antoine2021}, used extensively in quantum theory. Needless to say that the Dirac delta ``function'' is a distribution required for the properties of continuum states belonging to $\Phi^\times$ rather than $\Phi$. To cite Bohm \cite{bohm1978}: \textit{The difference between [the rigged] Hilbert space formulation and the usual [von Neumann] Hilbert space formulation appears to be minor from a physicists point of view, but is essential from a mathematical point of view and leads far to tremendous mathematical simplification; in fact it justifies the mathematically undefined operations that the physicists have been accustomed to in their calculations}. The RHS formalism can easily be generalized to a rigged Fock space formalism required in quantum field theory \cite{bogolubov1975book,Antoniou1998,Celeghini2019}.\footnote{As quantum operators (in first quantization) act on Hilbert spaces, quantum field operators act on Fock spaces.} 

To be more specific, the Gelfand triple for the spectrum of the Dirac operator is chosen as $\Phi\subseteq$dom$(H_{\rm D})\subseteq L_2\subseteq\Phi^\times$. Vectors in $\Phi$ will be complex linear, and the vectors in $\Phi^\times$ are complex antilinear continuous functionals compatible with the scalar product in the underlying Hilbert space, \eg $F: \psi\in\Phi\rightarrow \mathbb{C}$. For example, we define the action $F\in\Phi^\times$ on $\psi\in\Phi$ as an extension to the Hilbert space product $F(\psi)=\braket{\psi}{F}=\braket{F}{\psi}^*$. This action is linear to the right and antilinear to the left.\footnote{There is always another rigged Hilbert space $\Phi\subseteq\mathcal{H}\subseteq\Phi^*$, where $\Phi^*$ is the dual space of $\Phi$ containing the continuous, linear functionals over $\Phi$ \cite{Roberts1966,Madrid2012}. Dirac's bras and kets belong to $\Phi^*$ and 
$\Phi^\times$, respectively, and both spaces are isometrically isomorph.} Continuity is defined such that if $\psi_n\rightarrow \psi$ for $n\rightarrow\infty$, $\psi_n\in$ $\Phi, \psi\in L_2$ then $F(\psi_n)\rightarrow F(\psi)$ in $\mathbb{C}$.

States with given quantum numbers ${n,\kappa}$, when diving into the negative energy continuum, have complex eigenenergies \cite{kuleshov2015vs,Godunov2017} and therefore move out of the natural domain of the self-adjoint operator $H_D$. We need to give such states belonging to a subset of distributions in the space $\Phi^\times$ a physical interpretation, however, we need  first to interpret such generalized eigenfunctions (or eigenfunctionals) from a mathematical point of view. 

Let $A:\Phi\rightarrow\Phi$ a (closed) linear operator and $A^\times:\Phi^\times\rightarrow\Phi^\times$ its natural extension of the usual adjoint operator $A^\dagger$ such that $F(A\psi)=A^\times F(\psi)=\braket{A\psi}{F}=\braket{\psi}{A^\times F}$ for all $\psi\in\Phi, F\in\Phi^\times$.
According to the Riesz representation theorem, for every $F\in\Phi^\times$ and linear operator $A$ there exist a unique complex function $\phi$ with complex eigenvalue $\lambda$, $A \phi=\lambda \phi$ such that for all $\psi\in\Phi$ we have $F(A\psi)=\braket{A\psi}{F}=\braket{\psi}{A^\times \phi} =\lambda^* \braket{\psi}{\phi}$ \cite{bohm1978}. We call $\phi$ the generalized eigenfunction of $A$ with respect to $F$. These extended eigenstates can be used in the normal way using Dirac's notation keeping in mind that these may not be normalizable and, in general, have complex eigenvalues. Of course, the scalar product $\braket{\psi}{\phi}$ always needs to be finite which may require a specific (semi)norm definition or regularization of integrals. \\\\

\begin{acknowledgements}
We acknowledge financial support by the Program Hubert Curien Dumont d'Urville New Zealand - France Science \& Technology Support Program number 43245QC, and the Marsden Fund of the Royal Society of New Zealand. We thank Profs. Trond Saue (Toulouse), Ephraim Eliav (Tel Aviv), W. H. Eugen Schwarz (Siegen), James Avery (Copenhagen) and Vladimir Shabaev (St. Petersburg) for many stimulating and critical discussions. This material is also based upon work supported by the U.S.\ Department of Energy, Office of Science, Office of Nuclear Physics under award number DE-SC0013365.
P.S. thanks ENS-PSL Research University for providing a one month invited professor position during the course of this work.
P.I. is a member of the Allianz Program of the Helmholtz Association, contract no EMMI HA-216 ``Extremes of Density and Temperature: Cosmic Matter in the Laboratory''.
\end{acknowledgements}

\bibliographystyle{spphys}
\bibliography{dirac}

\end{document}